\documentclass[aps,prb, amsmath,amssymb, showpacs, twocolumn, superscriptaddress]{revtex4-1}
\usepackage{subfigure}
\usepackage{graphicx}
\usepackage{dcolumn}
\usepackage{bm}
\usepackage{color}
\usepackage{natbib}

\begin{document}

\title{Localization in one dimensional lattices with non-nearest-neighbor hopping: Generalized Anderson and Aubry-Andr\'e models}

\author{J. Biddle}
\affiliation{Condensed Matter Theory Center, Department of
Physics, University of Maryland, College Park, Maryland 20742,
USA}

\author{D. J. Priour Jr}
\affiliation{University of Missouri, Kansas City, Department of
Physics, Kansas City, Missouri, 64110, USA}
\affiliation{Condensed Matter Theory Center, Department of
Physics, University of Maryland, College Park, Maryland 20742,
USA}

\author{B. Wang}
\affiliation{Condensed Matter Theory Center, Department of
Physics, University of Maryland, College Park, Maryland 20742,
USA}

\author{S. Das Sarma}
\affiliation{Condensed Matter Theory Center, Department of
Physics, University of Maryland, College Park, Maryland 20742,
USA}


\begin{abstract}
We study the quantum localization phenomena of noninteracting particles in
one-dimensional lattices based on tight-binding models with
various forms of hopping terms beyond the nearest
neighbor, which are generalizations of the famous Aubry-Andr\'e
and noninteracting Anderson model. For the case with
deterministic disordered potential induced by a secondary
incommensurate lattice (i.e. the Aubry-Andr\'e model), we identify a class of self-dual models,
for which the boundary between localized and extended eigenstates
are determined analytically by employing a generalized
Aubry-Andr\'e transformation. We also numerically investigate the
localization properties of nondual models with
next-nearest-neighbor hopping, Gaussian, and power-law decay
hopping terms. We find that even for these non-dual models, the
numerically obtained mobility edges can be well
approximated by the analytically obtained condition for
localization transition in the self dual models, as
long as the decay of the hopping rate with respect to
distance is sufficiently fast. For the disordered
potential with genuinely random character, we examine scenarios
with next-nearest-neighbor hopping, exponential, Gaussian, and
power-law decay hopping terms numerically. We find that the
higher-order hopping terms can remove the symmetry in the
localization length about the energy band center compared to the
Anderson model. Furthermore, our results demonstrate that for the
power-law decay case, there exists a critical exponent below which mobility edges can be found.  Our theoretical results could, in principle, be directly tested in shallow atomic optical lattice systems enabling non-nearest-neighbor hopping.
\end{abstract}

\maketitle
\section{Introduction}
Quantum transport of matter waves in the presence of disorder has
long been a topic of interest for condensed-matter physicists. For
one-dimensional (1D) noninteracting systems, one of the oldest
and most extensively studied models for quantum transport is the
single-band, nearest-neighbor (nn) tight-binding model,
\begin{equation}
t_1(u_{n+1}+u_{n-1})+V_nu_n = E u_n, \label{eq:nnmodel}
\end{equation}
where $t_1$ is the nn hopping integral term and $V_n$ is the
on-site disordered potential \cite{Kramer93}. One of the main
merits of Eq. (\ref{eq:nnmodel}) is its simple form, which lends
itself to fast numerical analysis, as well as exact theoretical
statements on quantum transport in some cases.  Arguably the most
well known of the latter is when $V_n$ is bounded, uncorrelated
disorder (i.e., the noninteracting Anderson model \cite{Anderson}),
where it can be shown that all eigenstates of the system are
localized for any nonzero potential strength. Another well-known
example is the 1D incommensurate problem [in
particular, $V_n = V \cos(2\pi\alpha n + \delta)$,
where $\alpha$ is irrational] studied by Aubry and Andr\'e where
all eigenstates of the system are extended for potential strength
below a threshold value ($V_c=2t_1$) and localized
above this threshold \cite{Aubry}.  Conversely, this simple form
given by Eq. (\ref{eq:nnmodel}) makes it difficult to study
directly in solid-state systems, where the disorder is difficult
to control reliably and interactions can rarely be ignored.
However, recent advances in the manipulation of ultracold atoms
in optical lattices provide a powerful tool for directly examining
quantum transport in fundamental models such as Eq.
(\ref{eq:nnmodel}). This has been most notably demonstrated in
recent experiments conducted by Billy $et$ $al.$, where Anderson
localization was directly observed in a Bose-Einstein condensate
(BEC) subjected to a random laser speckle potential
\cite{Billy08}, and similarly in experiments conducted by Roati
$et$ $al.$, where Aubry-Andr\'e duality was directly observed in a
BEC loaded into a quasi-periodic optical lattice \cite{Roati08,*Modugno2009}.
These feats, which previously eluded experimental observation for decades,
illustrate the potential of ultracold atomic systems to
experimentally probe fundamental quantum phenomena.
Considering the degree of control afforded in
ultracold atomic systems, we can systematically relax, in a
controlled manner, basic assumptions inherent in Eq.
(\ref{eq:nnmodel}) and directly study their influence on quantum
transport and how it differs from the well-known Anderson and
Aubry-Andr\'e results. It is this potential in ultracold atomic
systems that motivates our present work, where we examine quantum
transport in tight-binding, non-interacting models that are
extensions of Eq. (\ref{eq:nnmodel}). In particular, we relax the
nn tight-binding assumption and theoretically examine transport in
models with long and short-range hopping schemes.  Such models
should be representative of diffuse gases of ultracold atoms
loaded into fairly shallow optical lattices.

We can go beyond the nn coupling assumption while still remaining
in the tight-binding framework by including higher-order hopping
terms. The general form of such a model with on-site disorder is
given by
\begin{equation}
\sum_{m} t_m u_{n+m} + V_n u_n = E u_n, \label{eq:genmodel}
\end{equation}
where the tight-biding terms $t_{m}$ may assume a
variety of forms. There is a small, but growing body of numerical
and analytical work examining transport in the context
of this generalized model \cite{biddle, Boers07, Riklund86, Moura,
Malshev, Xiong, Rodriguez03, DasSarma86}. In an effort to extend this growing
body of work, we wish to investigate transport through various
forms of Eq. (\ref{eq:genmodel}) with both incommensurate and
random on-site potentials. In particular, we study transport in
tight-binding models with next-nearest-neighbor (nnn) hopping (the $t_1-t_2$ model) and models incorporating an infinite number of  hopping
terms that decay by an exponential, Gaussian, or inverse
power law. Since we examine both deterministic
bichromatic potentials and random potentials,
this report is divided into two sections. The
first section examines the case of incommensurate potentials by
first studying the exponential hopping model, which has been shown to have an analytical mobility edge\cite{biddle-2009}, then
approximately extrapolating these results to predict the mobility
edges in the nnn, Gaussian, and inverse-power-law hopping models \cite{[{Mobility edges or the existence of extended states in disordered 1D systems are not unique to tight-binding models with higher order hopping terms. See, for example, }] Soukoulis82, *DasSarma88, *XieDasSarma88, *Scarola06}. The
second section numerically examines Eq. (\ref{eq:genmodel}) with nnn,
exponential, Gaussian, and inverse-power-law hopping for randomly
disordered potentials and highlights how localization in these
models is markedly different compared to
what is seen in the case of the nn Anderson model.

\section{Incommensurate Potentials}

\subsection{Self Dual models}
One of the first models to examine quantum transport in 1D
incommensurate potentials is the so-called Aubry-Andr\'e (AA)
model\cite{Aubry}. In this model, the on-site term in Eq.
(\ref{eq:nnmodel}) is a cosine with frequency incommensurate with
the primary lattice: $V_n = V \cos(2\pi\alpha n + \delta)$ where
$\alpha$ is an irrational number and $\delta$ is an
arbitrary phase. It has been shown that this model is self-dual
under the transformation,
\begin{equation}
u_n = \sum_{m} f_m e^{im(2\pi\alpha n+\delta)}e^{i\beta n},
\label{eq:AAtran}
\end{equation}
when $V = 2t_1$ (i.e. $f_m$ satisfies the same eigenvalue equation
as $u_n$). Also, under Eq. (\ref{eq:AAtran}), if the eigenstate $u_n$
is spatially localized, then the eigenstate of the dual problem,
$f_m$, is spatially extended and vice versa. Using this property
and the Thouless formula for incommensurate potentials
\cite{Thouless72}, it is argued that all eigenstates
are localized for $V>2t_1$ and extended for $V<2t_1$. The case
where $V=2t_1$ is especially interesting and has been shown to
yield a singular continuous eigenspectrum, which forms a Cantor
set in the thermodynamic limit \cite{Bellissard}. Furthermore AA duality can be shown to have a more general form \cite{Sokoloff85}. Consider the model,
\begin{equation}
\sum_{m}t_m u_{n+m} + V_0 v_n u_n = E u_n. \label{eq:AAgen}
\end{equation}
If the on-site potential and the hopping terms satisfy the
relation,
\begin{equation}
v_n = \sum_{m} t_m e^{i m (2\pi\alpha n + \delta)},
\label{eq:AAgencond}
\end{equation}
then this model also possesses an AA-like duality.

Other models have been shown to possess self-duality similar to the AA model \cite{DasSarma86, biddle-2009}. The particular model we consider here, given by
\begin{equation}
Eu_n = \sum_{n'\ne n} te^{-p|n-n'|}u_{n'}+V\cos(2\pi\alpha n +
\delta)u_n, \label{eq:model}
\end{equation}
is especially interesting because its self-duality condition naturally predicts energy-dependent mobility edges, in contrast to the AA model\cite{biddle-2009}.  To see this we now reiterate the results given in the previous work on this model to show that Eq. (\ref{eq:model}) does have a self-dual condition and we expand on the previous work to show that this self-dual condition does indeed define a mobility edge.

We begin by defining the parameter, $p_0>0$ such that
\begin{eqnarray}
(E+t)-V\cos(2\pi\alpha n + \delta) = \Omega T_n, \label{eq:param} \\
T_n = \frac{\cosh(p_0)-\cos(2\pi\alpha n + \delta)}{\sinh(p_0)}, \label{eq:Tn} \\
\Omega = \sqrt{(E+t)^2-V^2}\label{eq:w}.
\end{eqnarray}
Then it follows that $(E+t)/V = \cosh(p_0)$ and we can rewrite
Eq. (\ref{eq:model}) as
\begin{equation}
\Omega T_n u_n = \sum_{n'}te^{-p|n-n'|}u_{n'}. \label{eq:compact}
\end{equation}
If we now consider the transformation
\begin{equation}
\tilde{u}_m = \sum_{n}e^{im(2\pi\alpha n + \delta)}T_nu_n,
\label{eq:tran}
\end{equation}
and note that for $p>0$ we have the identity
\begin{equation}
T_n^{-1} = \sum_{m}e^{-p|m|}e^{im(2\pi\alpha n + \delta)},
\label{eq:Tnexpand}
\end{equation}
then it follows that the state $\tilde{u}_m$ satisfies the
equation
\begin{equation}
\Omega\tilde{T}_m\tilde{u}_m =
\sum_{m'}te^{-p_0|m-m'|}\tilde{u}_{m'}, \label{eq:dualmodel}
\end{equation}
where $\tilde{T}_m$ is given by
\begin{equation}
\tilde{T}_m = \frac{\cosh(p)-\cos(2\pi\alpha m +
\delta)}{\sinh(p)}.
\end{equation}
We see that Eq. (\ref{eq:compact}) is self-dual under the
transformation Eq. (\ref{eq:tran}) when $p=p_0$, or equivalently
$\cosh(p)=\cosh(p_0)$ for $p,p_0>0$. Thus, the duality condition
for Eq. (\ref{eq:model}) is given by
\begin{equation}
\cosh(p) = \frac{E+t}{V}. \label{eq:condition}
\end{equation}
Since the transformation given by Eq. (\ref{eq:tran})
transforms localized states to extended states and vice versa
[similar to Eq. (\ref{eq:AAtran})], we expect that the
eigenstates of the system are critical (neither localized nor
extended) when Eq. (\ref{eq:condition}) is
satisfied.

Similar to the arguments made by Aubry and Andr\'e for the AA model, we
now expand on the conjecture made in the previous work\cite{biddle-2009} and argue that the eigenstates of Eq. (\ref{eq:compact}) are localized
for $p>p_0$ and extended for $p<p_0$ [i.e. that Eq. (\ref{eq:condition}) does, indeed, define a mobility edge]. Since the Thouless formula
used by Aubry and Andr\'e was derived for models of the form of
Eq. (\ref{eq:nnmodel}), we can not use it for our particular model.
Therefore, our first step is to generalize the idea of the
Thouless formula for the long-range hopping model.
To do so, we treat $\Omega$ as the eigenvalue
and consider the Green's matrix,
\begin{eqnarray}
G(\Omega)_{m,n} = (\Omega \overline{\mathrm{I}} - \overline{\mathrm{H}})_{m,n}^{-1} \\
= \frac{\mathrm{cofactor}(\Omega \overline{\mathrm{I}} -
\overline{\mathrm{H}})_{m,n}} {\prod_{\beta}(\Omega -
\Omega_{\beta})}, \label{eq:Gnm}
\end{eqnarray}
where the cofactor is the appropriately signed determinant with
the $m$th row and $n$th column removed and $\overline{\mathrm{H}}$
is the Hamiltonian corresponding to the eigenvalue equation given
in Eq. (\ref{eq:compact}) where we have set $t=1$ without loss of
generality; $\overline{\mathrm{I}}$ is the identity
matrix. Assuming a nondegenerate eigenspectrum, the Green's
matrix has a simple pole for each eigenvalue, $\Omega_{\mu}$.
Since, by definition, the residue of $G(\Omega^{(\mu)})_{m,n}$ is
the product of the $m$th and $n$th elements of the
eigenvector (i.e. $Res(G(\Omega)_{m,n}, \Omega_{\mu}) =
u_m^{(\mu)}u_n^{(\mu)}$)\cite{Thouless72}, we have for the product of the
first and last elements of the eigenvector
\begin{equation}
u_1^{(\mu)}u_N^{(\mu)} = \frac{\mathrm{cofactor}(\Omega_{\mu}
\overline{\mathrm{I}} - \overline{\mathrm{H}})_{1,N}}
{\prod_{\beta \ne \mu}(\Omega_{\mu} - \Omega_{\beta})}.
\label{eq:product}
\end{equation}
If the state is exponentially localized about the site $n'$, then
we expect $u_n \sim \exp(-\gamma |n'-n|)$, where $\gamma \geq 0$ is
the characteristic (or Lyapunov) exponent. Therefore the product
$u_1 u_N \sim \exp(-\gamma (N-1))$. Thus, the characteristic
exponent for large $N$ is given by
\begin{eqnarray}
\gamma(\Omega_{\mu}) &=& \lim_{N\to\infty} -(N-1)^{-1}\ln|u_1^{(\mu)}u_N^{(\mu)}| \nonumber\\
&=&\lim_{N\to\infty} (N-1)^{-1}(\sum_{\beta\ne\mu}
\ln|\Omega_{\mu}-\Omega_{\beta}| \nonumber\\ &-&
\ln|\mathrm{cofactor}(\Omega_{\mu} \overline{\mathrm{I}} -
\overline{\mathrm{H}})_{1,N}|). \label{eq:genthf}
\end{eqnarray}
This is the generalized Thouless relation for the characteristic
exponent of a wavefunction. For the case where
$\overline{\mathrm{H}}$ is given by Eq. (\ref{eq:compact}), the cofactor
takes on the form:
\begin{equation}
\mathrm{cofactor}(\Omega_{\mu}
\overline{\mathrm{I}}-\overline{\mathrm{H}})_{1,N} =
\Omega_{\mu}^{N-2} e^{-(N-1)p}T_N^{-1}. \label{eq:cofact}
\end{equation}
Then we have for the characteristic exponent,
\begin{equation}
\gamma(\Omega_{\mu}) = p - \ln|\Omega_{\mu}| + \lim_{N\to\infty}
(N-1)^{-1}\sum_{\beta\ne\mu} \ln|\Omega_{\mu}-\Omega_{\beta}|. \label{eq:exp}
\end{equation}

We now compare the characteristic exponents of the eigenvectors of
Eq. (\ref{eq:compact}), which we denote as $\gamma(\Omega)$, with the
exponents of the dual problem Eq. (\ref{eq:dualmodel}), denoted as
$\tilde{\gamma}(\Omega)$. Since the eigenvalue $\Omega$ is not
changed by the transformation given by Eq. (\ref{eq:tran}), we expect the
summation term on the right-hand side of Eq. (\ref{eq:exp}) to be equal for both
Eq. (\ref{eq:compact}) and Eq. (\ref{eq:dualmodel}). Therefore, the
characteristic exponents have the following relation:
\begin{equation}
\gamma(\Omega) = \tilde{\gamma}(\Omega) + (p - p_0).
\label{eq:expcon}
\end{equation}
Considering the case when $p>p_0$, since
$\tilde{\gamma}(\Omega)\geq0$, it follows that
$\gamma(\Omega) > 0$ and therefore the eigenstate, $u_n$ is
localized while the dual state, $\tilde{u}_n$ is extended.
Similarly, when $p_0>p$, we can argue that
$\tilde{\gamma}(\Omega)>0$ and therefore the dual state,
$\tilde{u}_n$, is localized while $u_n$ is extended. Therefore,
returning to the original problem given by Eq. (\ref{eq:model}) and
using the fact that $\cosh(p_0)$ is a monotonically increasing
function of $p_0 >0$, it follows that the eigenstates are
localized for $(E+t)/V<\cosh(p)$ and extended for
$(E+t)/V>\cosh(p)$.

\begin{figure*}
\subfigure[]{
\includegraphics[type=pdf,ext=.pdf,read=.pdf,width=.47\textwidth]{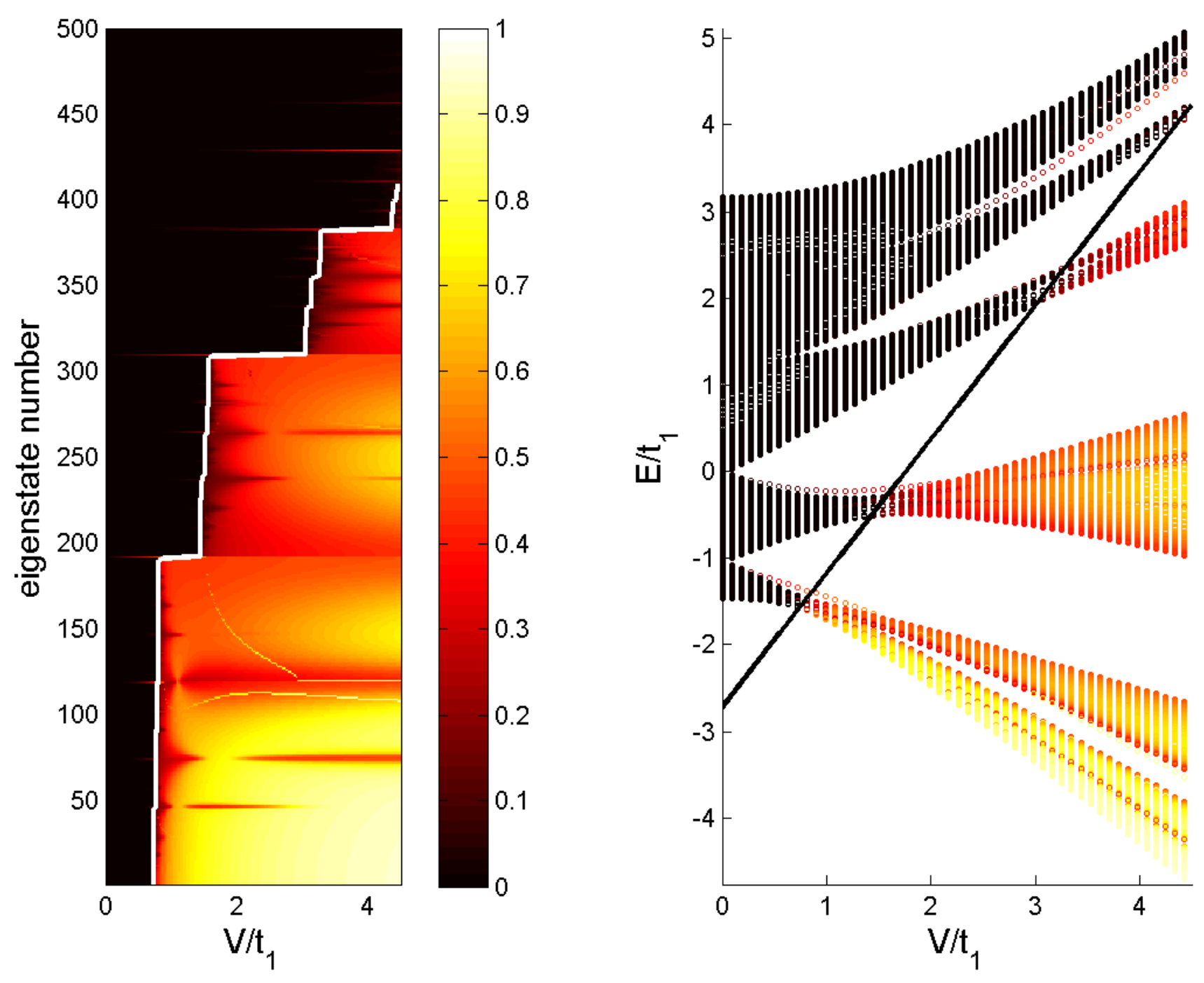}
\label{fig:subipr1} } \subfigure[]{
\includegraphics[type=pdf,ext=.pdf,read=.pdf,width=.47\textwidth]{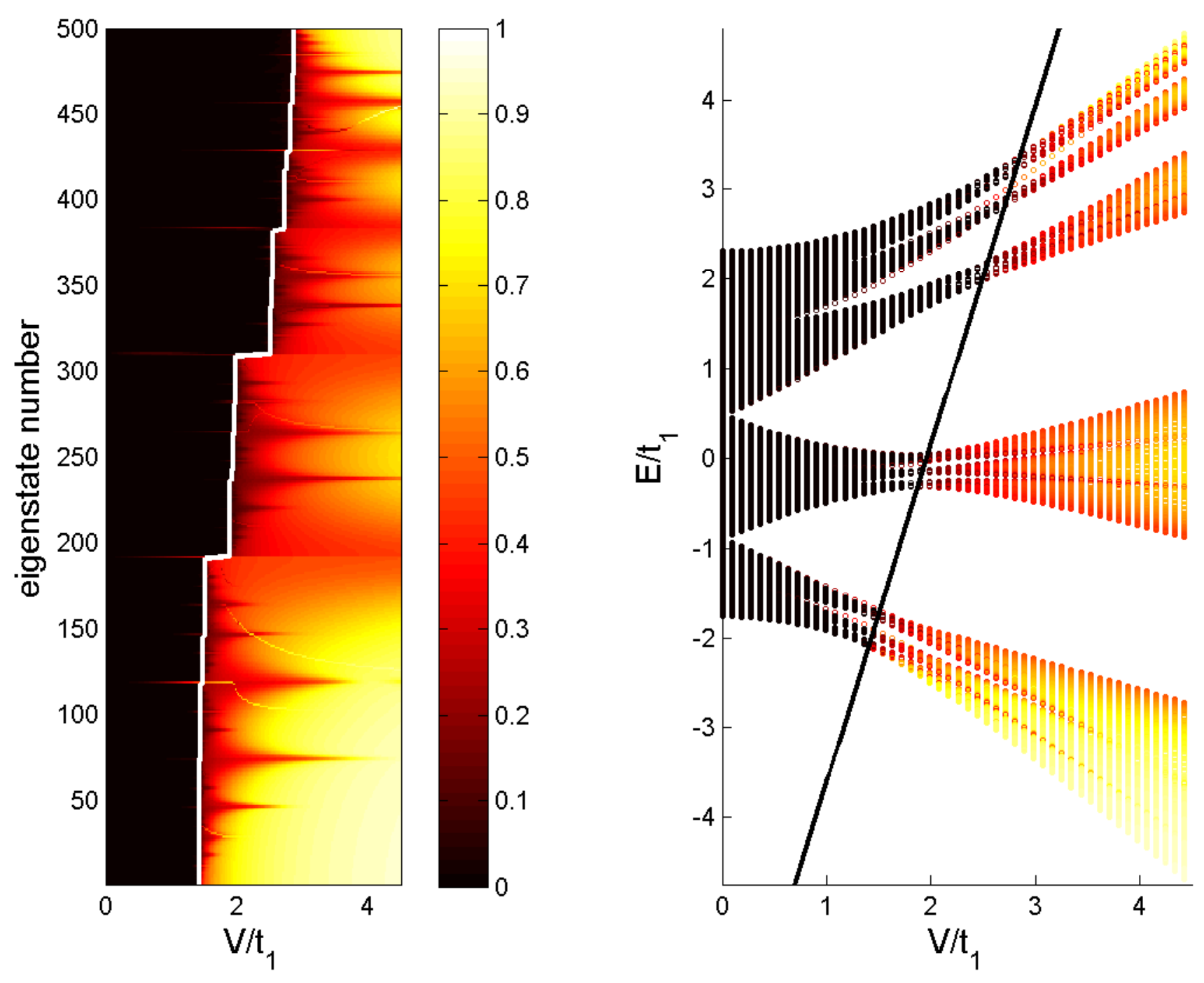}
\label{fig:subipr2} } \subfigure[]{
\includegraphics[type=pdf,ext=.pdf,read=.pdf,width=.47\textwidth]{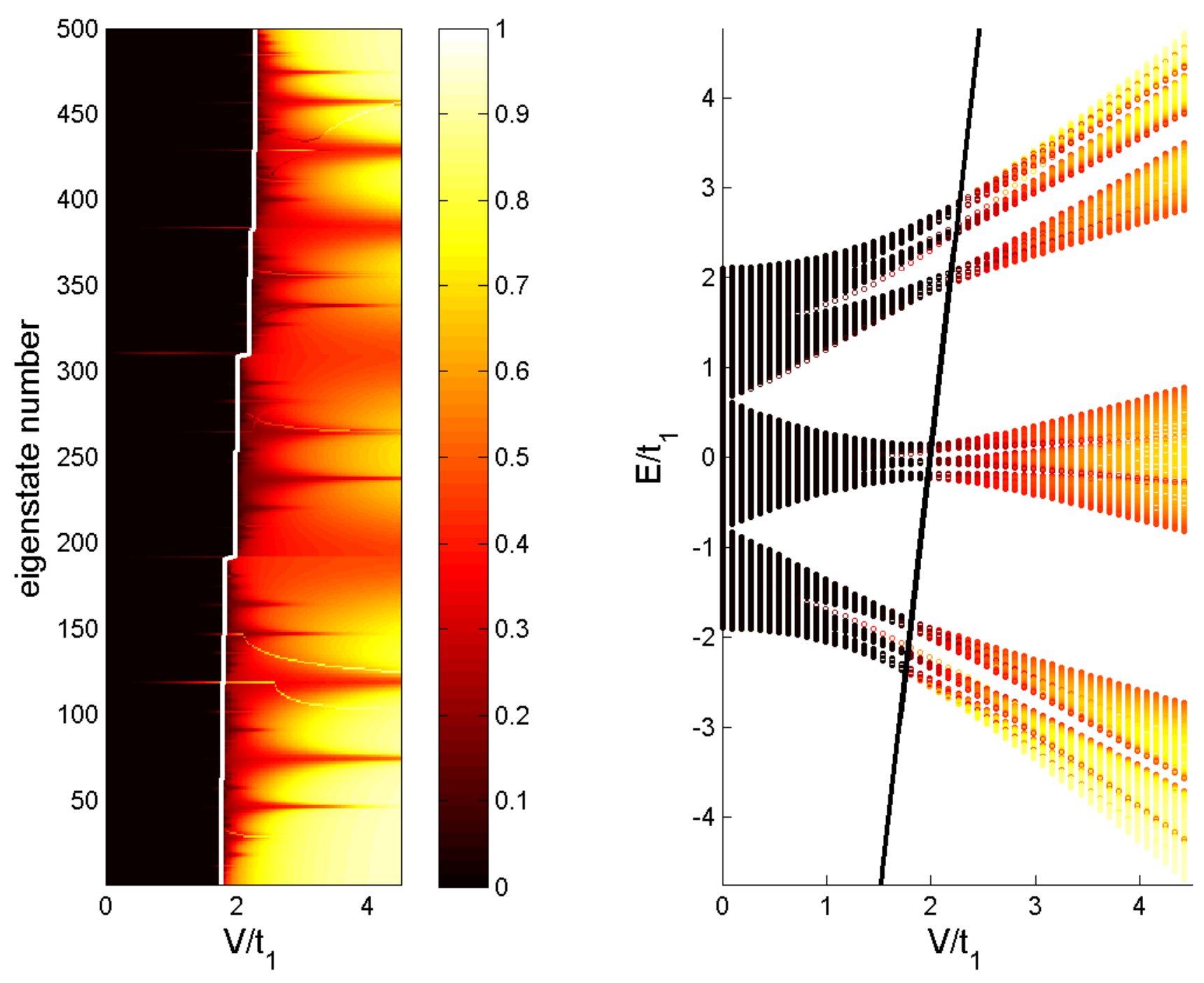}
\label{fig:subipr3} } \subfigure[]{
\includegraphics[type=pdf,ext=.pdf,read=.pdf,width=.47\textwidth]{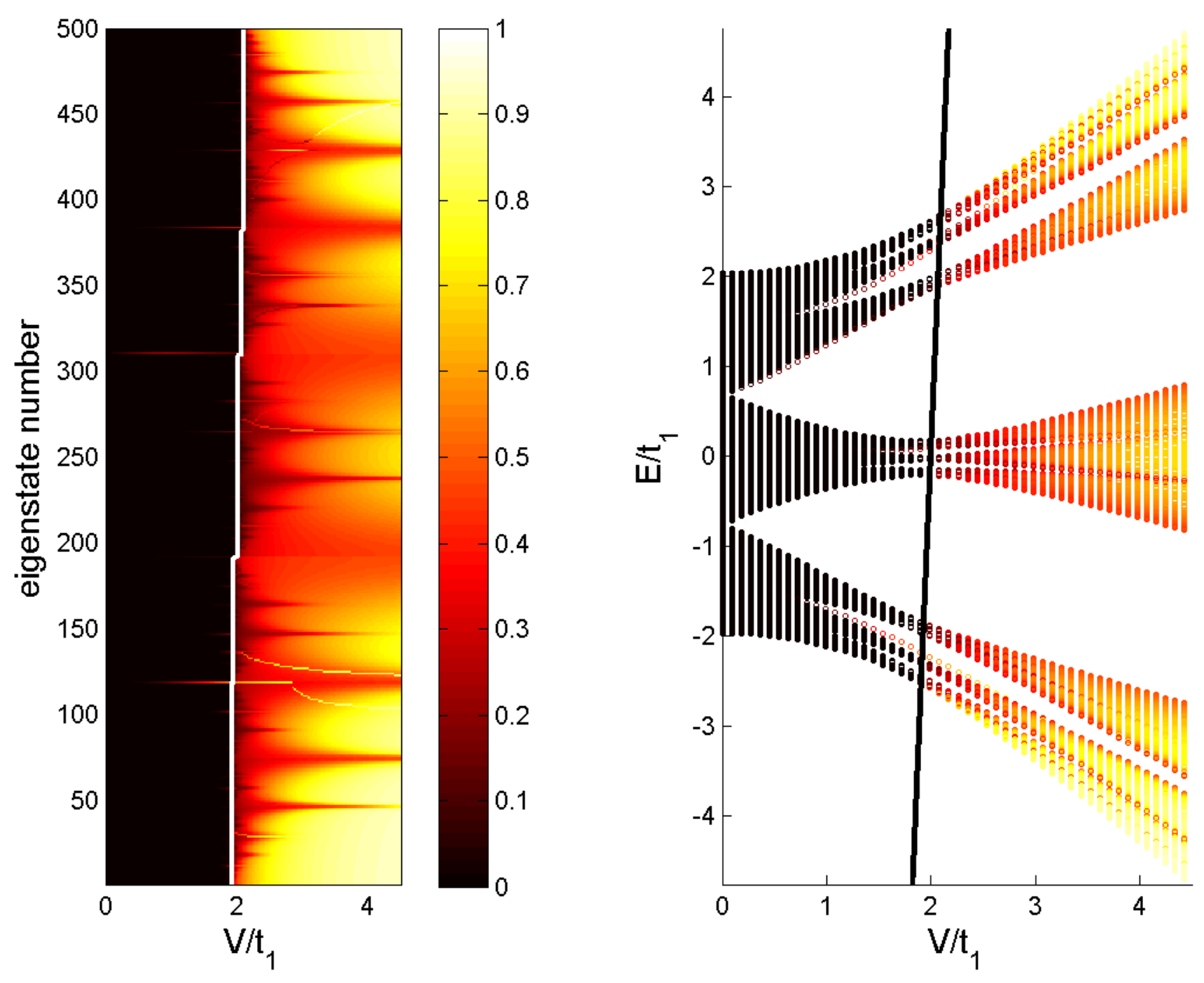}
\label{fig:subipr4} } \caption{\label{fig:IPRtbm} Energy
eigenvalues and inverse participation ratios of Eq. (\ref{eq:model})
with 500 lattice sites and $\alpha = (\sqrt{5}-1)/2$ for
\subref{fig:subipr1} $p=1$, \subref{fig:subipr2} $p=2$,
\subref{fig:subipr3} $p=3$, \subref{fig:subipr4} $p=4$. The solid
line represents the analytical boundary between spatially
localized and spatially extended states.}
\end{figure*}
\begin{figure}
\includegraphics[type=pdf,ext=.pdf,read=.pdf,width=.5\textwidth]{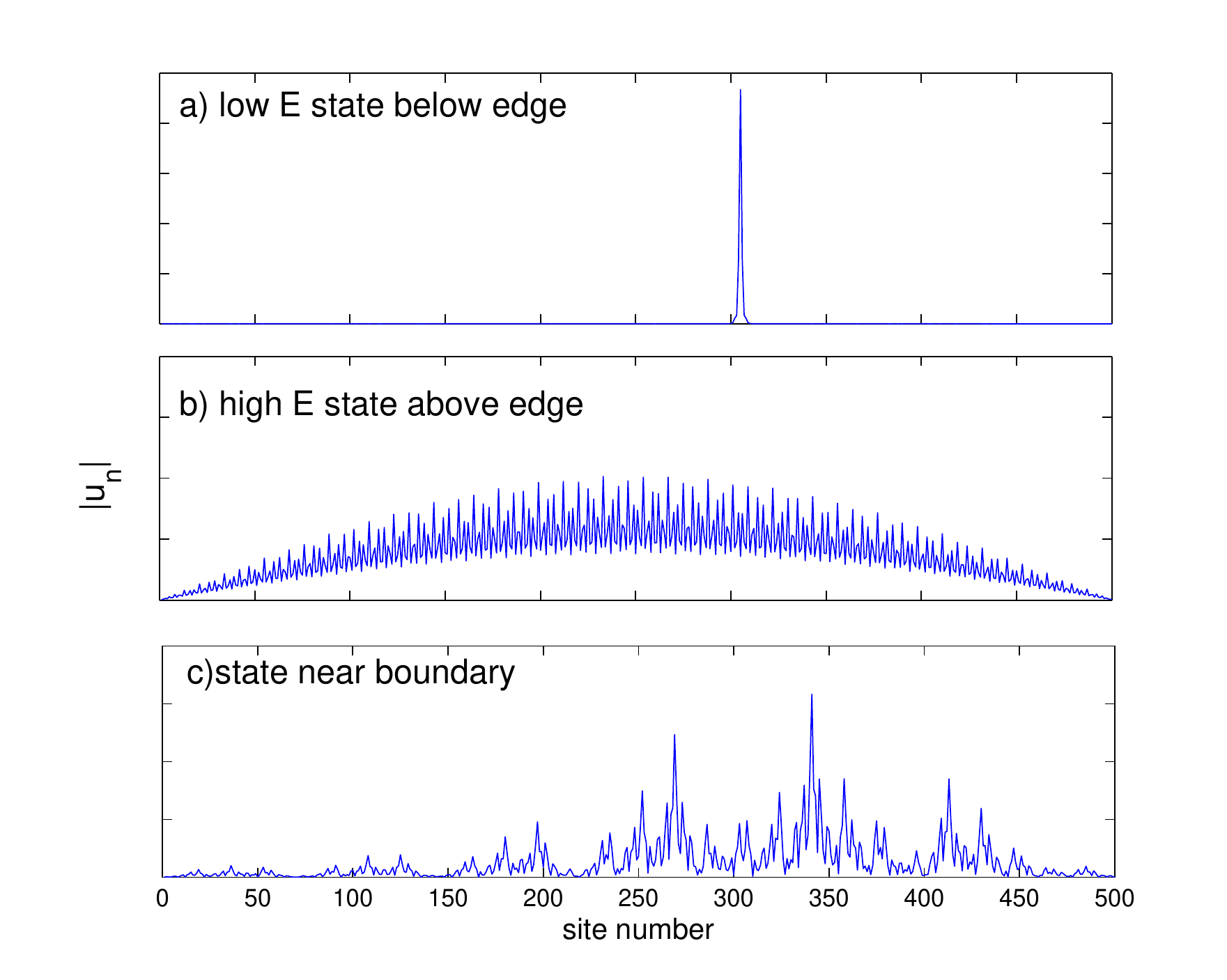}
\caption{\label{fig:statestbm} Eigenstates of Eq. (\ref{eq:model})
with 500 lattice sites, $\alpha = (\sqrt{5}-1)/2$, $V = 1.8$, and
$p=1.5$ for different energy eigenvalues: a) low energy localized
state below the mobility edge b) high energy extended state above
the mobility edge c) critical state near the mobility edge.}
\end{figure}

Similar to the AA model, the self-duality described above has a
general form, considering, again, a model of the form given in
Eq. (\ref{eq:AAgen}). The model will have this form of self-duality if the
on-site potential and the hopping terms satisfy the relation:
\begin{equation}
\frac{A}{B-v_n} = \sum_{m} t_m e^{i m (2\pi\alpha n + \delta)}
\label{eq:gencond}
\end{equation}
where $A$ and $B$ are constants. In particular, the constant $B$ gives the slope of the the duality condition (i.e., $B = E/V_0$). 

We now numerically examine localization in Eq. (\ref{eq:model}) [and
equivalently Eq. (\ref{eq:compact})] by calculating the inverse
participation ratio (IPR) of the wave functions, given by:
\begin{equation}
\text{IPR}^{(i)}=\frac{\sum_n|u^{(i)}_n|^4}{({\sum_n|u^{(i)}_n|^2})^2},
\label{eq:ipr}
\end{equation}
where the superscript $i$ denote the $i$ eigenstate. The IPR
approaches zero for spatially extended wave functions and is finite
for localized wave functions,
 and hence has a useful diagnostic role.
Fig. \ref{fig:IPRtbm} plots energy eigenvalues (or eigenstate
number) and the IPR of the corresponding wave functions for Eq.
(\ref{eq:model}) as a function of potential strength, $V$, with
$\alpha = (\sqrt{5}-1)/2$ and $p=1$, 2, 3, or 4. The solid
curves in the figures represent the boundary given
in Eq. (\ref{eq:condition}). From the figure we see that IPR
values are approximately zero for energies above the boundary and
are finite for energies below the boundary. This supports our
assertion that the mobility edge is, indeed, given by Eq.
(\ref{eq:condition}).

In Fig. \ref{fig:statestbm}, we directly examine sample eigenstates in
each regime (i.e., localized, extended, and near the mobility edge) for
$p=1.5$ and $V = 1.8$. We see that the wavefunction is localized
for low energies [Fig \ref{fig:statestbm}a], extended for high
energies [Fig \ref{fig:statestbm}b], and critical near the
boundary [Fig \ref{fig:statestbm}c].
\begin{figure}
\subfigure[]{
\includegraphics[type=pdf,ext=.pdf,read=.pdf,width=.47\textwidth]{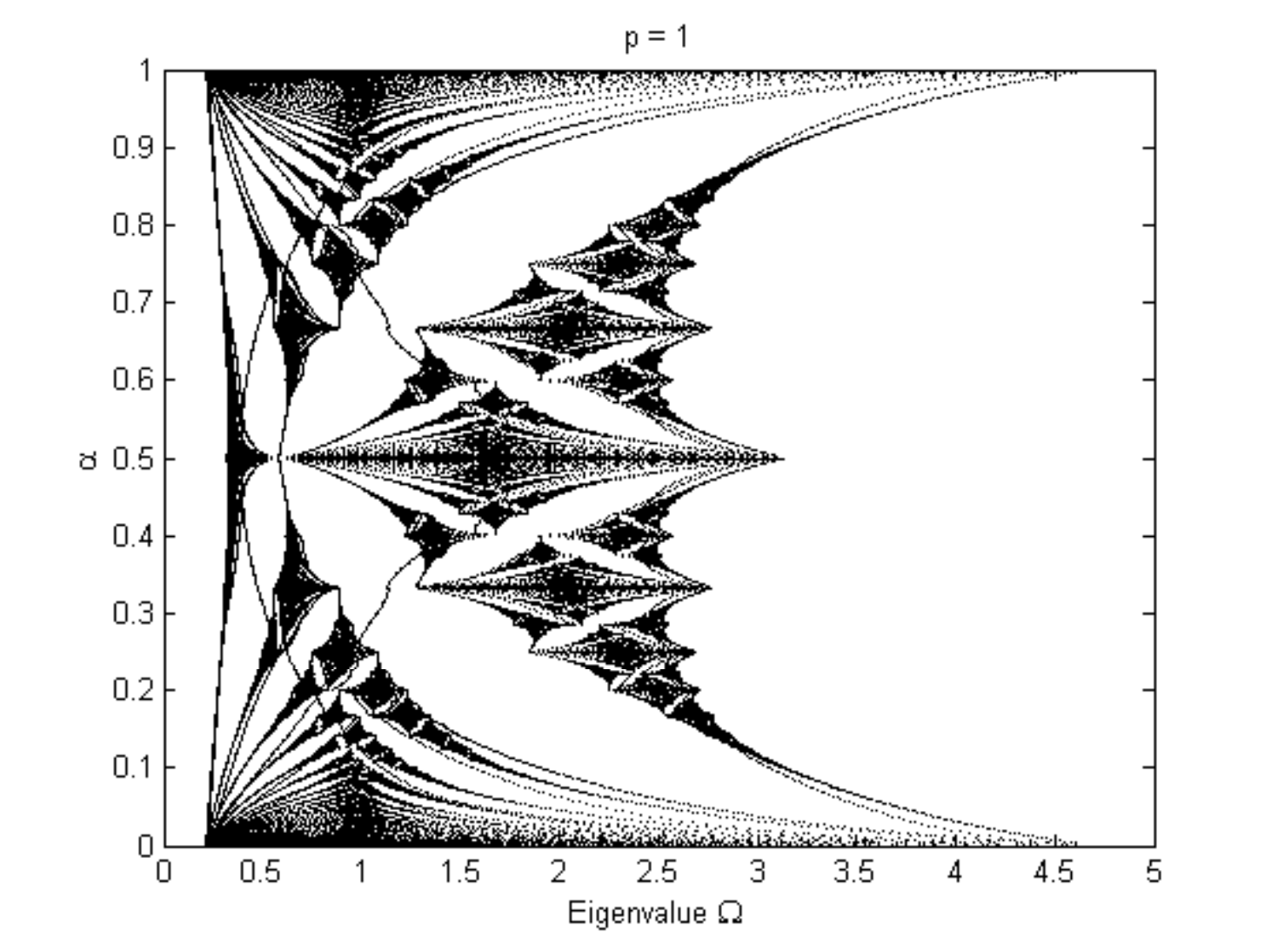}
\label{fig:subbfly1} } \subfigure[]{
\includegraphics[type=pdf,ext=.pdf,read=.pdf,width=.47\textwidth]{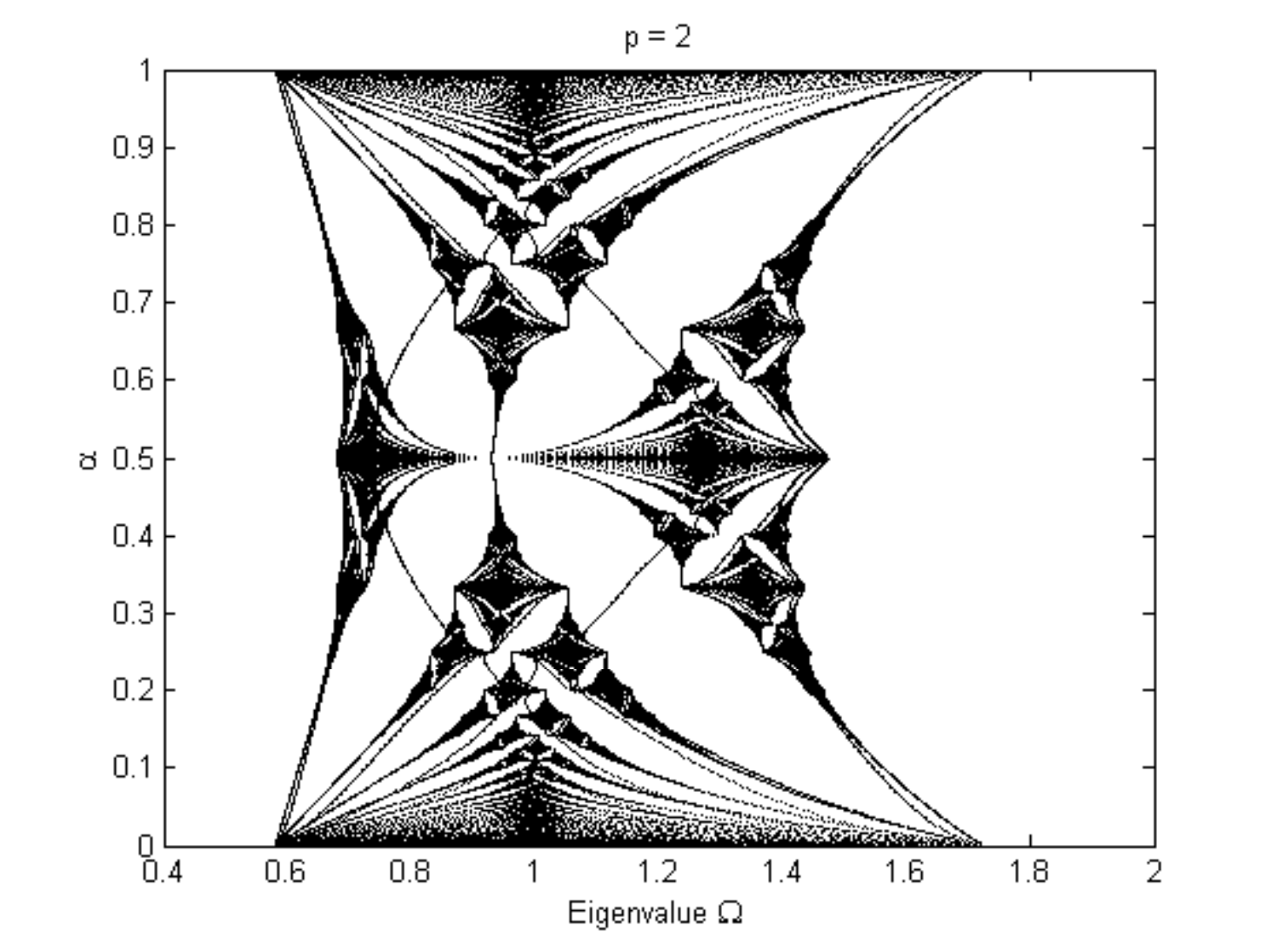}
\label{fig:subbfly2} } \subfigure[]{
\includegraphics[type=pdf,ext=.pdf,read=.pdf,width=.47\textwidth]{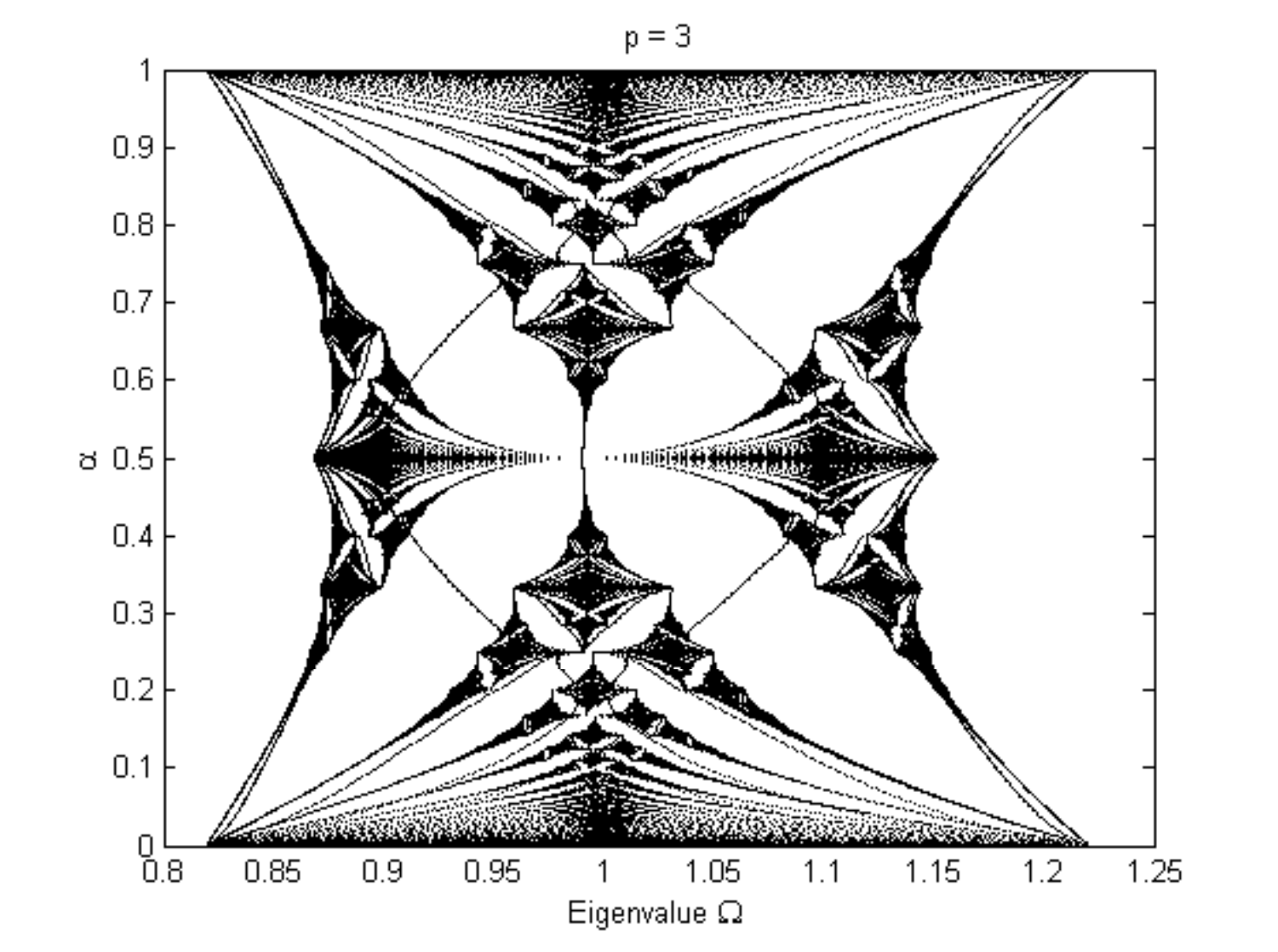}
\label{fig:subbfly3} } \caption{\label{fig:butterfly}
Eigenspectrum of Eq. (\ref{eq:compact}) with varying $\alpha$ for
\subref{fig:subbfly1} $p = 1$ , \subref{fig:subbfly2} $p = 2$, and
\subref{fig:subbfly3} $p = 3$.}
\end{figure}

Finally, we examine the eigenvalues of Eq. (\ref{eq:compact}) for
different values of $\alpha$ at the duality point ($p=p_0$) where
we expect the eigenspectrum to form a fractal set for large $N$. The
results of this are given in Fig. \ref{fig:butterfly}. In the
figure, we see that for large values of $p$, the eigenspectrum
closely resembles the well-known Hofstadter's butterfly which
results from the solutions of Harper's equation \cite{Harper55, Hofstadter76}. For smaller
values of $p$, however, we see a generalized form of Hofstadter's
butterfly that is not symmetrical about the the band center, but
skewed toward lower eigenvalues.  The self-similarity in the
figure suggests that the eigenspectrum does, indeed, form a Cantor
set at the duality point in the thermodynamic limit.

\subsection{Nondual models}

\begin{figure*}
\subfigure[]{
\includegraphics[type=pdf,ext=.pdf,read=.pdf,width=.47\textwidth]{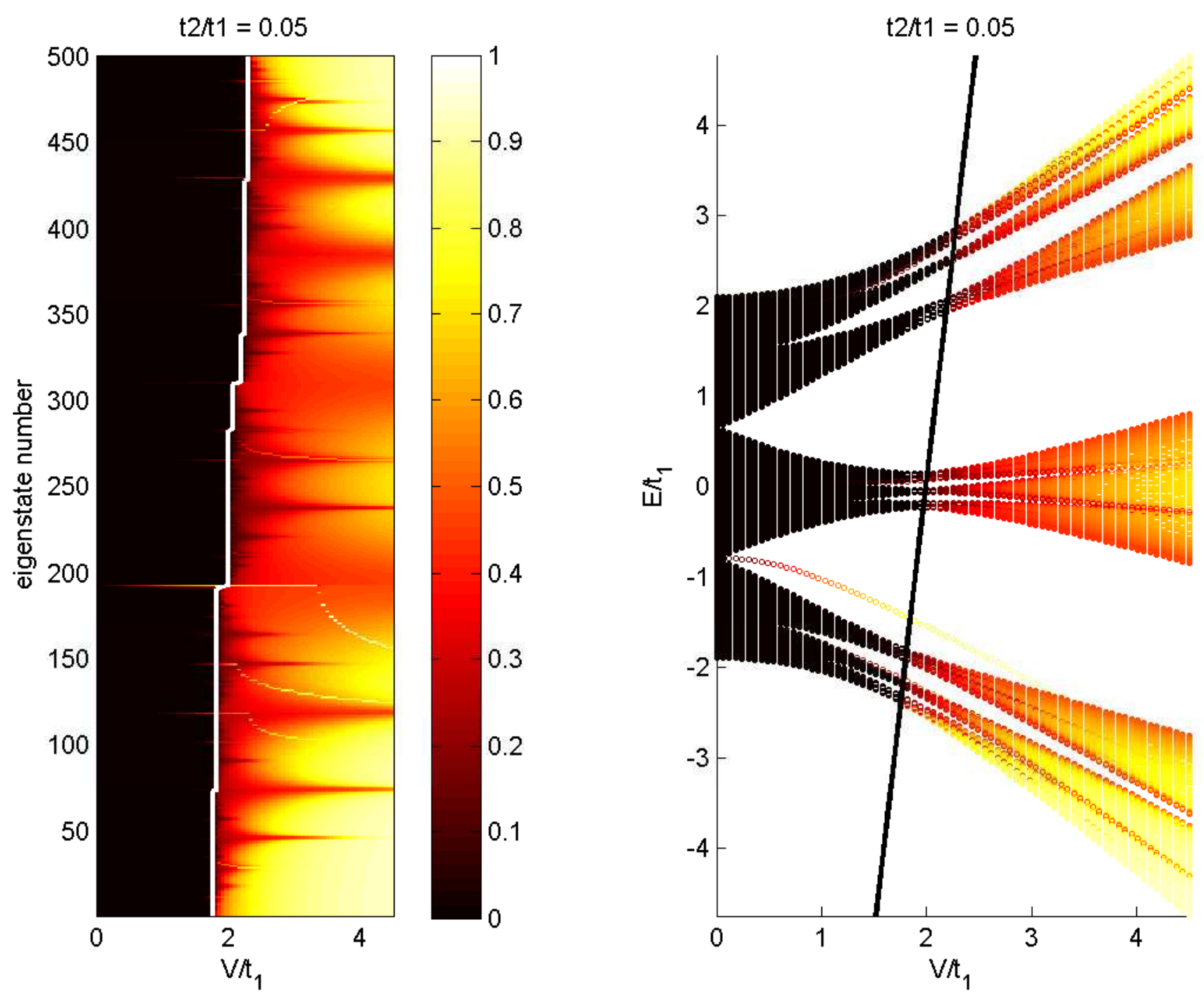}
\label{fig:subnnn1}

} \subfigure[]{
\includegraphics[type=pdf,ext=.pdf,read=.pdf,width=.47\textwidth]{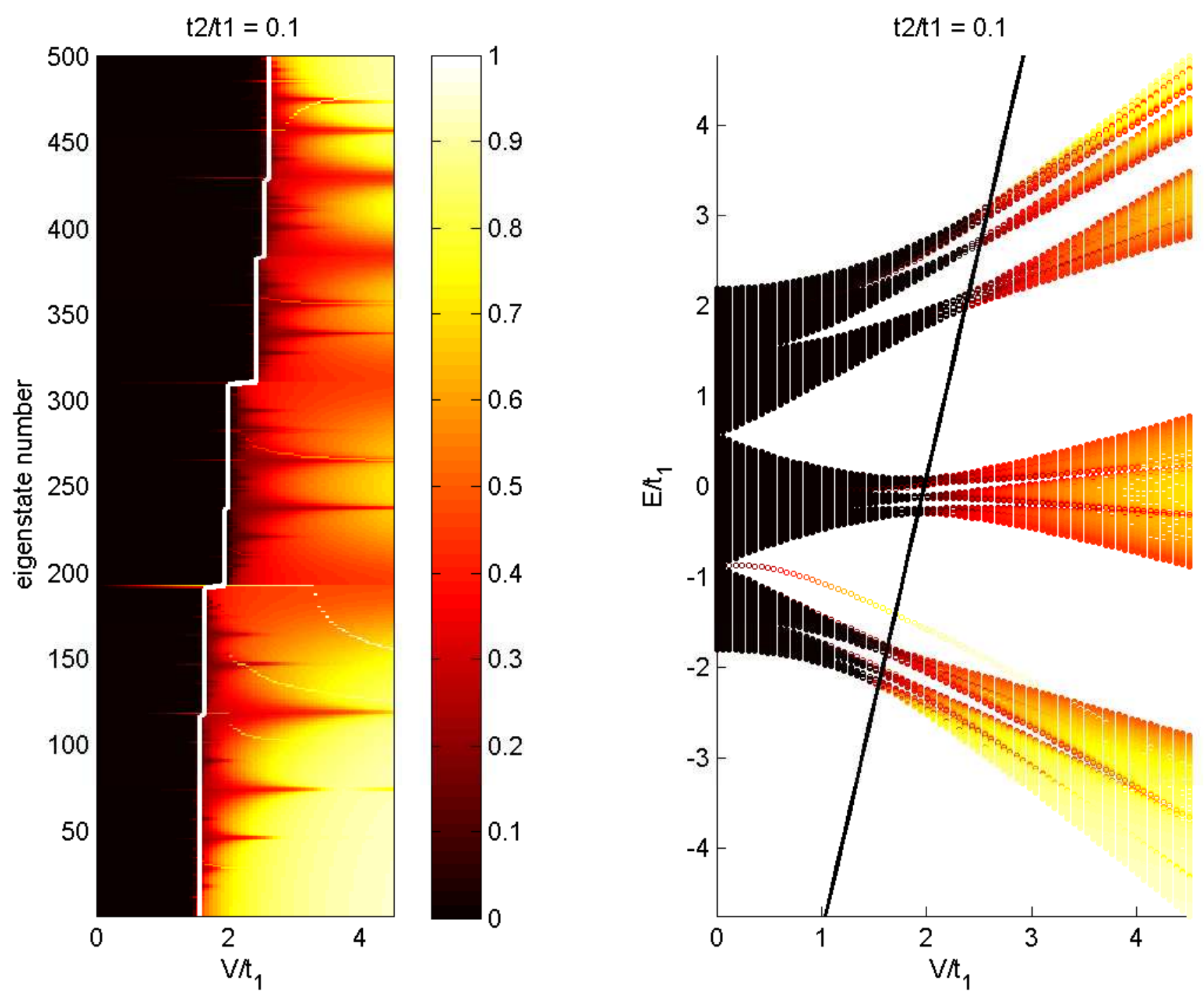}
\label{fig:subnnn2}

} \subfigure[]{
\includegraphics[type=pdf,ext=.pdf,read=.pdf,width=.47\textwidth]{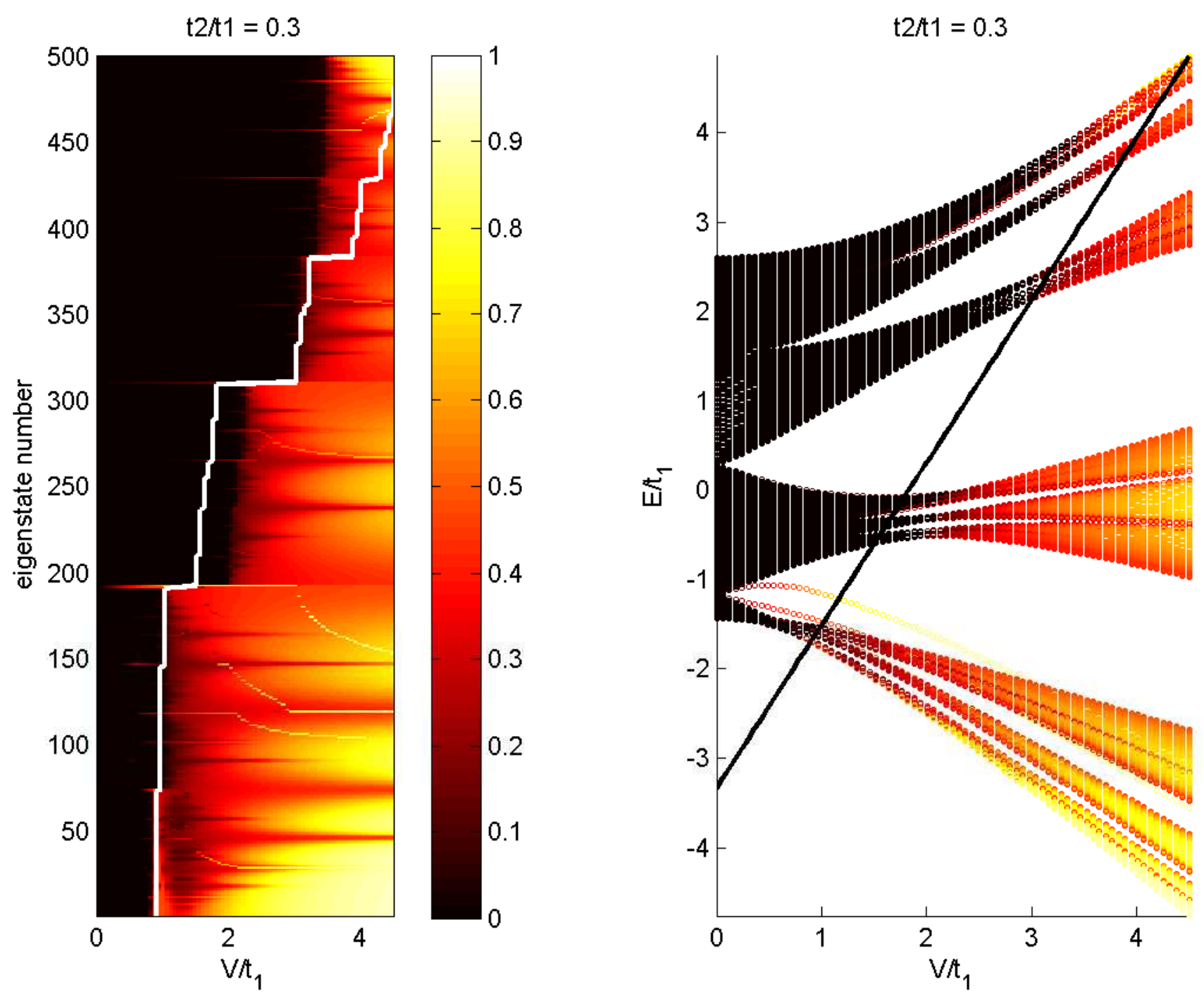}
\label{fig:subnnn3}

} \subfigure[]{
\includegraphics[type=pdf,ext=.pdf,read=.pdf,width=.47\textwidth]{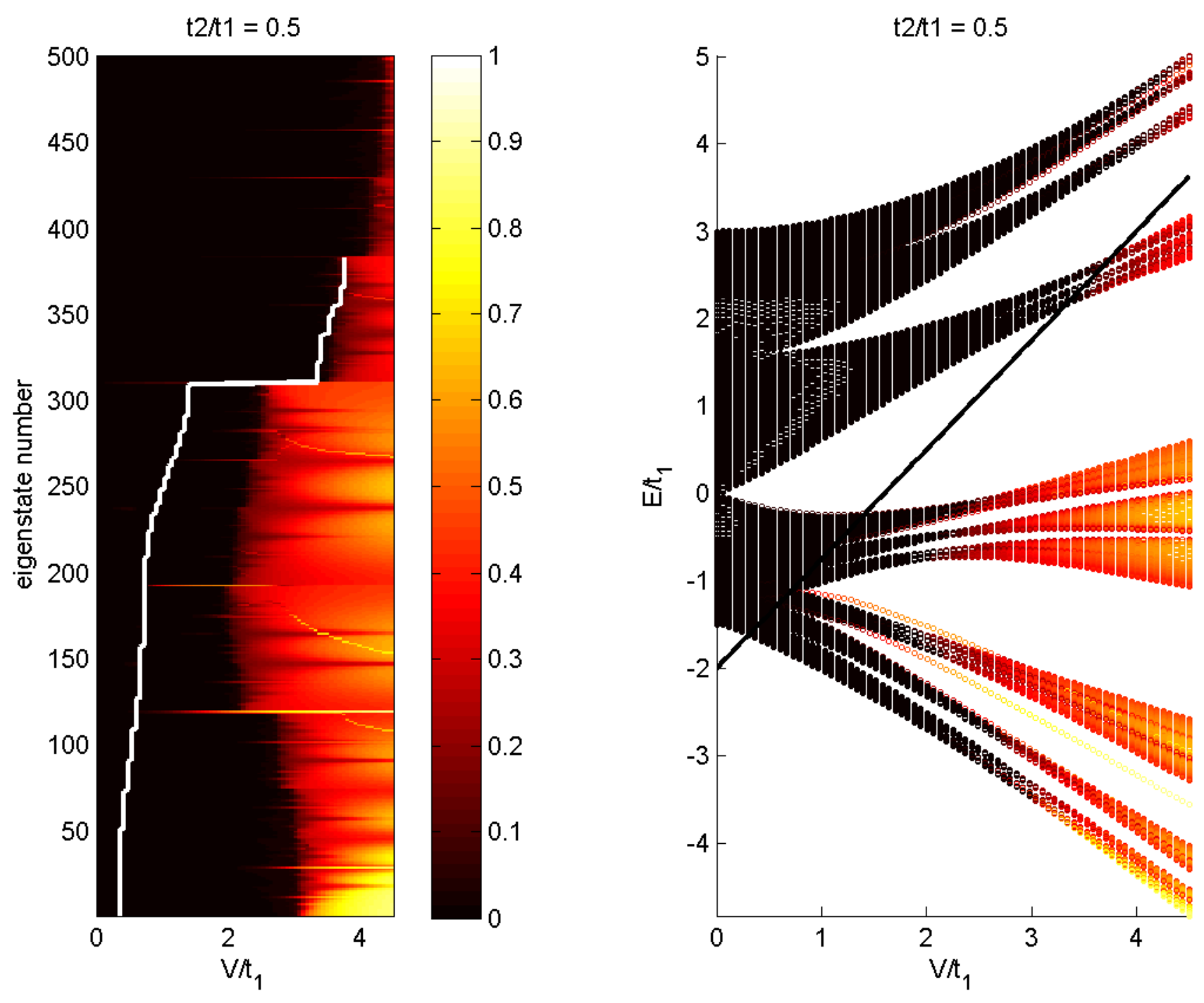}
\label{fig:subnnn4} }

\caption{\label{fig:IPRnnn} Inverse participation ratios and
energy eigenvalues of all eigenstates of Eq. (\ref{eq:t2model}) with
500 lattice sites and $\alpha = (\sqrt{5}-1)/2$ for $t_2/t_1=$
\subref{fig:subnnn1} 0.05, \subref{fig:subnnn2} 0.1,
\subref{fig:subnnn3} 0.3, \subref{fig:subnnn4} 0.5. The solid
curves represent the approximate boundary between spatially
localized and spatially extended states.}
\end{figure*}

General realizations of Eq. (\ref{eq:genmodel}) with an incommensurate
potential should not be expected to satisfy either
Eq. (\ref{eq:AAgencond}) or Eq. (\ref{eq:gencond}). Thus, the mobility
edges in non-dual incommensurate problems may not be discernible
by theoretical means. However, approximate theoretical statements
can be made for some non-dual models with hopping terms that fall
off in some general manner. If this fall off is fast enough, then
the localization transitions in these models are largely
determined by the ratio $t_2/t_1$. This ratio can be used to
determine an approximately equivalent model of the form of
Eq. (\ref{eq:model}), which, as shown in the above section, has an
exact theoretical localization boundary. To show this possibility,
we numerically examine the eigenstates of several non-dual models.
In particular, we examine the $t_1-t_2$ model, the Gaussian hopping
model, and the inverse-power-law hopping model with an incommensurate
potential and see how closely the approximate localization
boundary matches with the numerically observed one.
\begin{figure*}
\subfigure[]{
\includegraphics[type=pdf,ext=.pdf,read=.pdf,width=.47\textwidth]{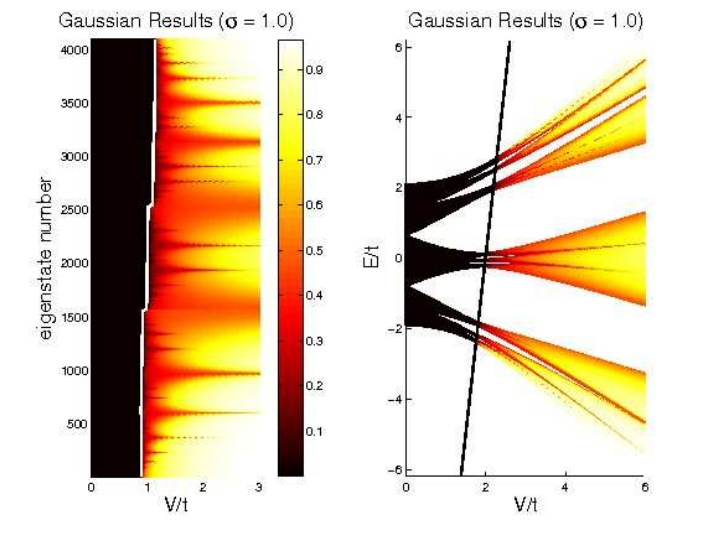}
\label{fig:subsig1p0}

} \subfigure[]{
\includegraphics[type=pdf,ext=.pdf,read=.pdf,width=.47\textwidth]{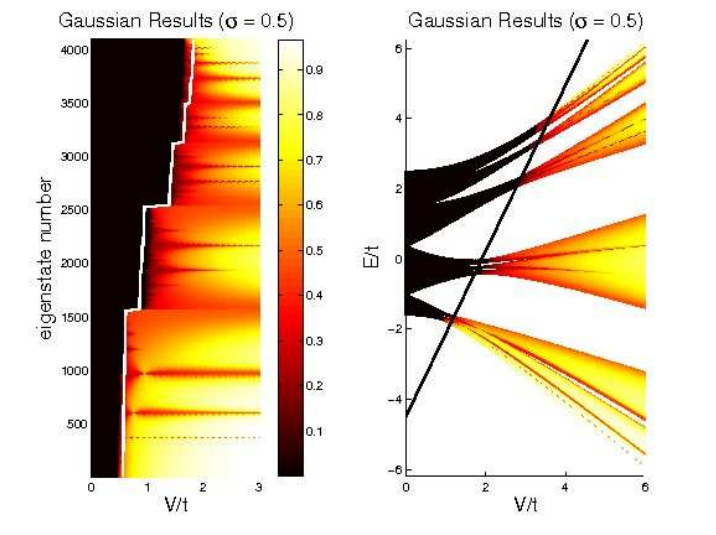}
\label{fig:subsigp5}

} \subfigure[]{
\includegraphics[type=pdf,ext=.pdf,read=.pdf,width=.47\textwidth]{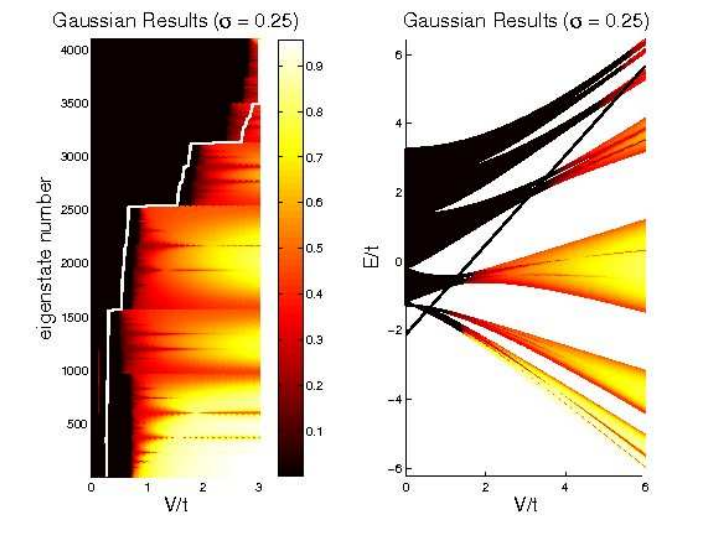}
\label{fig:subsigp25}

} \subfigure[]{
\includegraphics[type=pdf,ext=.pdf,read=.pdf,width=.47\textwidth]{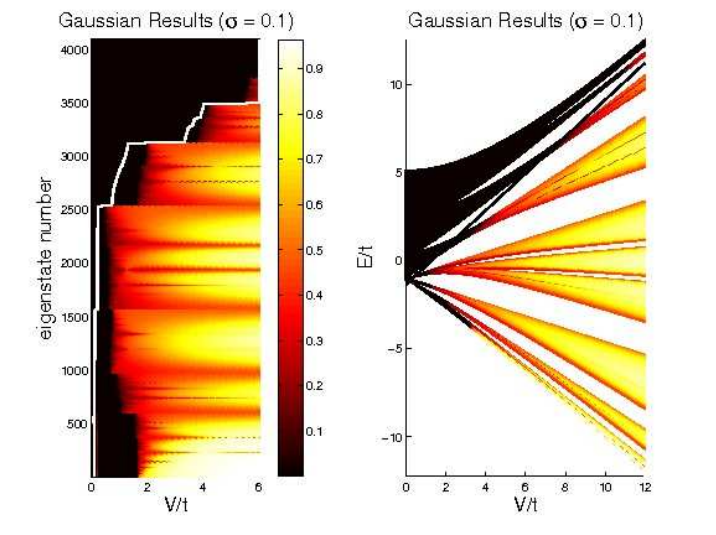}
\label{fig:subsigp1} }
 \caption{\label{fig:IPRgauss} Inverse participation ratios and
energy eigenvalues of all eigenstates of Eq. (\ref{eq:gaussmodel})
with 4096 lattice sites and $\alpha = (\sqrt{5}-1)/2$ for $\sigma
=$ (a) 1.0, (b) 0.5, (c) 0.25, and (d) 0.10. The solid curves
represent the approximate boundary between spatially localized and
spatially extended states.}
\end{figure*}

The $t_1-t_2$ model is the nnn extension of the
AA model and is given by
\begin{equation}
\left( \begin{array}{c}
t_2(u_{n+2}+u_{n-2})+t_1(u_{n+1}+u_{n-1}) \\+V\cos(2\pi\alpha n +
\delta) u_n \end{array} \right ) = E u_n. \label{eq:t2model}
\end{equation}
The parameters of the approximately equivalent exponential hopping
model [Eq. (\ref{eq:model})] are given by $p = \ln(t_1/t_2)$ and $t =
t_1e^p$. Using Eq. (\ref{eq:condition}), we can approximate the
boundary between localized and extended states. To examine this
approximation we calculate the IPR of the eigenstates of
Eq. (\ref{eq:t2model}). The results are given in Fig. \ref{fig:IPRnnn}
for 500 lattice sites, $\alpha = (\sqrt{5}-1)/2$ and various
values of the ratio $t_2/t_1$. The solid lines in the figure give
the approximate mobility edge given by Eq. (\ref{eq:condition}). From
the figure, we see that for small values of $t_2/t_1$, the
approximate boundary is in good qualitative agreement with the
numerical IPR results. For larger values ($t_2/t_1
\gtrsim 0.3$), however  the boundary differs considerably from the linear
condition in Eq. (\ref{eq:condition}).

The tight-binding incommensurate model with Gaussian hopping has
the form:
\begin{equation}
\sum_{n'\ne n}e^{-\sigma |n-n'|^{2}}u_{n'} + V\cos(2\pi\alpha n +
\delta)u_n = E u_n. \label{eq:gaussmodel}
\end{equation}
Similar to the $t_1-t_2$ model, the approximately equivalent
exponential hopping model can be determined from the ratio
$t_2/t_1$, which yields $p = 3\sigma$. The IPR results for this
model are given in Fig. \ref{fig:IPRgauss} [again, $\alpha =
(\sqrt{5}-1)/2$]. In this figure, we see that the approximate
boundary is in good qualitative agreement with the numerical
results for larger values of $\sigma$. Small values of $\sigma$,
however, result in very interesting energy-dependent mobility
edges that are not linear in potential strength, which is similar
to the $t_1-t_2$ results for large values of $t_1/t_2$.
\begin{figure*}
\subfigure[$r=1/3$]{
\includegraphics[type=pdf,ext=.pdf,read=.pdf,width=.47\textwidth]{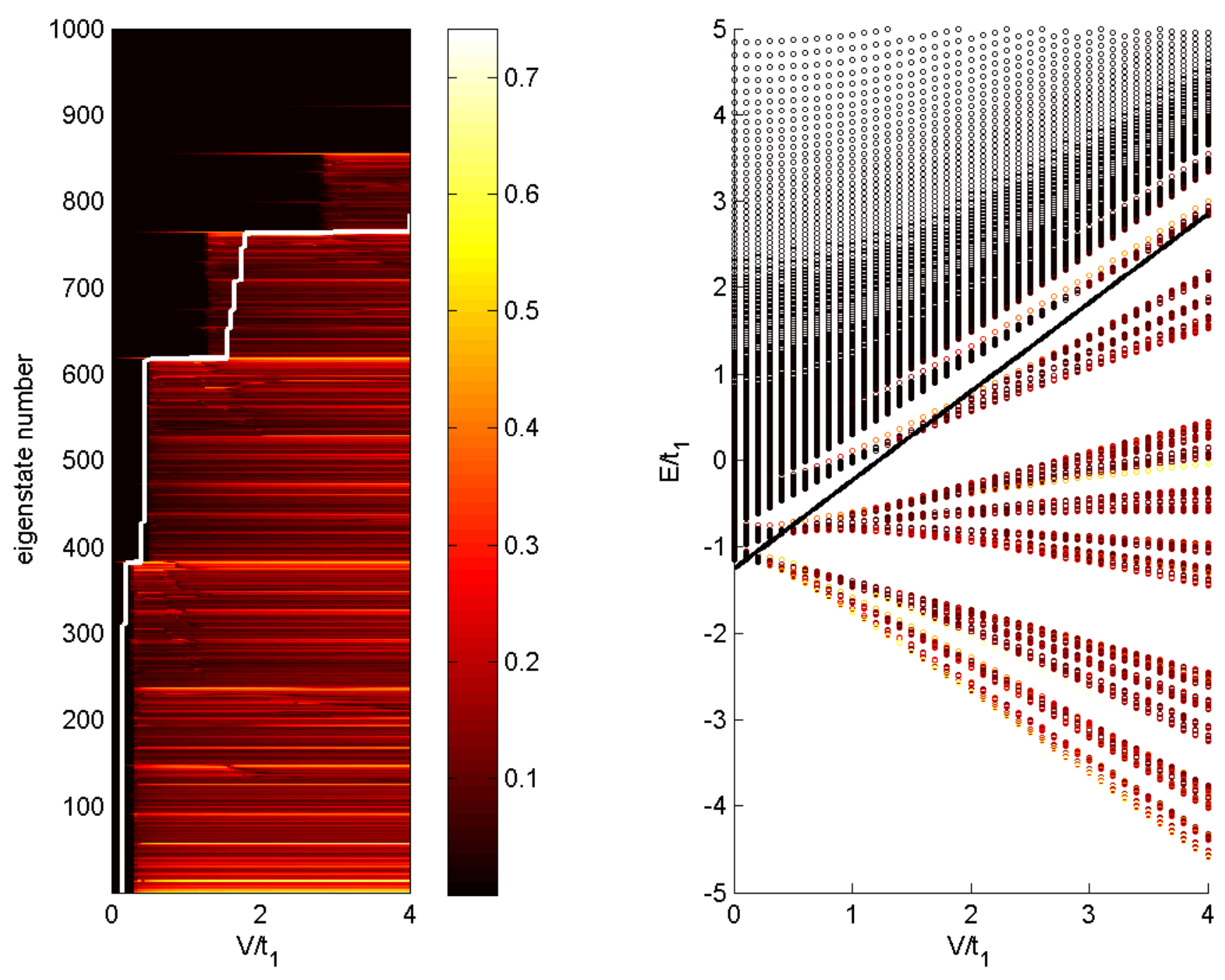}
\label{fig:subpoly1}

} \subfigure[$r=2$]{
\includegraphics[type=pdf,ext=.pdf,read=.pdf,width=.47\textwidth]{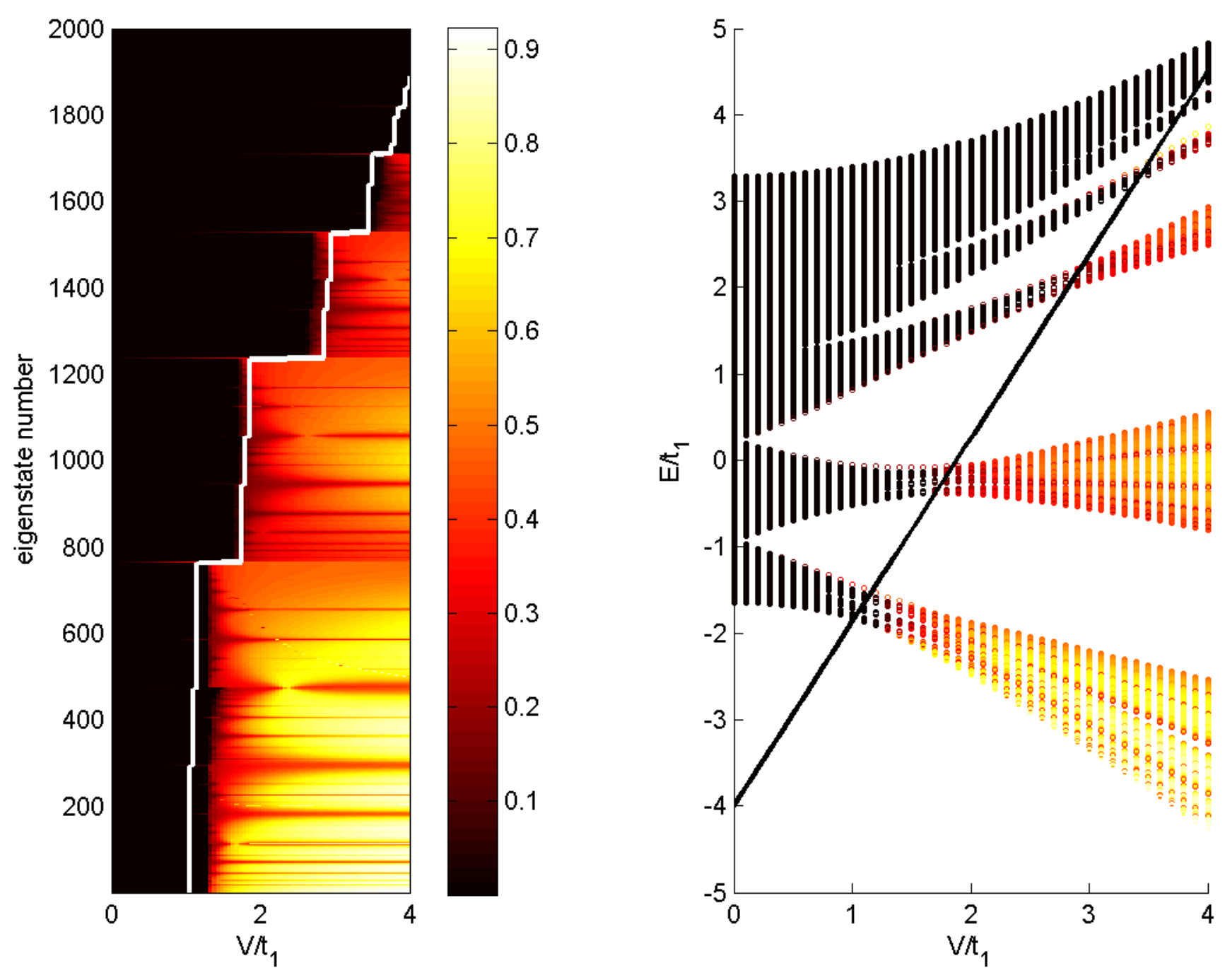}
\label{fig:subpoly2}

} \subfigure[$r=3$]{
\includegraphics[type=pdf,ext=.pdf,read=.pdf,width=.47\textwidth]{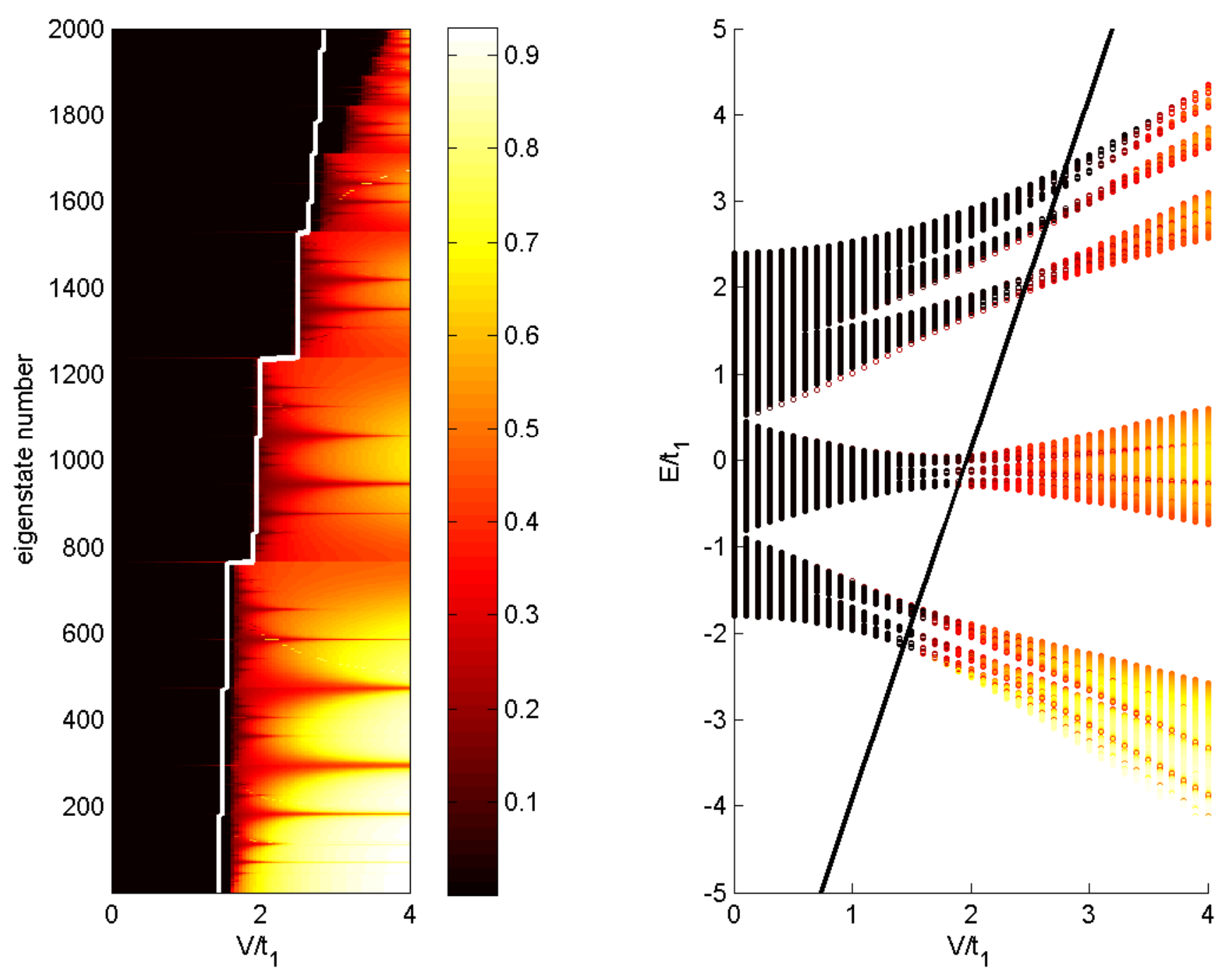}
\label{fig:subpoly3}

}

\caption{\label{fig:IPRpoly} Inverse participation ratios and
energy eigenvalues of all eigenstates of Eq. (\ref{eq:polymodel}) with
2000 lattice sites and $\alpha = (\sqrt{5}-1)/2$ for various
values of $r$. The solid curves represent the approximate boundary
between spatially localized and spatially extended states.}
\end{figure*}

For power law decay in the hopping terms, we examine the model:
\begin{equation}
\sum_{n' \ne n}\frac{u_{n'}}{|n-n'|^r} + V\cos(2\pi\alpha n +
\delta) = E u_n. \label{eq:polymodel}
\end{equation}
In this case, the exponential coefficient, $p$ is given by $p =
r\ln(2)$. Fig \ref{fig:IPRpoly} gives IPR results for this model
with $\alpha = (\sqrt{5}-1)/2$ and $r = 1/3$ [Fig
\ref{fig:subpoly1}], $r = 2$ [Fig \ref{fig:subpoly2}], and $r = 3$
[Fig \ref{fig:subpoly3}]. In each of these cases, the approximate
localization boundary is in good qualitative agreement with the
numerical results.

From the above numerical results, we see that the localization
boundary for the exponential hopping model gives a good
qualitative agreement for the Gaussian and inverse-power-law hopping
models with large-enough decay coefficients (i.e., $r$ and $\sigma
\gtrsim 1$). Thus, we believe that the localization boundary for
any tight-binding model with hopping terms that decay fast enough
can be approximated by the results of the exponential hopping
model. In general, however, we see that the energy-dependent
mobility edges in the non-dual models are not linear in $V$ as with
the case in the exponential hopping model. Whether there is an
exact theoretical statement to describe these peculiar mobility
edges is still an open question.

\section{Random disorder}

We examine the case of a random potential in the
tight binding framework in the context of both the
nnn model (i.e. the $t_1-t_2$ model) and
extended schemes where the coupling may be short ranged in the
sense of decaying exponentially (or more rapidly, as in a Gaussian
decay) or long-ranged with a power law decay.

We examine the effects of a random potential directly in the
context of characteristics of the eigenstates, by studying the IPR,  which provides information as to the extent of localization. We calculate the IPR versus eigenstate number for a
range of random potential strengths, which will tend to localize
states.

We produce surface graphs of the IPR to show the characteristics of the
eigenstates with respect to localization. In addition, one may
calculate histograms of the IPR instead of preparing surface plots
with respect to eigenstate number and the width of the potential.
The former complement the latter by showing how the statistical weight for a
particular IPR value evolves with increasing system size.

We find in both the surface plots and in the histograms a
convergence toward bulk behavior, where self-averaging in
sufficiently large systems reduces the differences in the
characteristics of the eigenstates corresponding to independently
generated random potential realizations. In the thermodynamic
limit, the type of tight-binding coupling scheme and
the overall statistical characteristics of the random potential
 are the factors which determine the distribution of the
 properties of the eigenstates.

\begin{figure*}
\subfigure[$t_2=0$]{\includegraphics[type=pdf,ext=.pdf,read=.pdf,width=.47\textwidth]{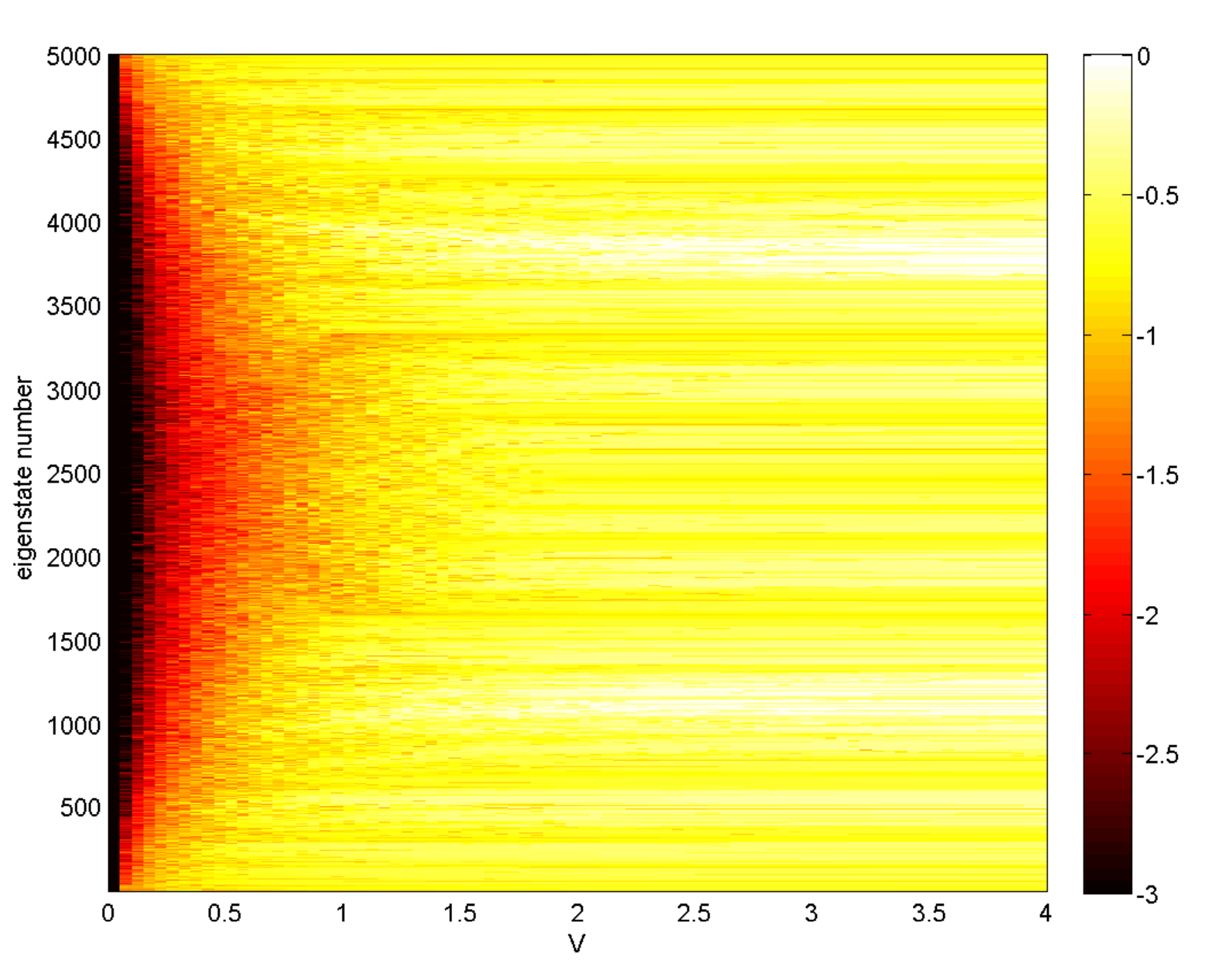}\label{fig:subt2bin0}}
\subfigure[$t_2=0.2$]{\includegraphics[type=pdf,ext=.pdf,read=.pdf,width=.47\textwidth]{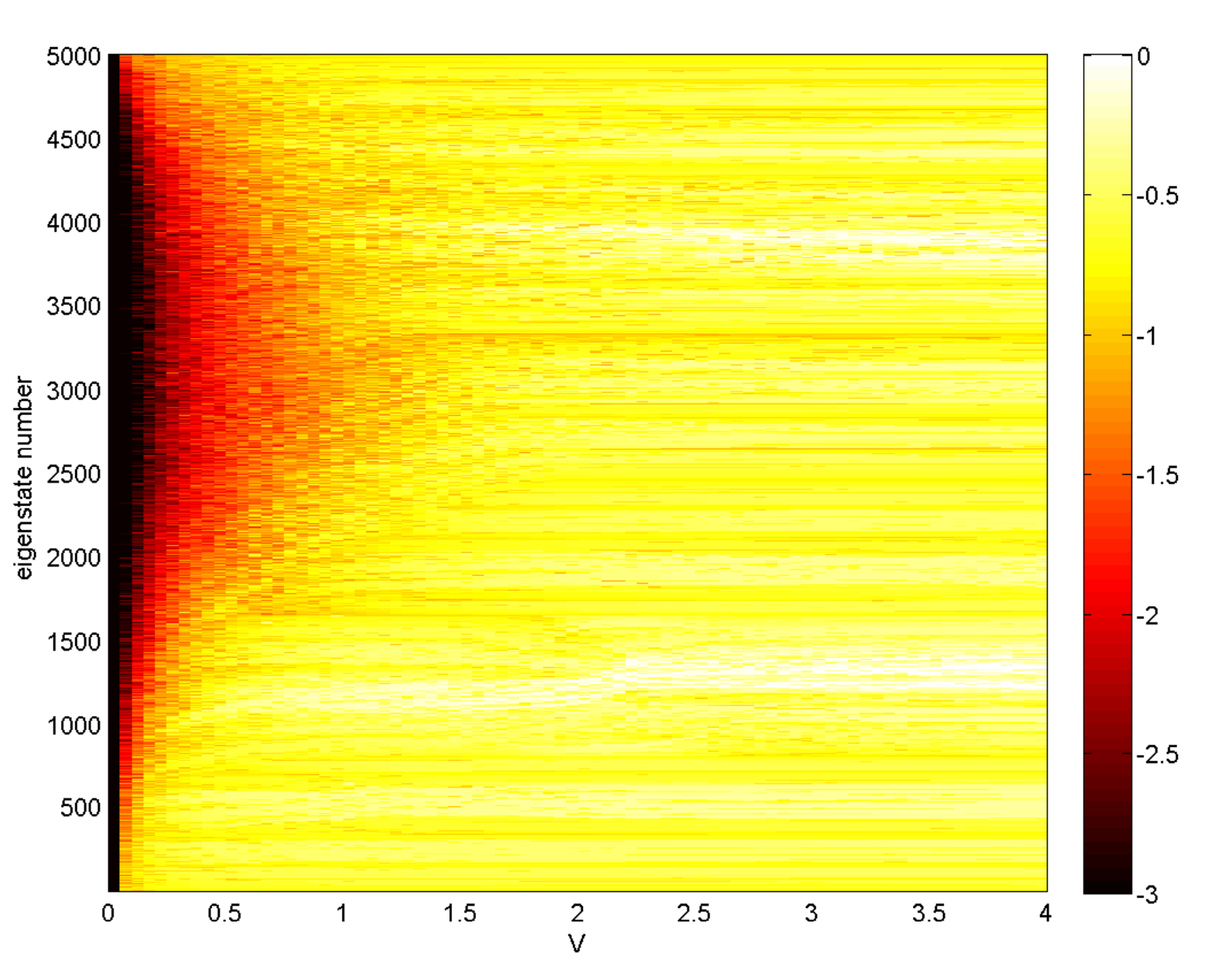}\label{fig:subt2binp2}}
\subfigure[$t_2=0.4$]{\includegraphics[type=pdf,ext=.pdf,read=.pdf,width=.47\textwidth]{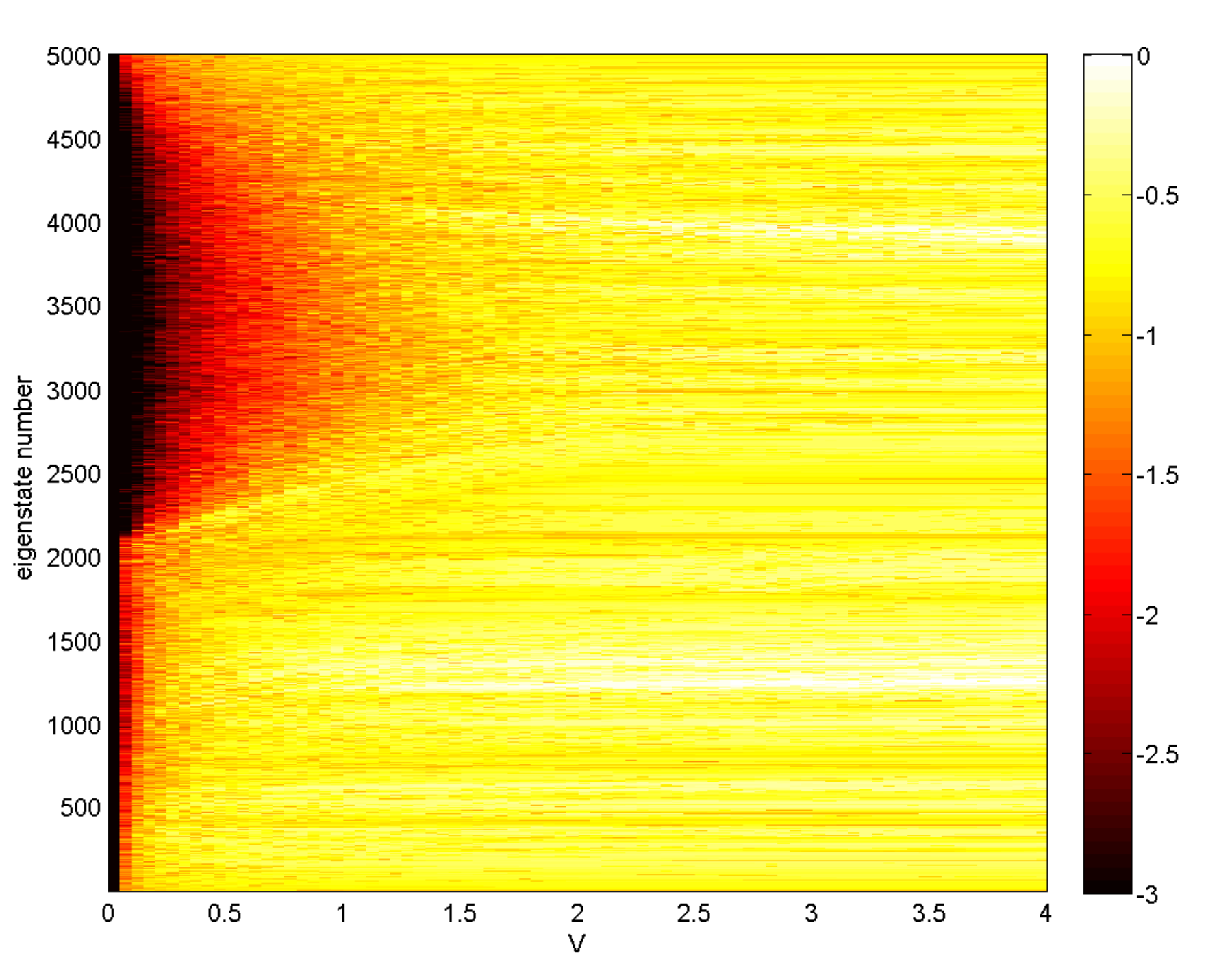}\label{fig:subt2binp4}}
\subfigure[$t_2=0.8$]{\includegraphics[type=pdf,ext=.pdf,read=.pdf,width=.47\textwidth]{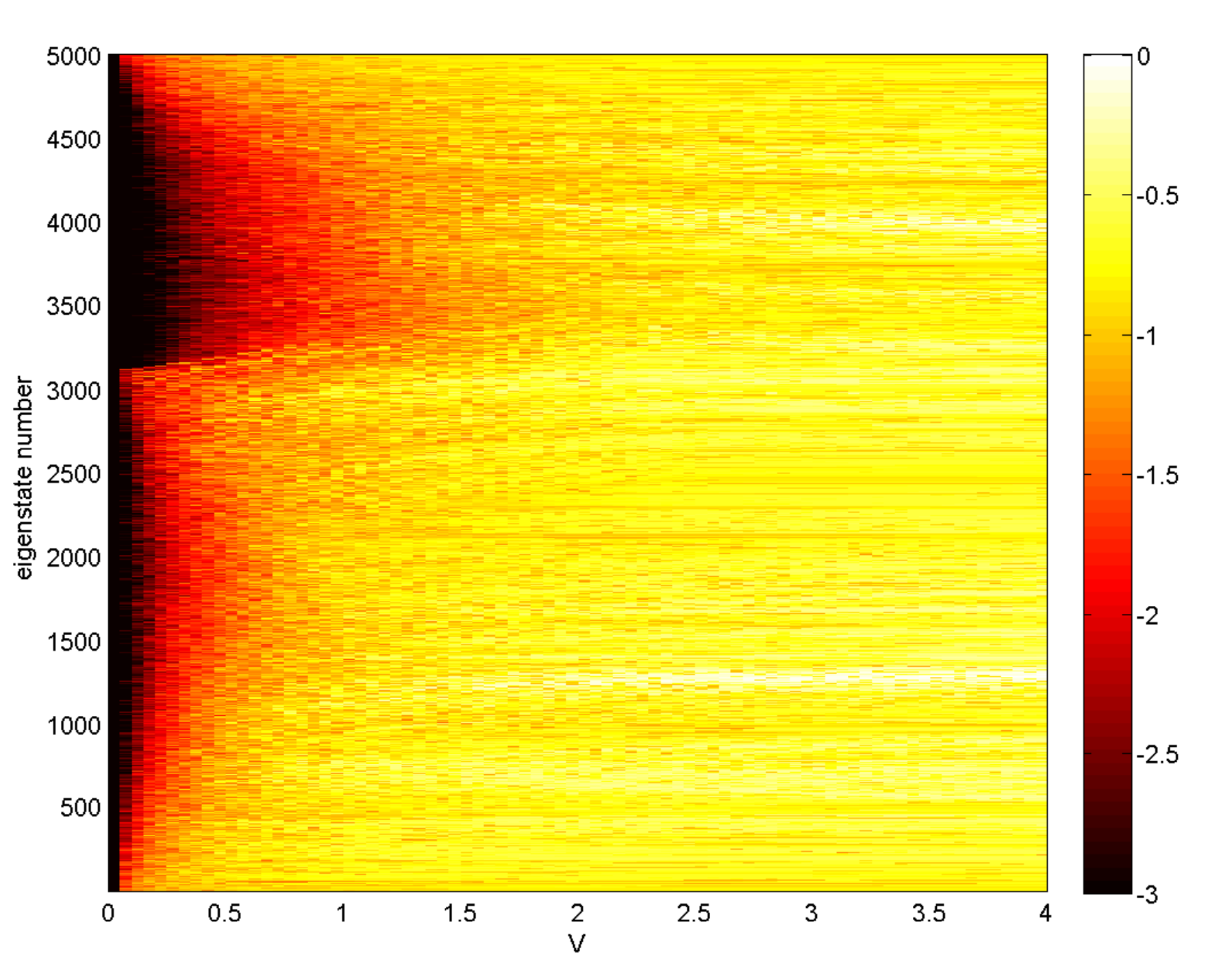}\label{fig:subt2binp8}}
\caption{\label{fig:t1t2binary} (Color online) Suface plot of the base 10 logarithm of the IPRs for the $t_1-t_2$ model with a random, binary potential for 5000 lattice sites.}
\end{figure*}

\begin{figure}
\subfigure[$t_2=0$]{\includegraphics[type=pdf,ext=.pdf,read=.pdf,width=.47\textwidth]{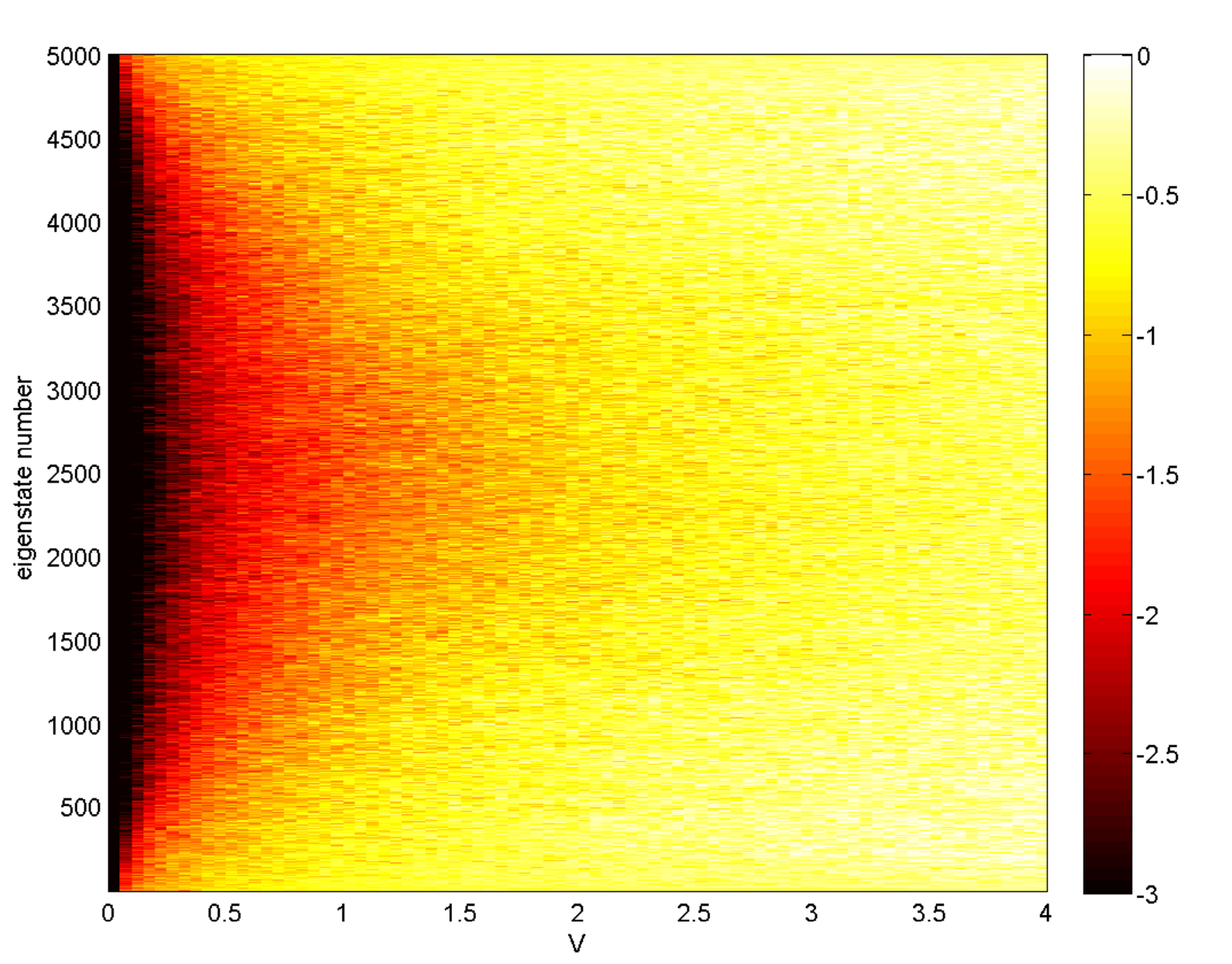}\label{fig:subt2uni0}}
\subfigure[$t_2=0.2$]{\includegraphics[type=pdf,ext=.pdf,read=.pdf,width=.47\textwidth]{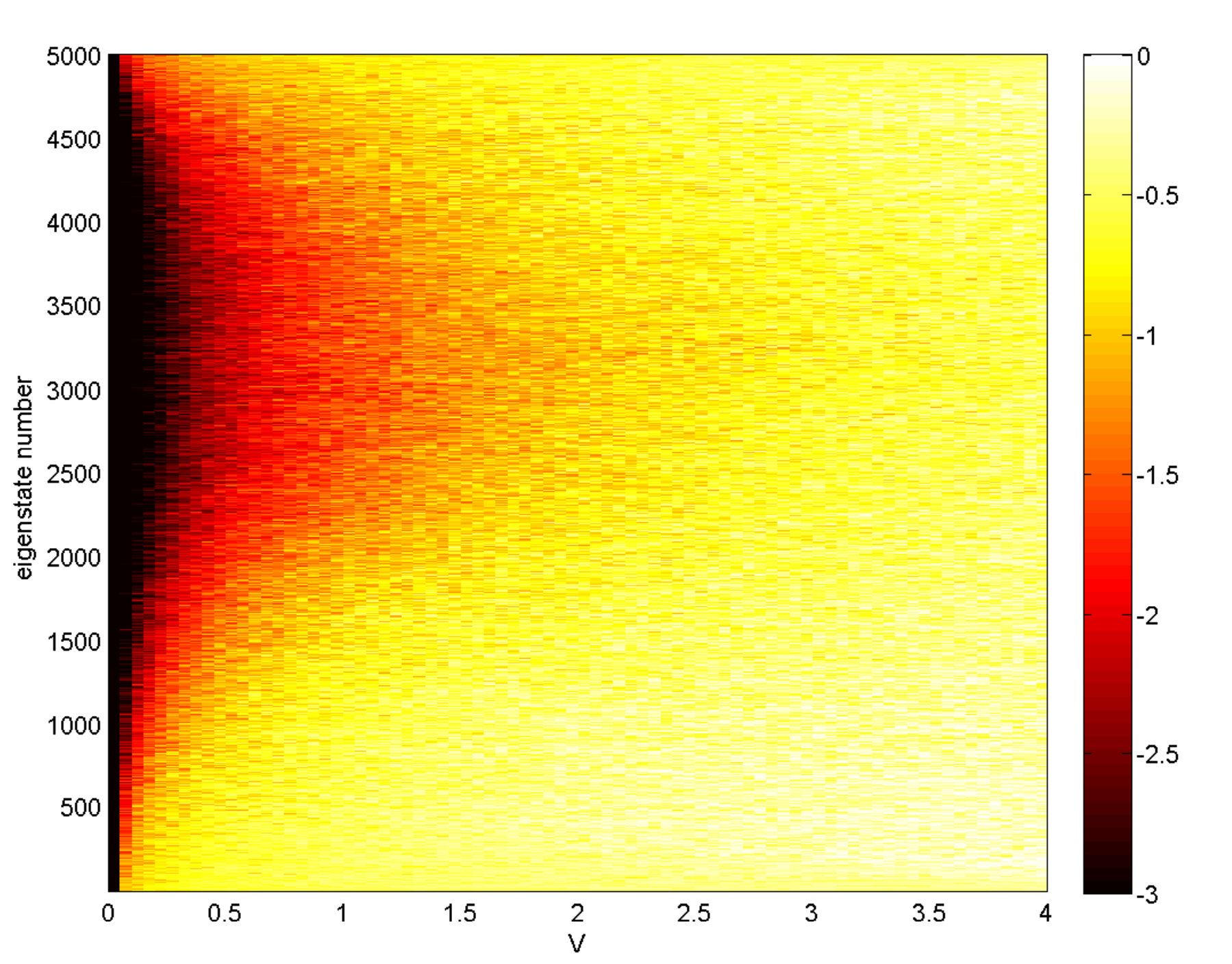}\label{fig:subt2unip2}}
\caption{\label{fig:t1t2uniform} (Color online) Surface plots of the base 10 logarithm of the IPRs for the $t_1-t_2$ model with a random, uniformly distributed potential for 5000 lattice sites.}
\end{figure}

\begin{figure}
\subfigure[$t_2=0$]{\includegraphics[type=pdf,ext=.pdf,read=.pdf,width=.47\textwidth]{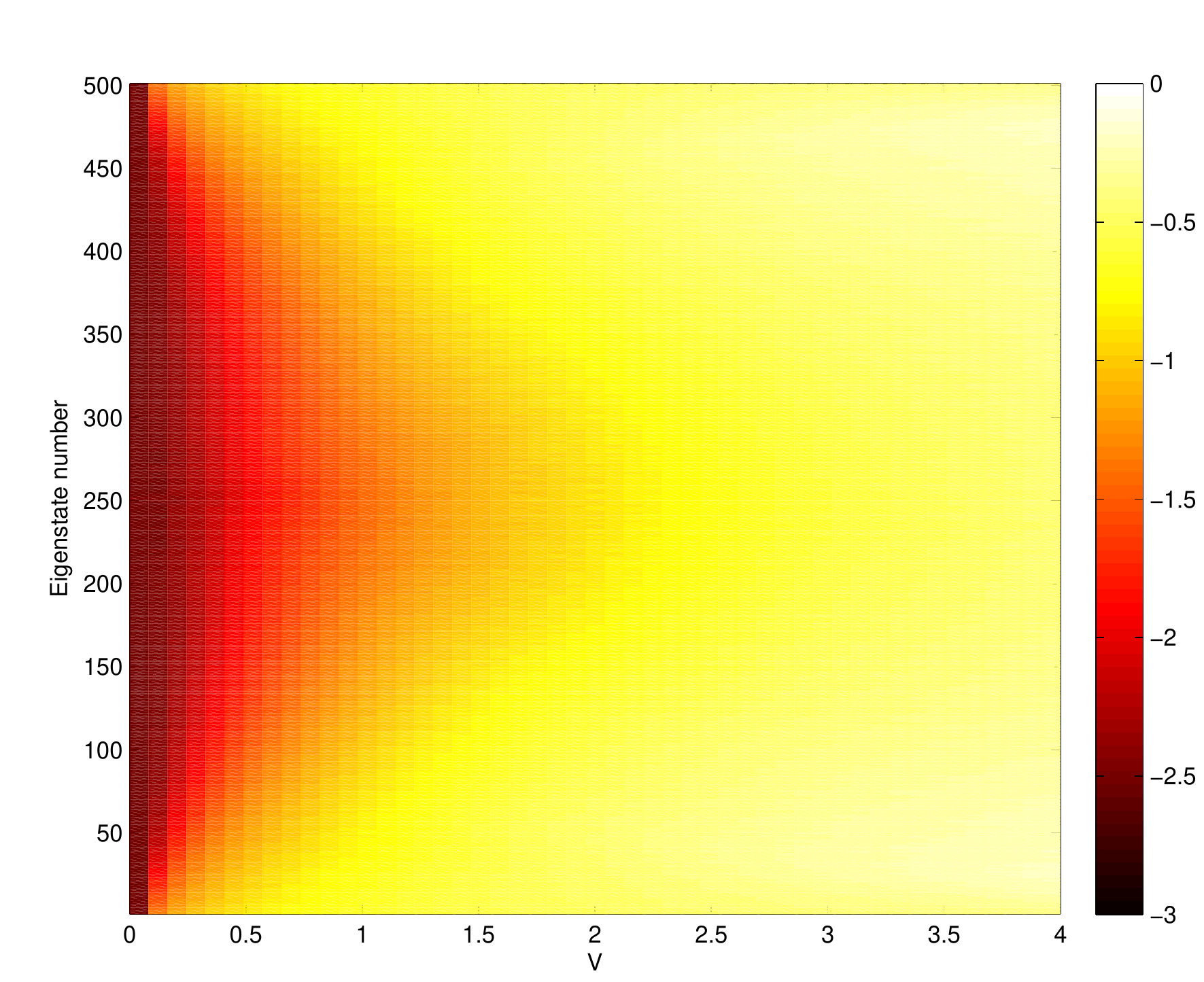}\label{fig:subt2avg0}}
\subfigure[$t_2=0.2$]{\includegraphics[type=pdf,ext=.pdf,read=.pdf,width=.47\textwidth]{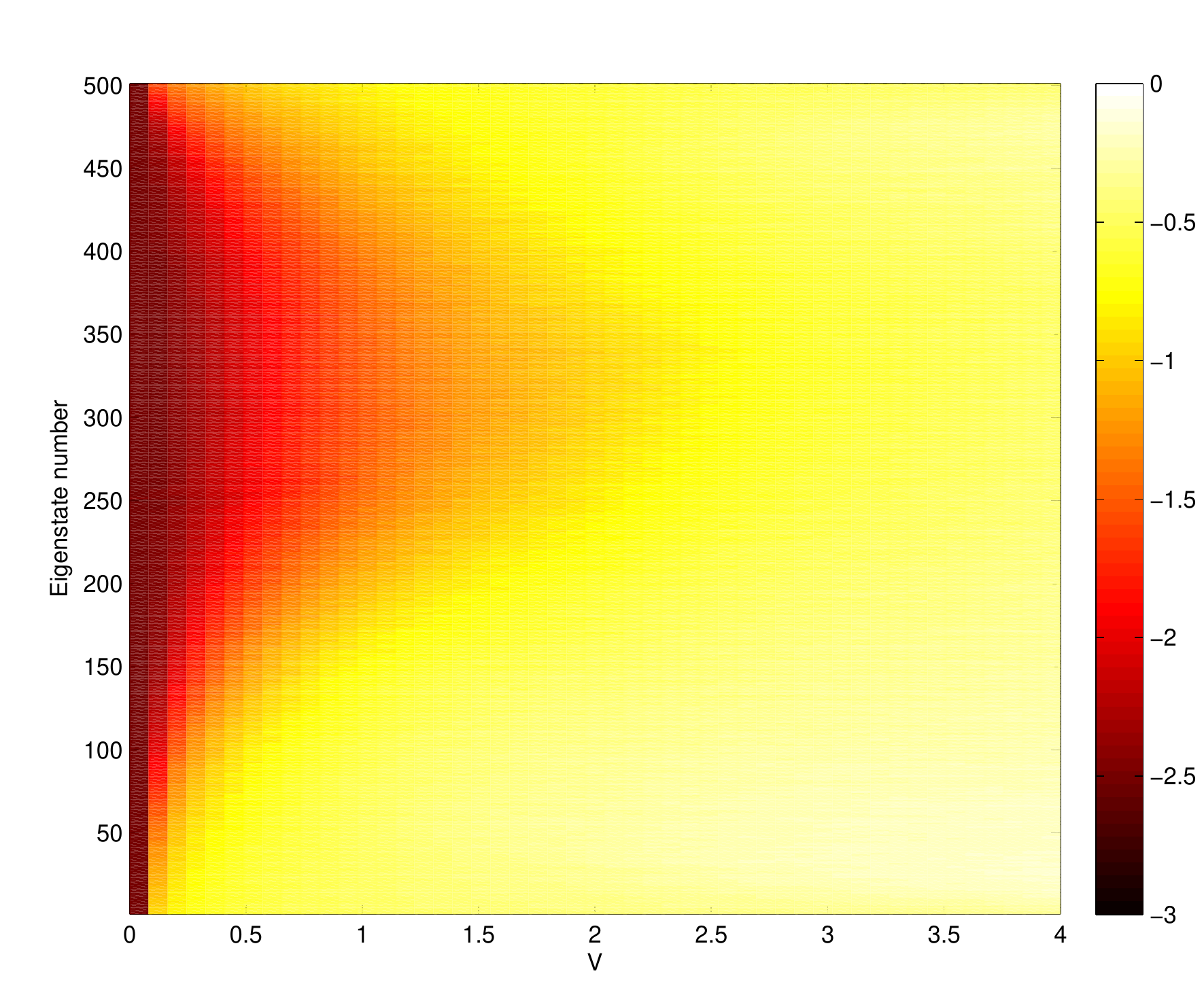}\label{fig:subt2avgp2}} 
\caption{\label{fig:t1t2avg} (Color online) Surface plots of the base 10 logarithm of the IPRs for the $t_1-t_2$ model averaged over 100 realizations of random, uniformly distributed potentials with 500 lattice sites.}
\end{figure}

\begin{figure}
\subfigure[$\sigma=0.25$]{\includegraphics[type=pdf,ext=.pdf,read=.pdf,width=.47\textwidth]{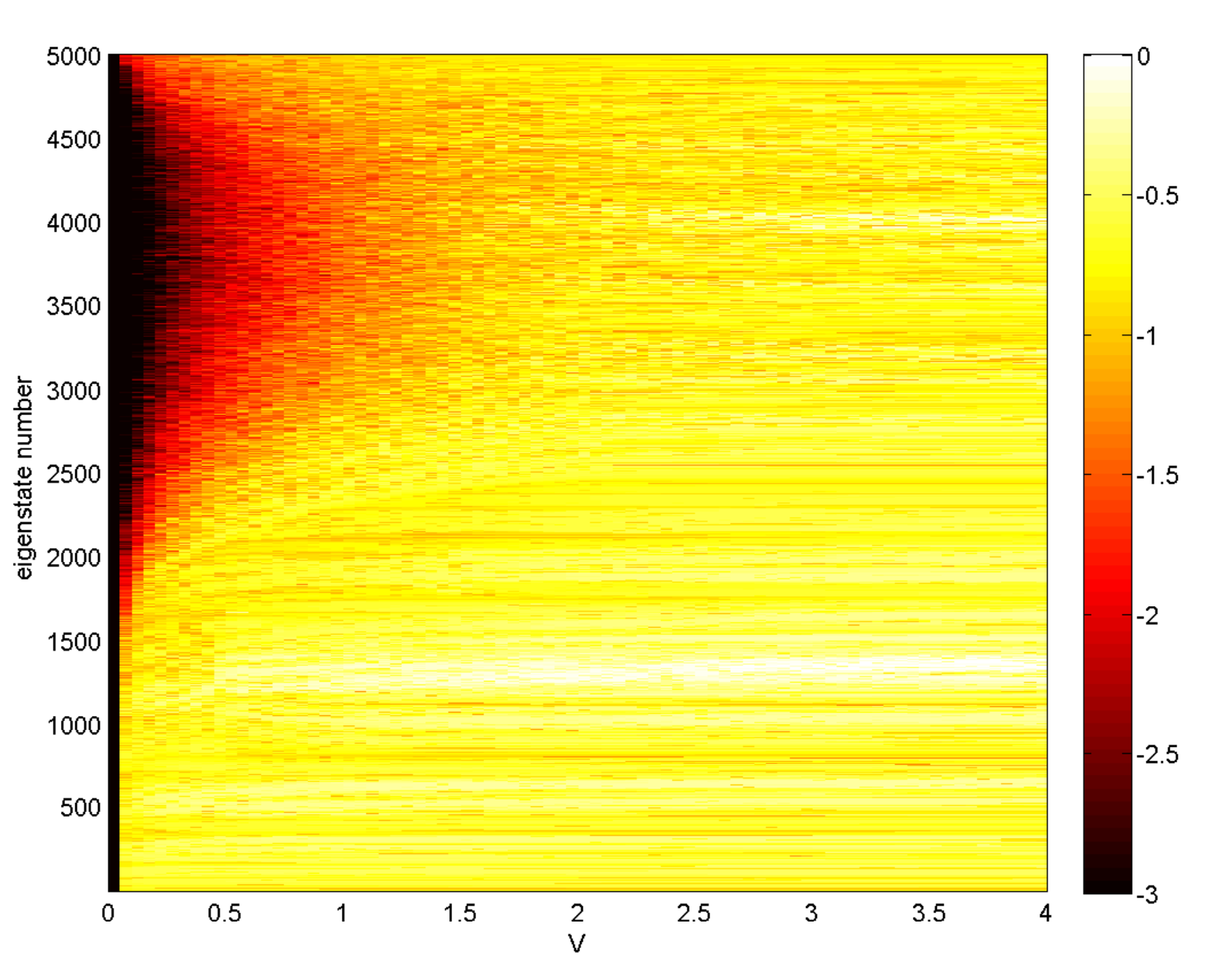}\label{fig:subgbinp25}}
\subfigure[$\sigma=1$]{\includegraphics[type=pdf,ext=.pdf,read=.pdf,width=.47\textwidth]{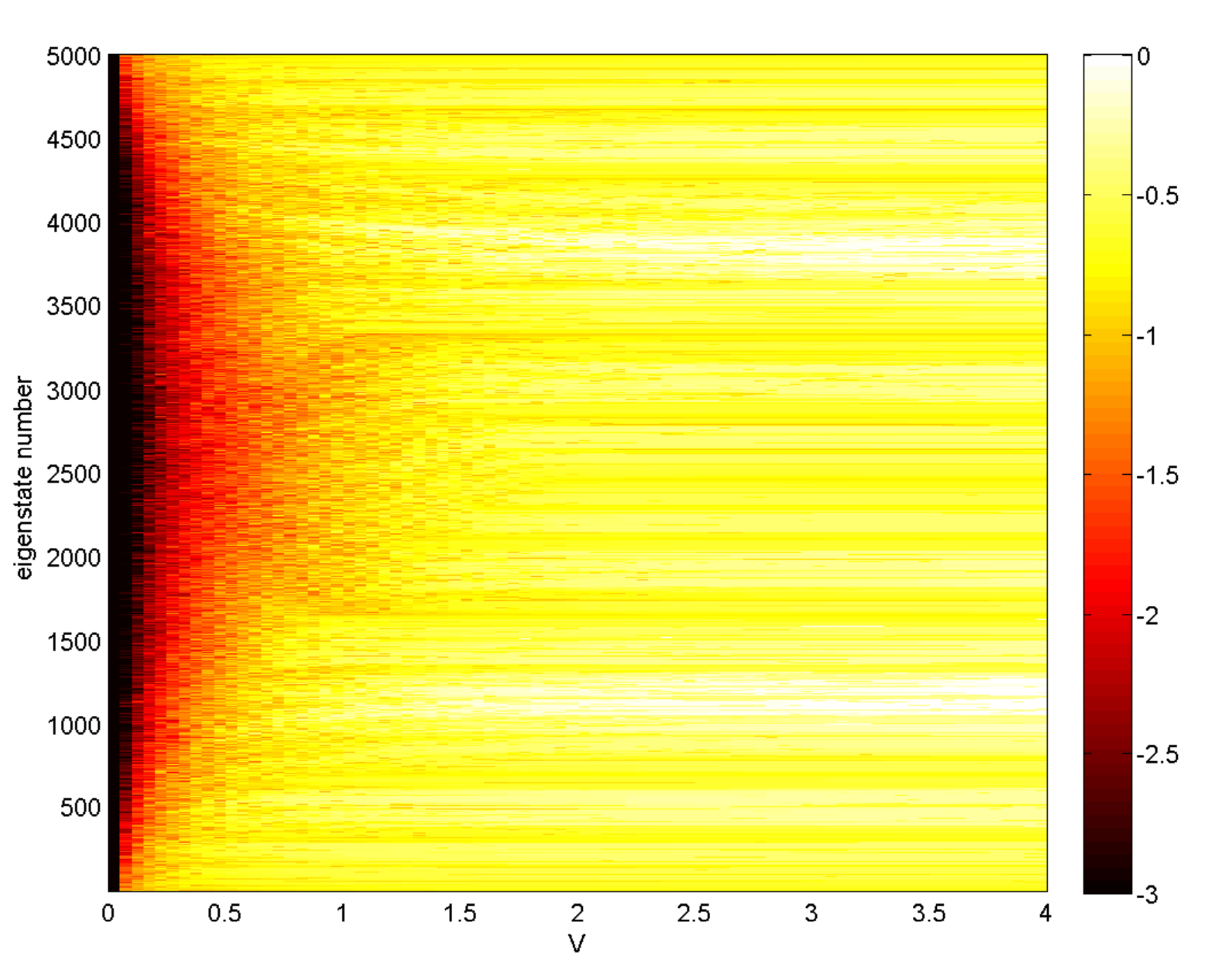}\label{fig:subgbin1}}
\caption{\label{fig:gaussbinary} (Color online) Surface plots of the base 10 logarithm of the IPRs for the Gaussian hopping model with a random, binary potential for 5000 lattice sites.}
\end{figure}

\begin{figure}
\subfigure[$\sigma=0.25$]{\includegraphics[type=pdf,ext=.pdf,read=.pdf,width=.47\textwidth]{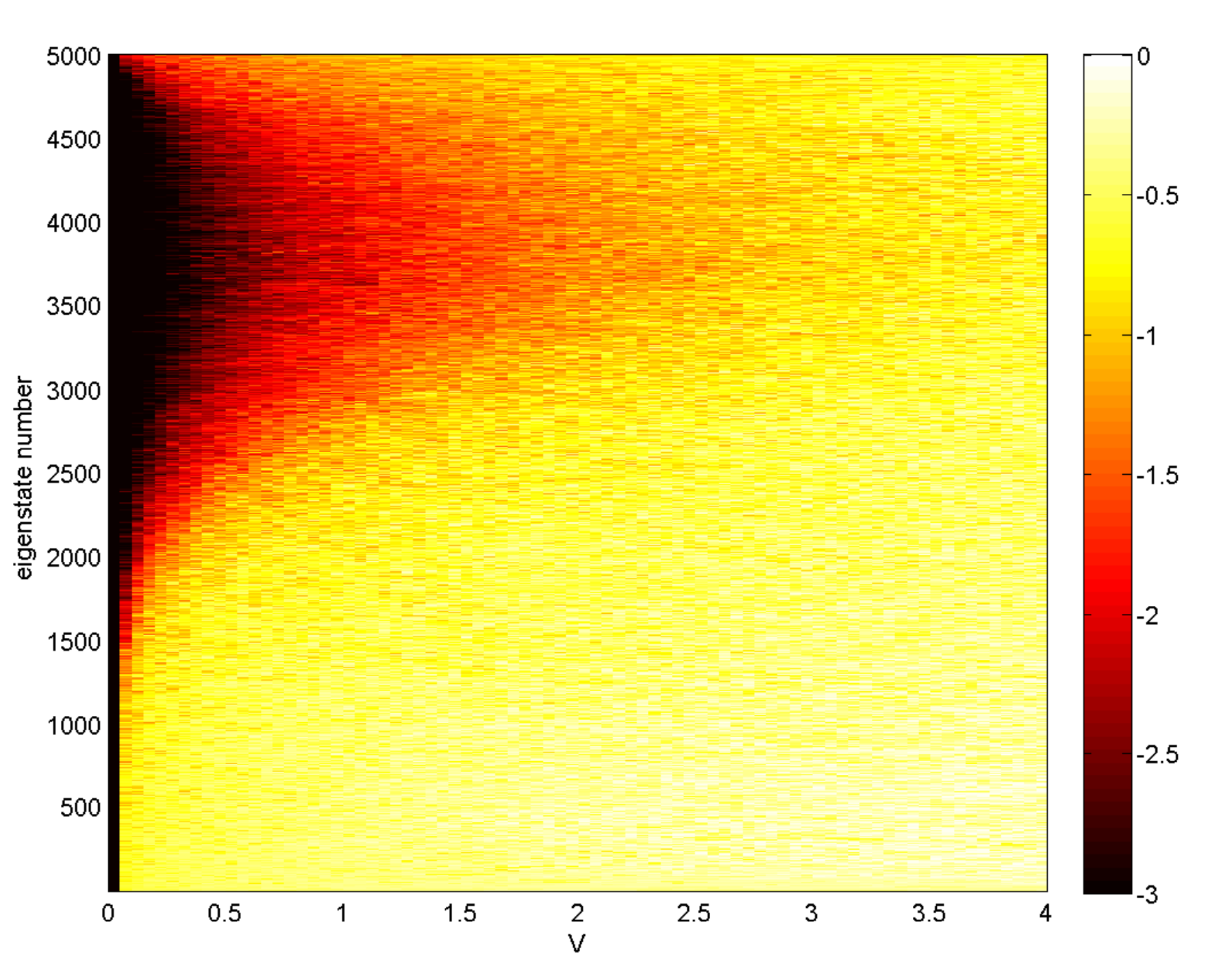}\label{fig:subgunip25}}
\subfigure[$\sigma=1$]{\includegraphics[type=pdf,ext=.pdf,read=.pdf,width=.47\textwidth]{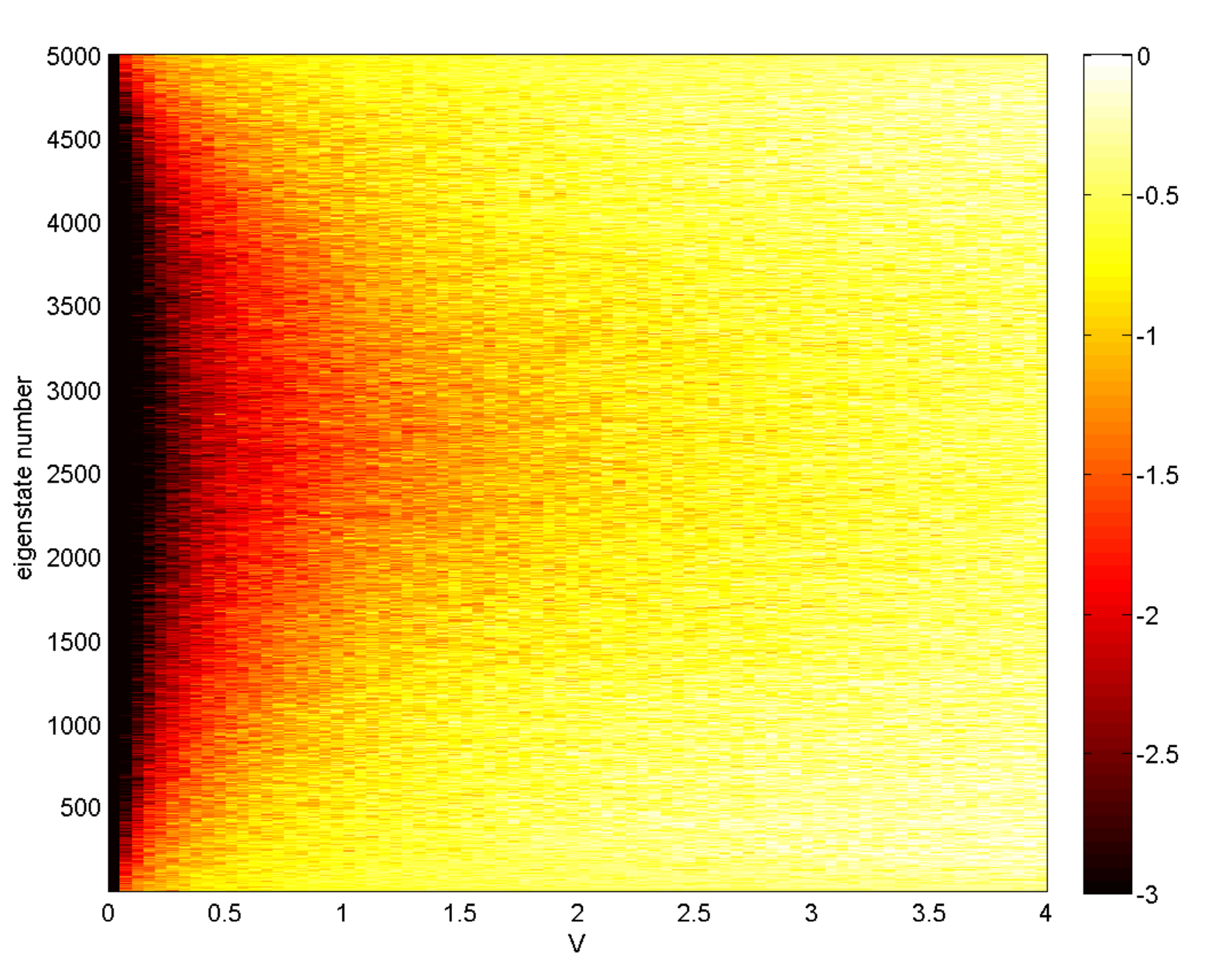}\label{fig:subguni1}}
\caption{\label{fig:gaussuni} (Color online) Surface plots of the base 10 logarithm of the IPRs for the Gaussian hopping model with a random, uniformly distributed potential for 5000 lattice sites.}
\end{figure}

\begin{figure}
\subfigure[$\sigma=0.25$]{\includegraphics[type=pdf,ext=.pdf,read=.pdf,width=.47\textwidth]{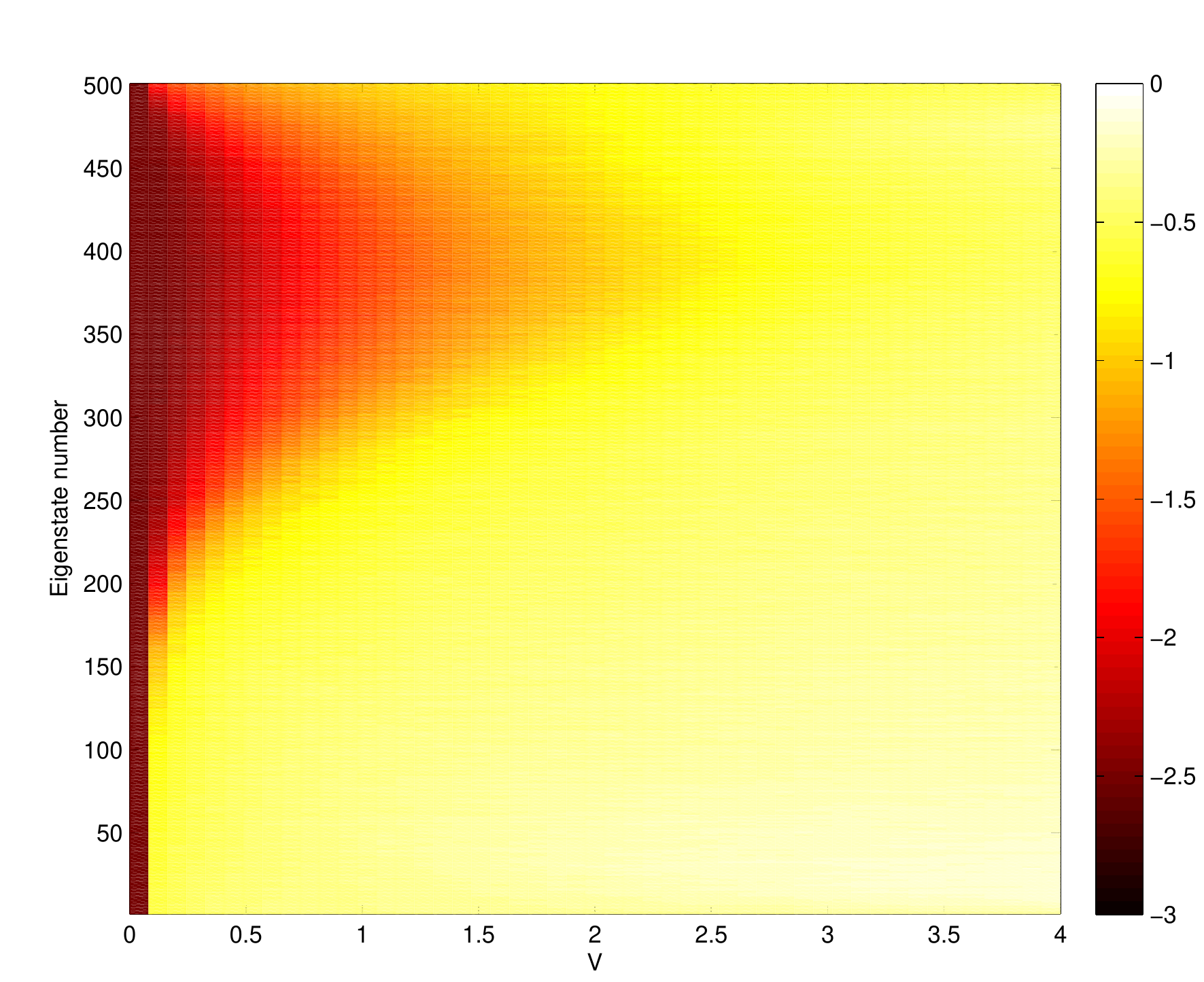}\label{fig:subgavgp25}}
\subfigure[$\sigma=1$]{\includegraphics[type=pdf,ext=.pdf,read=.pdf,width=.47\textwidth]{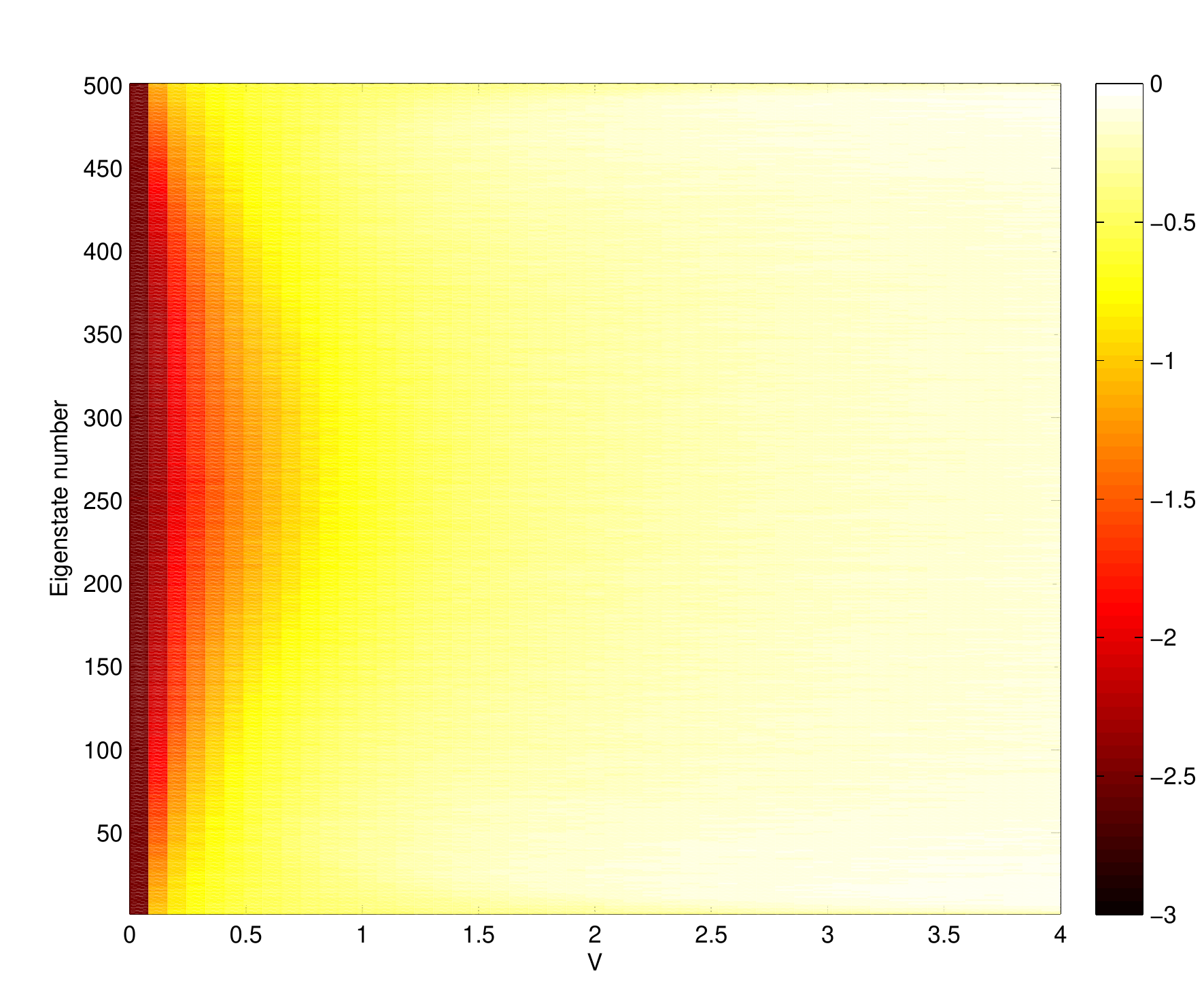}\label{fig:subgavg1}}
\caption{\label{fig:gaussavg} (Color online) Surface plots of the base 10 logarithm of the IPRs for the Gaussian hopping model averaged over 100 realizations of random, uniformly distributed potentials with 500 lattice sites.}
\end{figure}

\begin{figure}
\subfigure[$p=1$]{\includegraphics[type=pdf,ext=.pdf,read=.pdf,width=.47\textwidth]{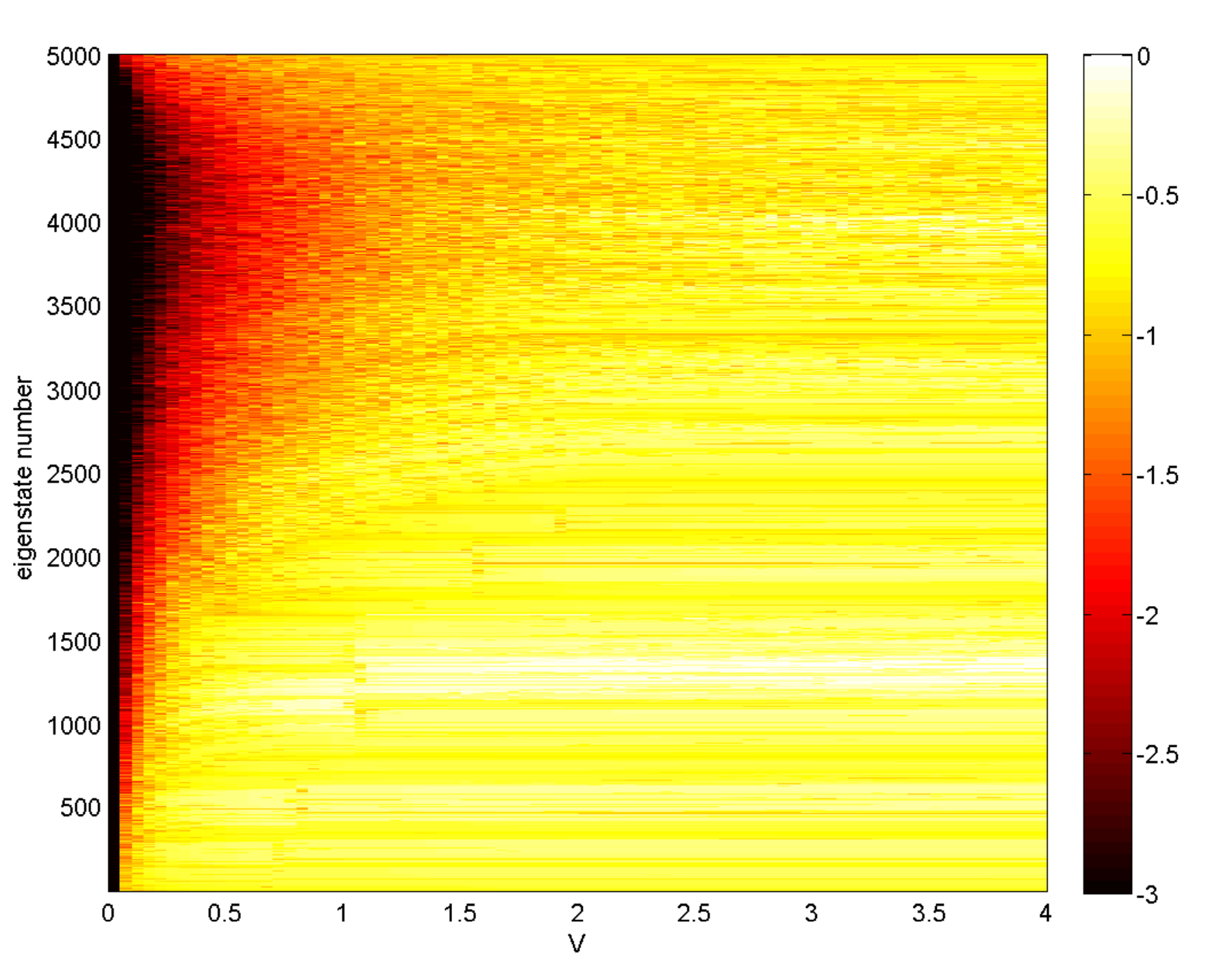}\label{fig:subebin1}}
\subfigure[$p=2$]{\includegraphics[type=pdf,ext=.pdf,read=.pdf,width=.47\textwidth]{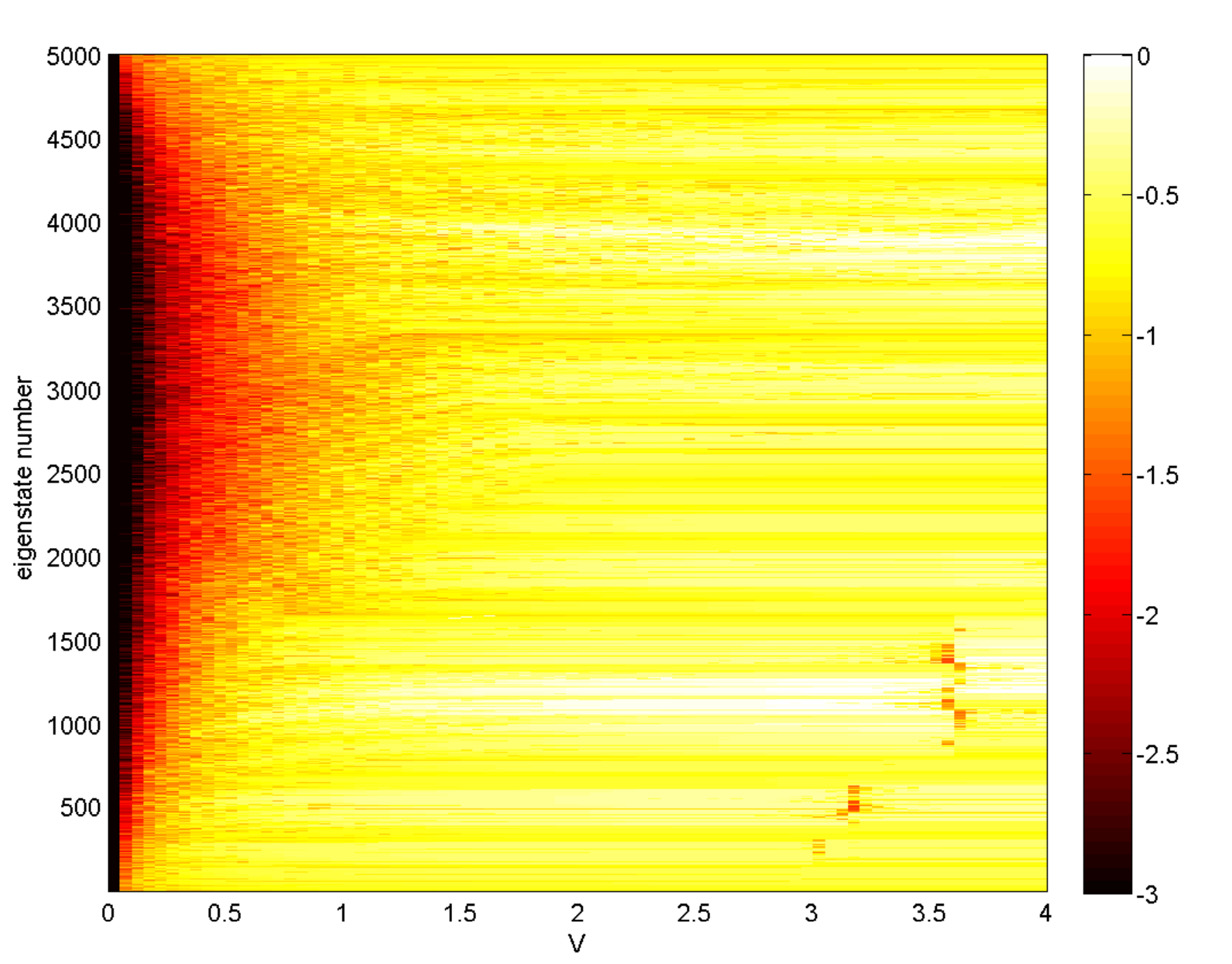}\label{fig:subebin2}}
\caption{\label{fig:expbin} (Color online) Surface plots of the base 10 logarithm of the IPRs for the exponential hopping model with a random, binary potential for 5000 lattice sites.}
\end{figure}

\begin{figure}
\subfigure[$p=1$]{\includegraphics[type=pdf,ext=.pdf,read=.pdf,width=.47\textwidth]{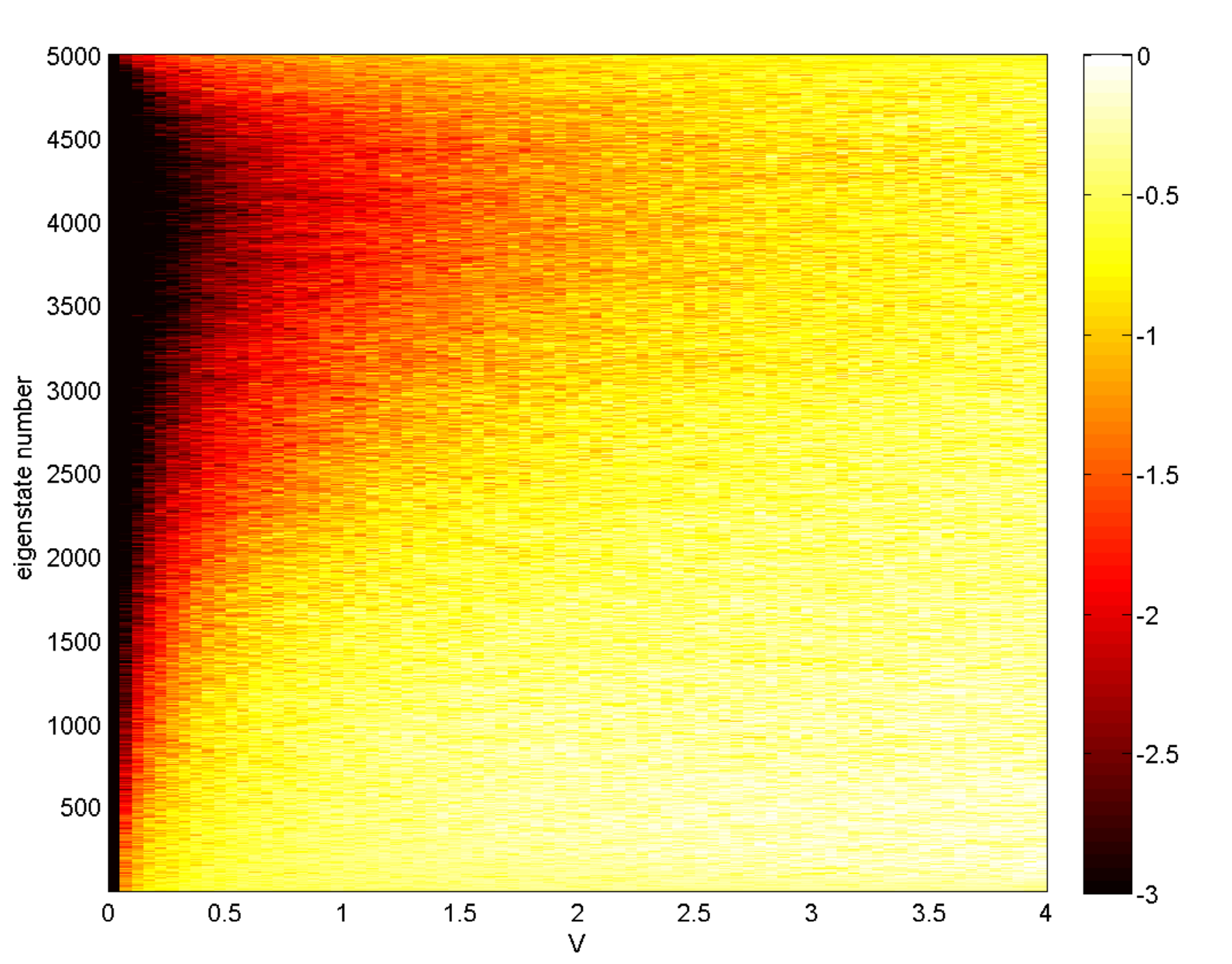}\label{fig:subeuni1}}
\subfigure[$p=2$]{\includegraphics[type=pdf,ext=.pdf,read=.pdf,width=.47\textwidth]{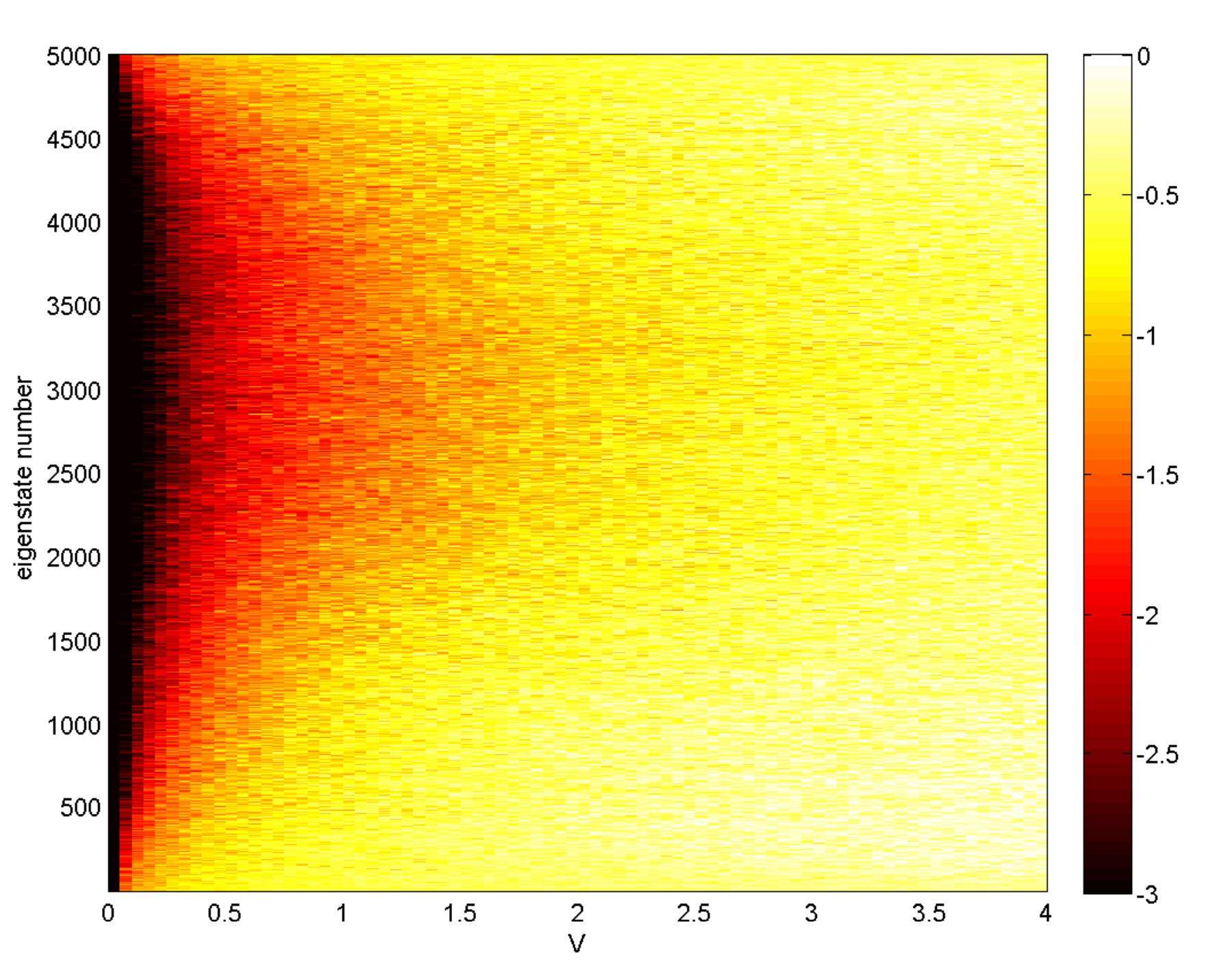}\label{fig:subeuni2}}
\caption{\label{fig:expuni} (Color online) Surface plots of the base 10 logarithm of the IPRs for the exponential hopping model with a random, uniformly distributed potential for 5000 lattice sites.}
\end{figure}

\begin{figure}
\subfigure[$p=1$]{\includegraphics[type=pdf,ext=.pdf,read=.pdf,width=.47\textwidth]{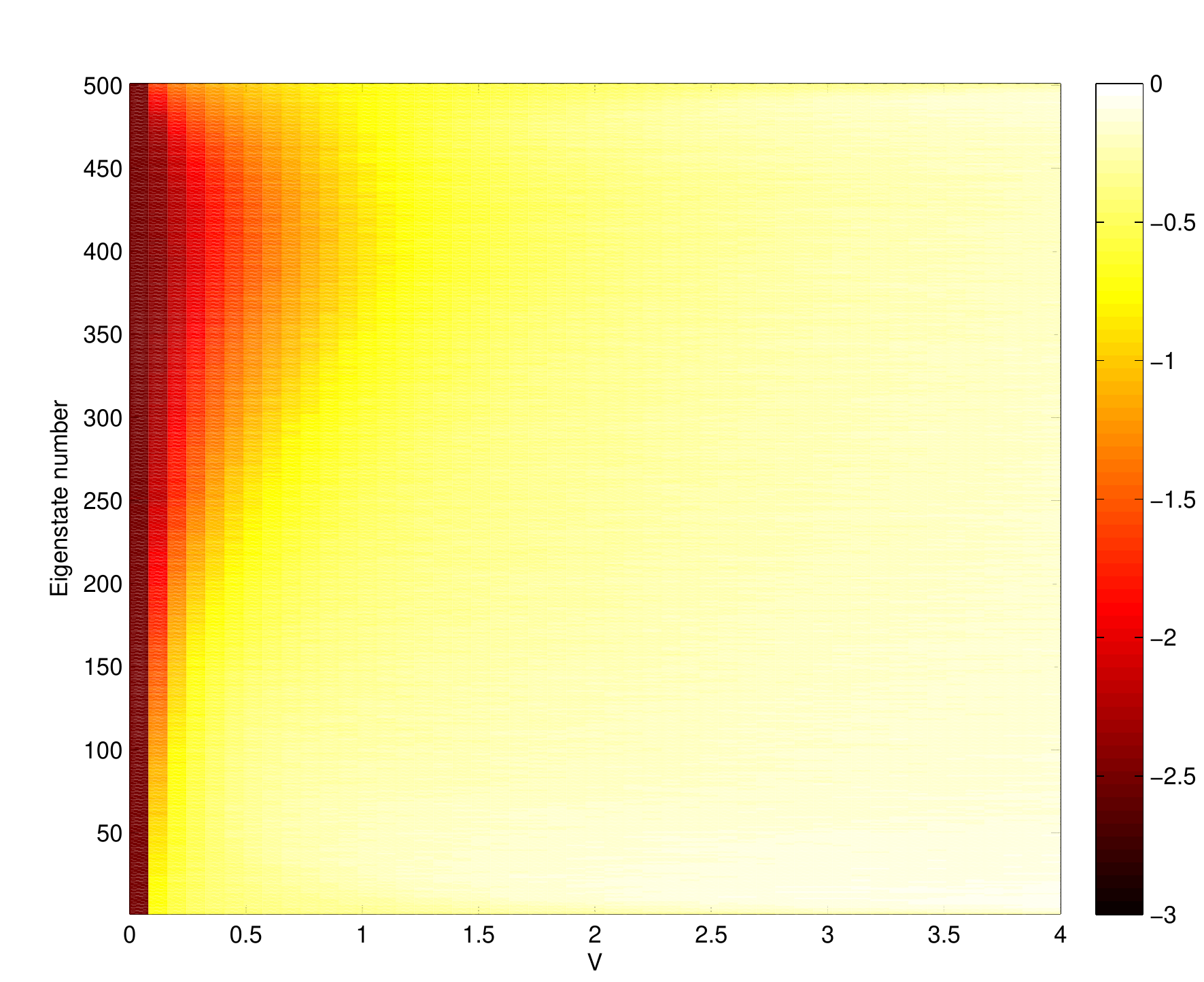}\label{fig:subeavg1}}
\subfigure[$p=2$]{\includegraphics[type=pdf,ext=.pdf,read=.pdf,width=.47\textwidth]{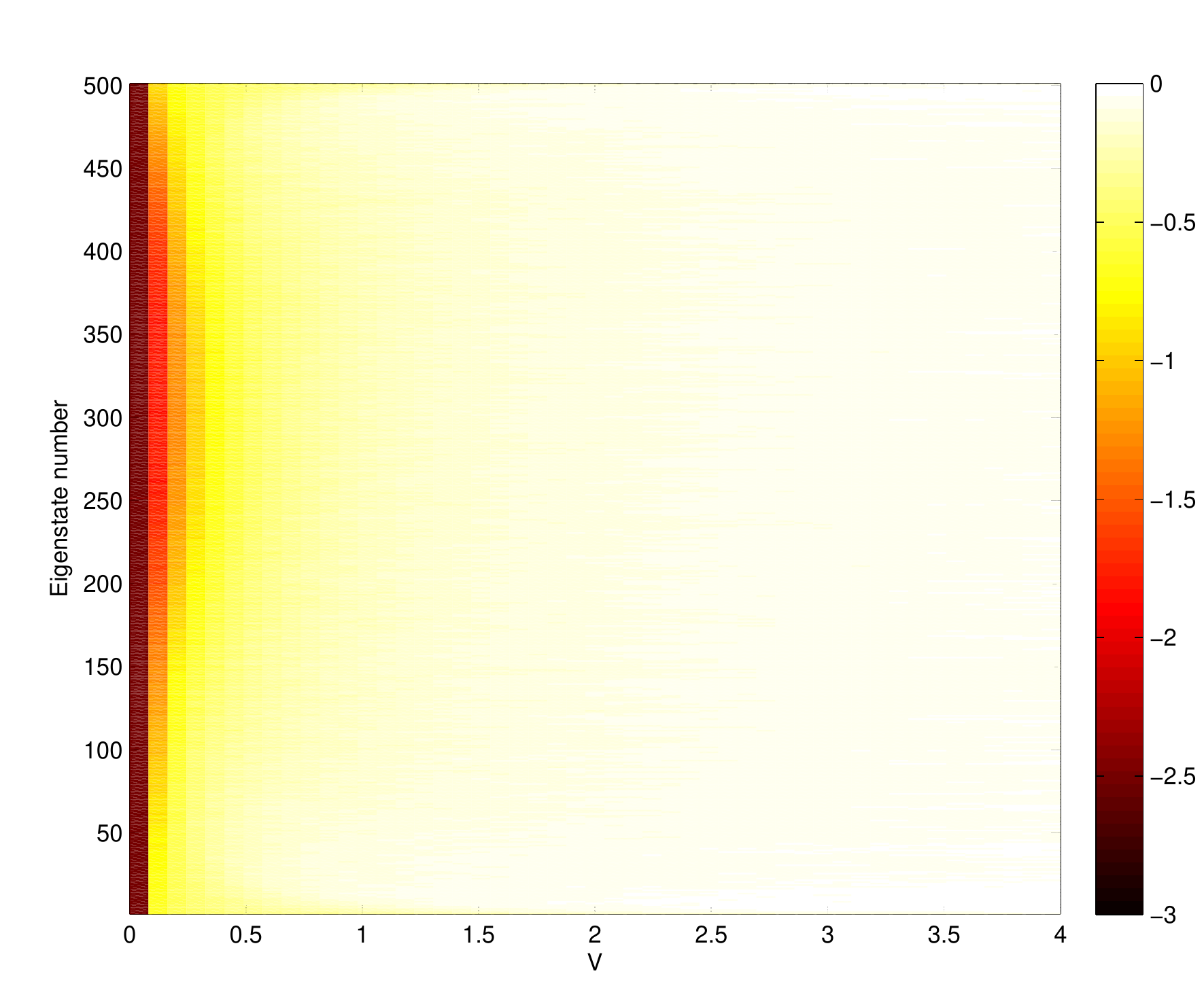}\label{fig:subeavg2}}
\caption{\label{fig:expavg} (Color online) Surface plots of the base 10 logarithm of the IPRs for the exponential hopping model averaged over 100 realizations of random, uniformly distributed potentials with 500 lattice sites.}
\end{figure}

\begin{figure}
\subfigure[$r=1/3$]{\includegraphics[type=pdf,ext=.pdf,read=.pdf,width=.47\textwidth]{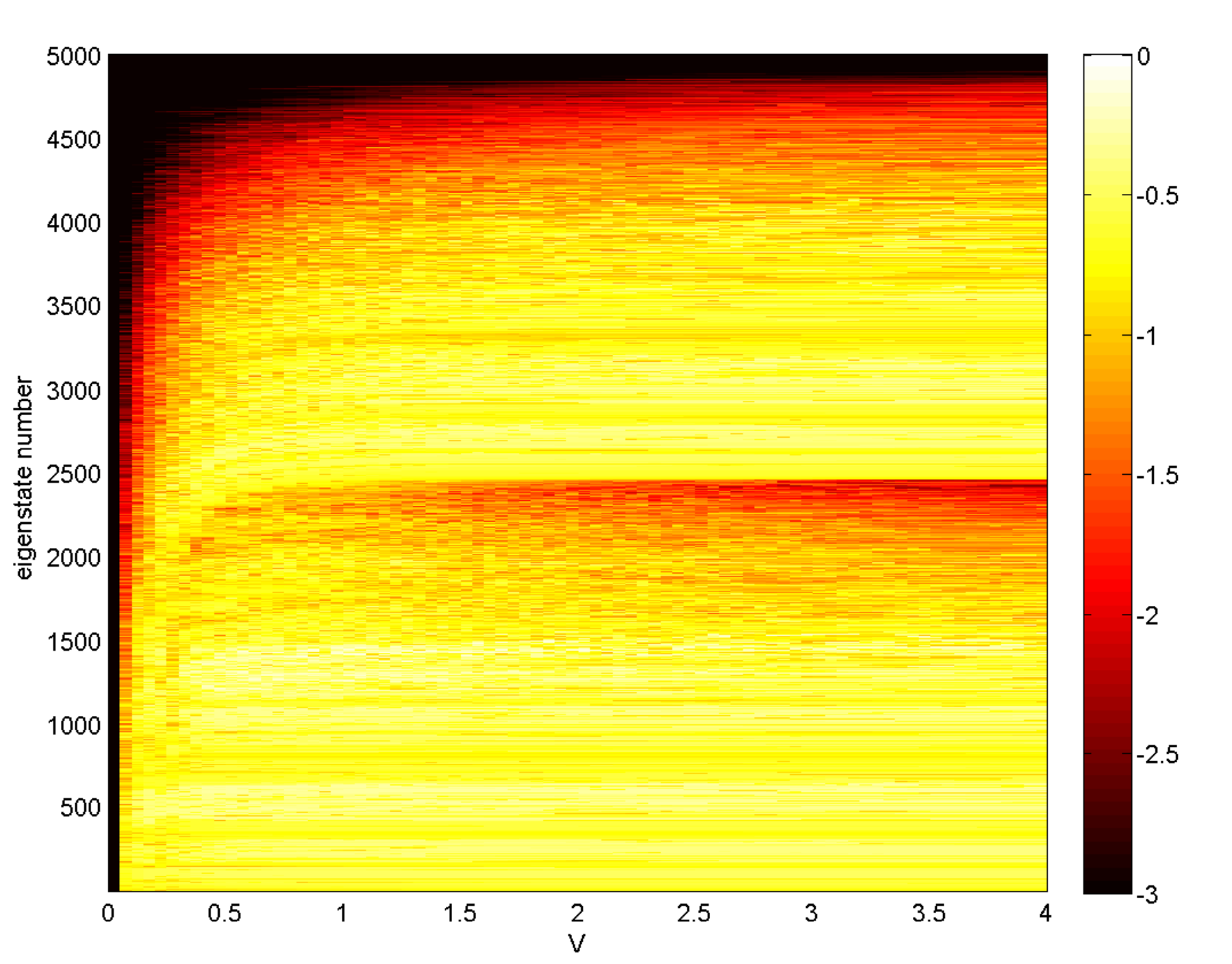}\label{fig:subpbinp33}}
\subfigure[$r=1$]{\includegraphics[type=pdf,ext=.pdf,read=.pdf,width=.47\textwidth]{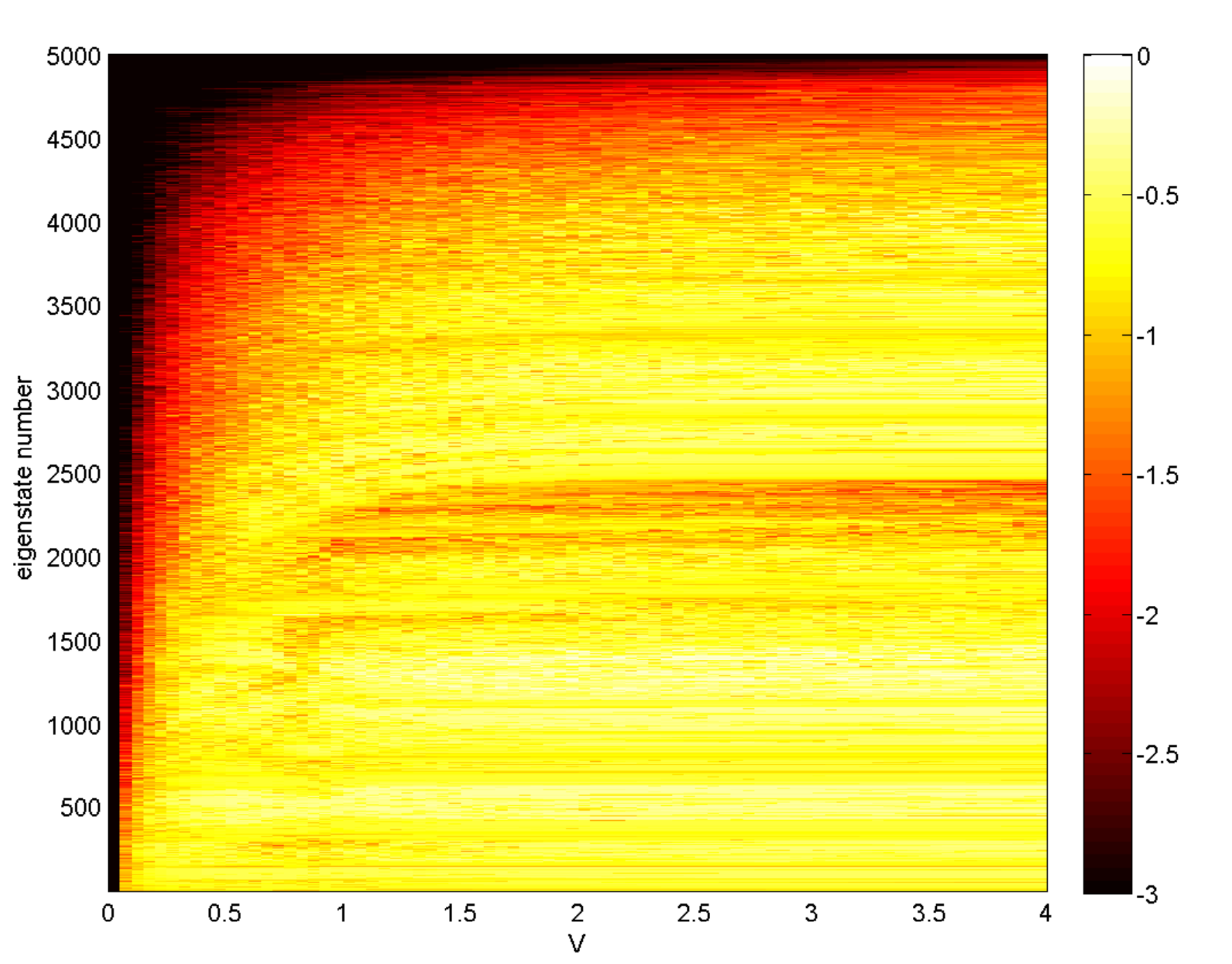}\label{fig:subpbin1}}
\subfigure[$r=2$]{\includegraphics[type=pdf,ext=.pdf,read=.pdf,width=.47\textwidth]{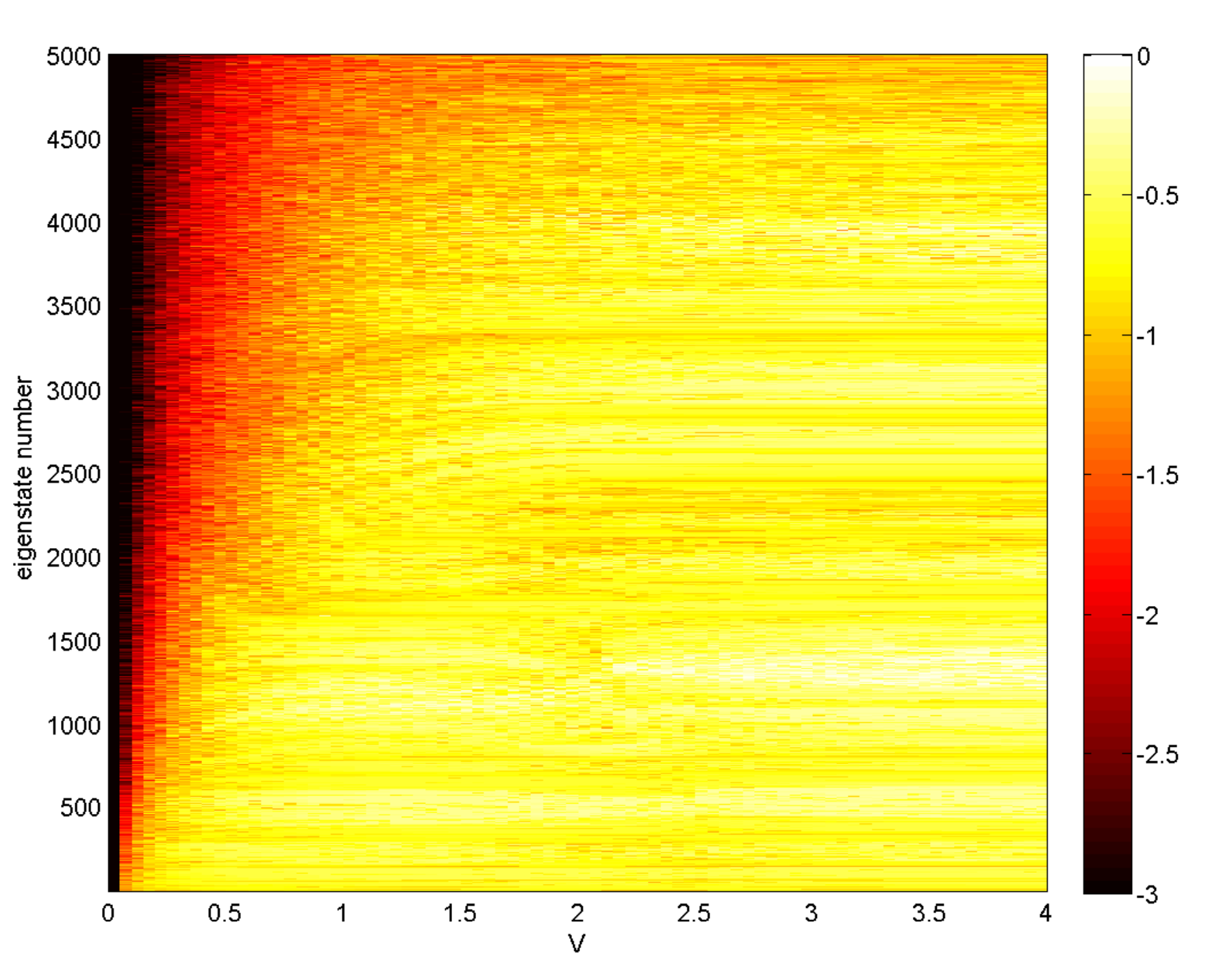}\label{fig:subpbin2}}
\caption{\label{fig:polybin} (Color online) Surface plots of the base 10 logarithm of the IPRs for the inverse-power-law model with a random, binary potential for 5000 lattice sites.}
\end{figure}

\begin{figure}
\subfigure[$r=1/3$]{\includegraphics[type=pdf,ext=.pdf,read=.pdf,width=.47\textwidth]{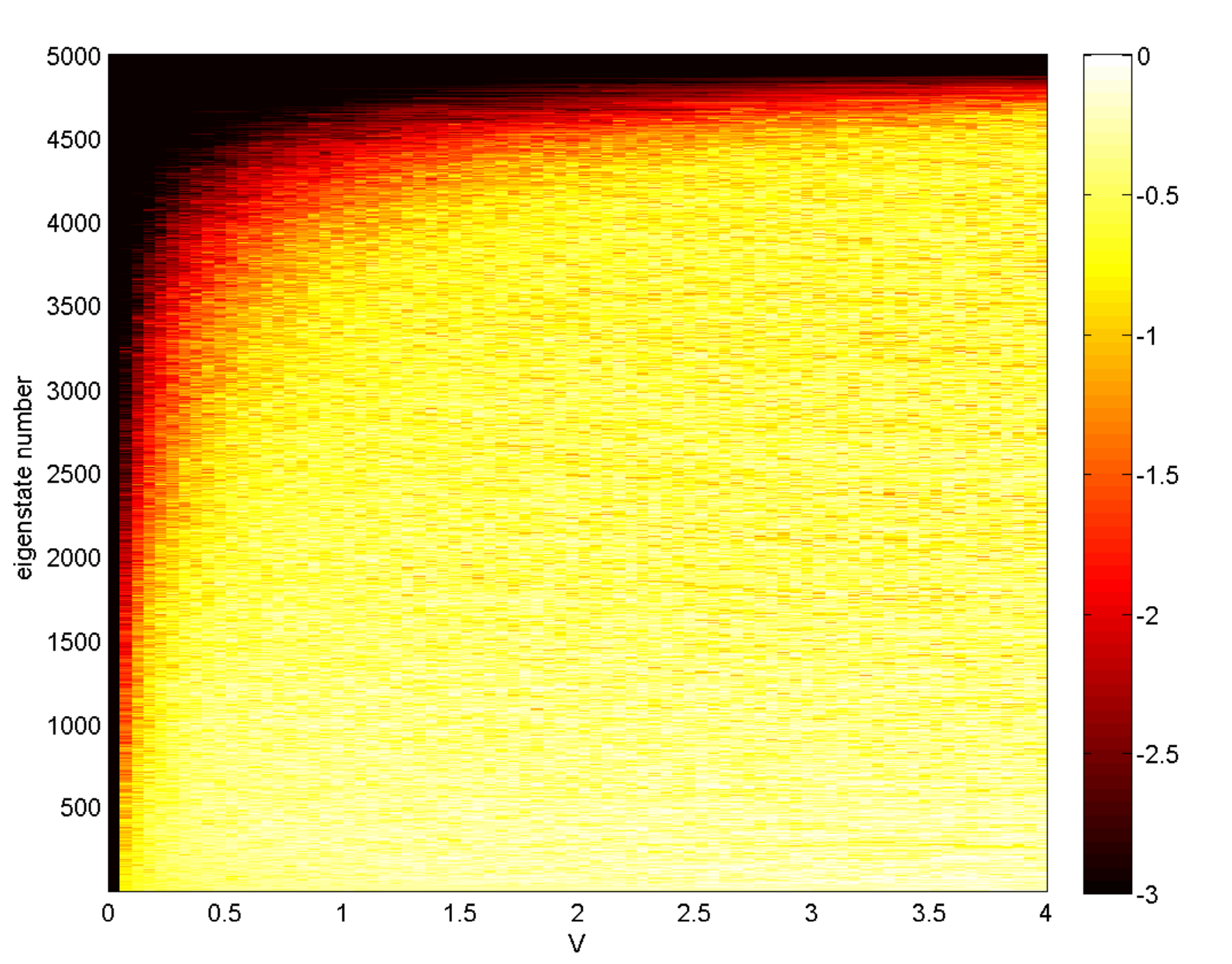}\label{fig:subpunip33}}
\subfigure[$r=1$]{\includegraphics[type=pdf,ext=.pdf,read=.pdf,width=.47\textwidth]{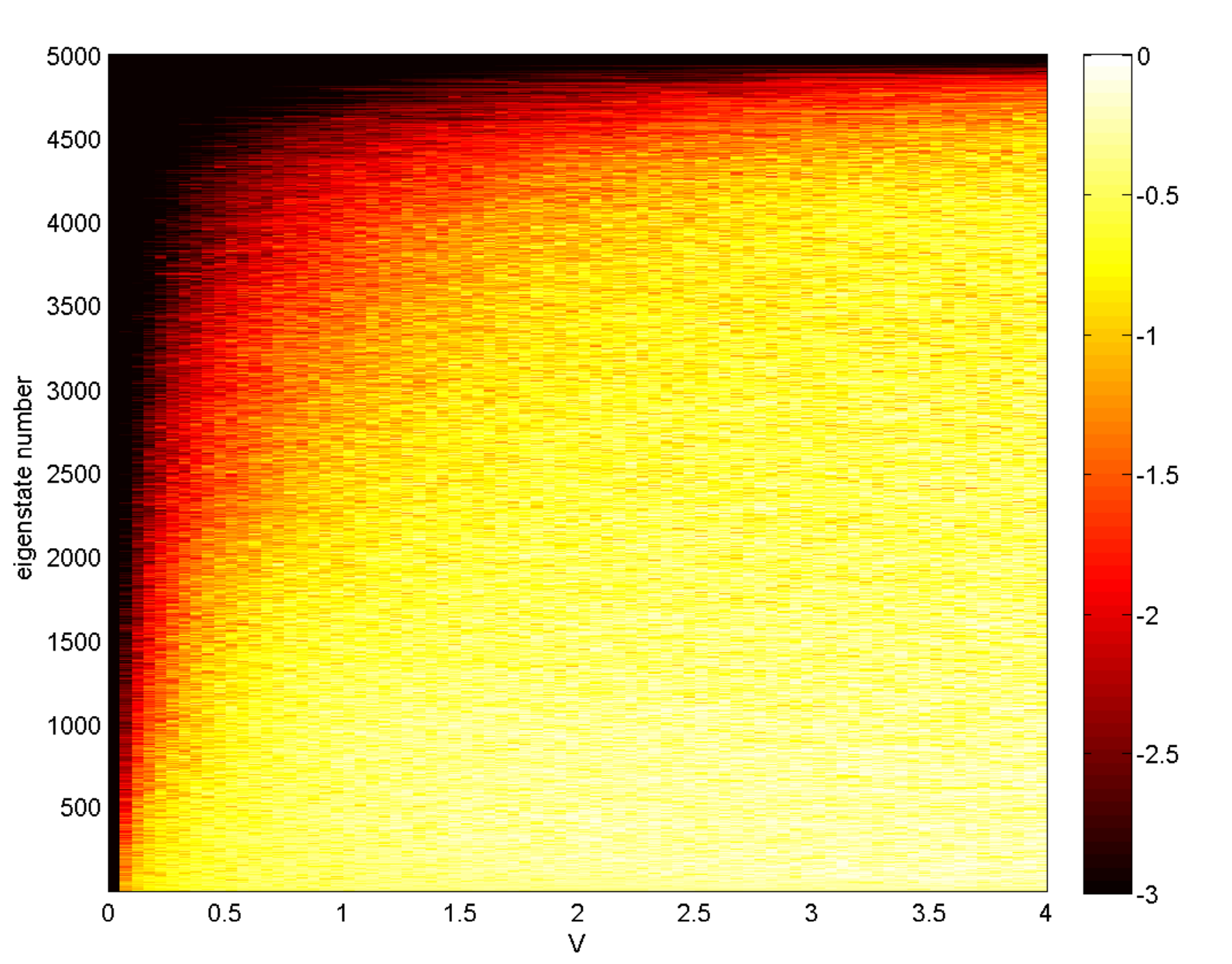}\label{fig:subpuni1}}
\subfigure[$r=2$]{\includegraphics[type=pdf,ext=.pdf,read=.pdf,width=.47\textwidth]{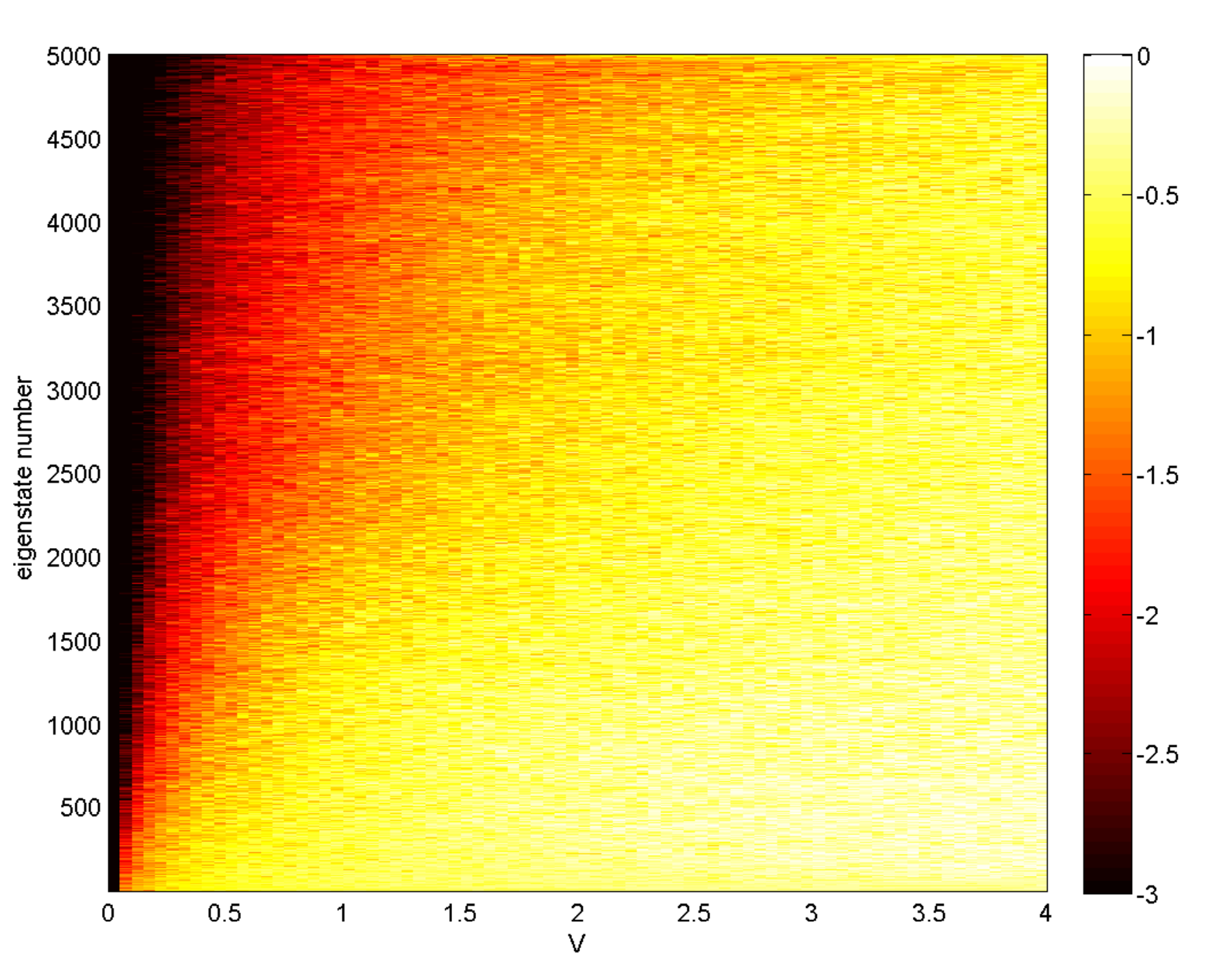}\label{fig:subpuni2}}
\caption{\label{fig:polyuni} (Color online) Surface plots of the base 10 logarithm of the IPRs for the inverse-power-law model with a random, uniformly distributed potential for 5000 lattice sites.}
\end{figure}

\begin{figure}
\subfigure[$r=1/3$]{\includegraphics[type=pdf,ext=.pdf,read=.pdf,width=.45\textwidth]{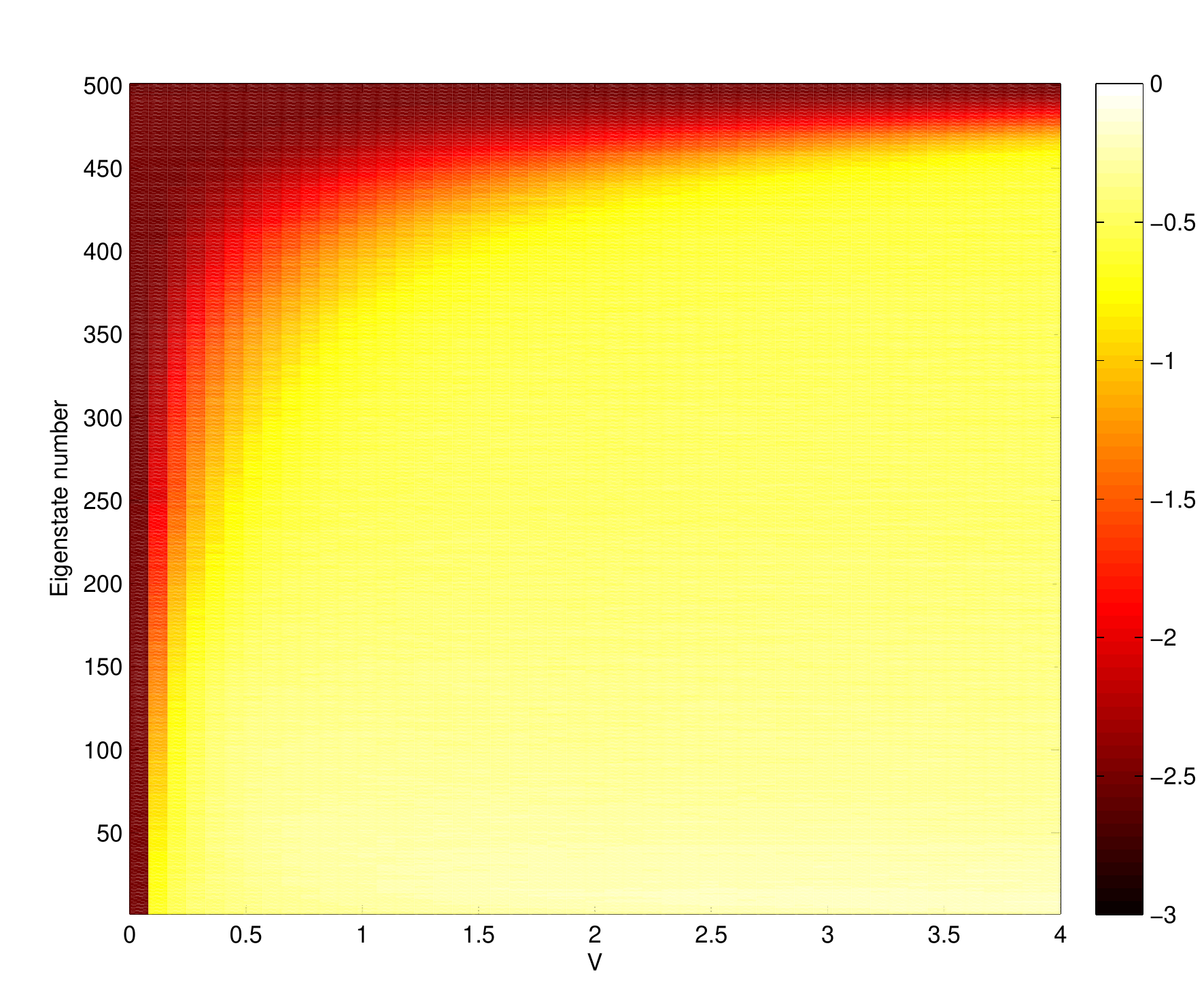}\label{fig:subpavgp33}}
\subfigure[$r=1$]{\includegraphics[type=pdf,ext=.pdf,read=.pdf,width=.45\textwidth]{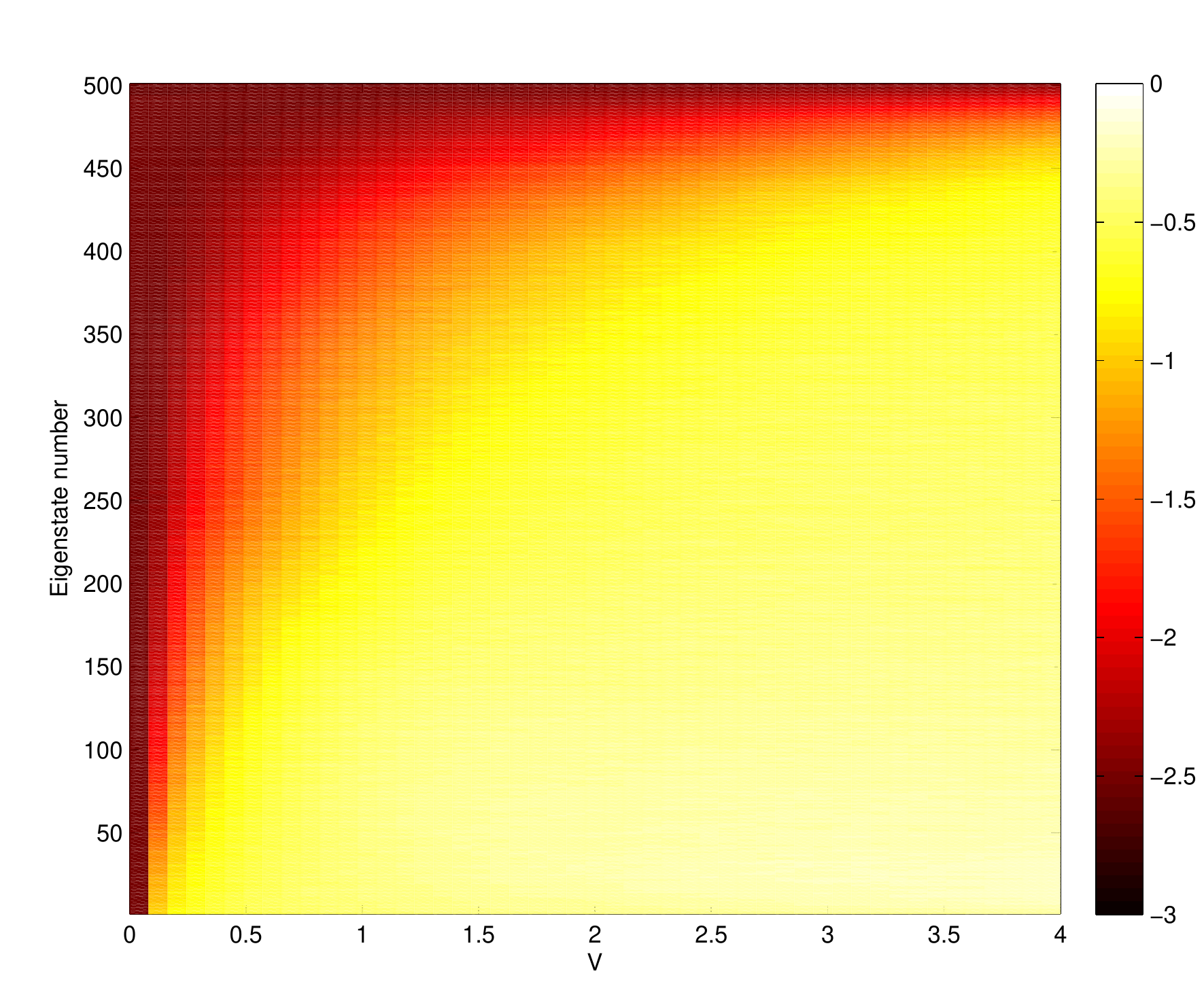}\label{fig:subpavg1}}
\subfigure[$r=2$]{\includegraphics[type=pdf,ext=.pdf,read=.pdf,width=.45\textwidth]{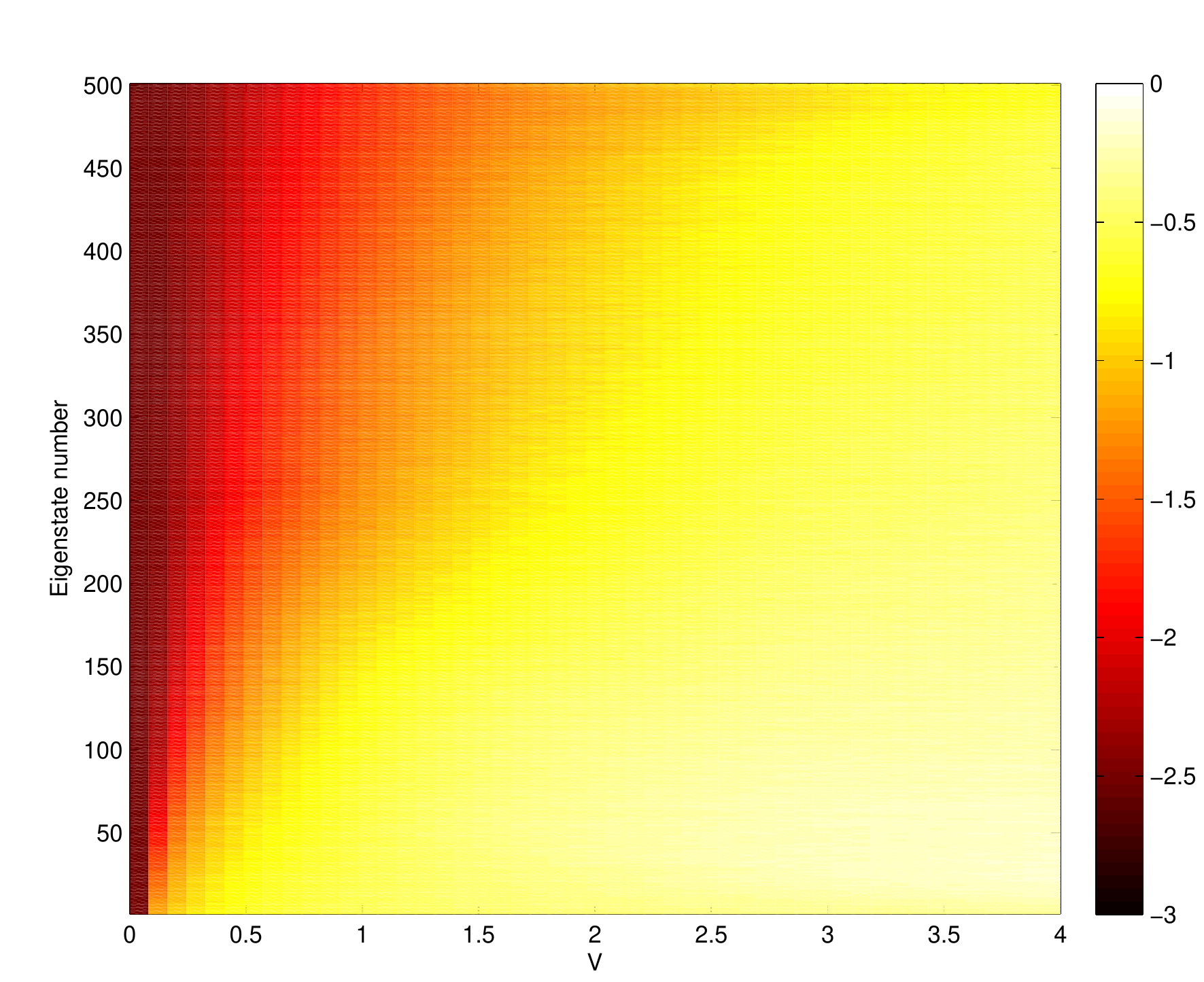}\label{fig:subpavg2}}
\caption{\label{fig:polyavg} (Color online) Surface plots of the base 10 logarithm of the IPRs for the inverse-power-law model averaged over 100 realizations of random, uniformly distributed potentials with 500 lattice sites.}
\end{figure}

In implementing the on-site random potential, we operate either in
terms of a binary potential where the potential at a particular
site assumes the value $V$ or $-V$ with equal probability or
a continuous one which is chosen with uniform
probability between the bounds $V$ and $-V$. In either case, the
strength of the random potential may be considered to be
parameterized by $V$. We begin by studying the $t_1-t_2$ model
with random on-site disorder. Fig. \ref{fig:t1t2binary} gives the
IPR as a function of the disorder strength, $V$, and
the eigenstate number, $i$, for the eigenstates of the $t_1-t_2$
tight-binding model with a binary random on-site
potential and 5000 lattice sites. In the figure we see that for $t_2=0$, we have the 1D Anderson model where localization is relatively weak at the
band center and comparatively strong near the band edges
\cite{Kramer93}.  However as we increase $t_2$, we see that the weakly localized states shift to higher energies toward the top band edge.
Moreover, as the relative value of $t_{2}$ is increased, cross
sections of the IPR along the eigenstate number axis develop a
bimodal profile.  In the case where $t_2=0.8$ [Fig.
\ref{fig:subt2binp8}] the bimodal structure is particularly stark.
In this case we begin to see two distinct regions of weakly
localized states, in contrast to the Anderson case where the localization is weakest in the band center, giving way to more strongly localized states at the band edges.  We also obtain qualitatively similar results when we consider a uniformly distributed, rather than binary,
random potential (Fig. \ref{fig:t1t2uniform}).

In addition to surface plots produced for a single realization of
the random potential, we also generate surface plots of the IPR
with respect to $V$ and eigenstate number where the results are
averaged over a large number of random configurations. In
particular, in Fig. \ref{fig:t1t2avg}, we average the IPR over 100
different realizations of an on-site, uniformly distributed random
potential for the $t_1-t_2$ model with 500 lattice sites.  The
qualitative similarities among Figures \ref{fig:t1t2binary},
\ref{fig:t1t2uniform}, and \ref{fig:t1t2avg} suggest
that the localization behavior is self-averaging in
the sense that statistical fluctuations play a small role in
determining the characteristics of the system if the system size
$N$ is sufficiently large.

We now study the randomly disordered tight-binding model with a
Gaussian decay in the hopping terms.
Fig.~\ref{fig:gaussbinary} and Fig.~\ref{fig:gaussuni} give the IPR
as a function of potential strength, $V$, and eigenstate number,
$i$, of the eigenstates of the Gaussian decay tight-binding model
[i.e. $t_m = \exp(-\sigma |m|^2)$] with a random binary and random
uniform on-site potential, respectively, and 5000 lattice sites.  In
both Fig.~\ref{fig:gaussbinary} and
Fig.~\ref{fig:gaussuni}, we see that for small $\sigma$, the
weakly localized states appear at higher energies in comparison to
the Anderson model (similar to the $t_1-t_2$
model for large $t_2$).  For larger $\sigma$, the IPR is
qualitatively similar to that of the Anderson
model case as would be expected
given the very rapid decay which strongly suppresses
hopping beyond the nearest neighbors for which the Anderson
model is an idealization with hopping confined strictly
to nearest neighbors.  We also average the IPR over 100 different realizations
of an on-site, uniformly distributed random potential
for the Gaussian decay hopping model with 500
lattice sites and report the results in Fig.
\ref{fig:gaussavg}. Again, the qualitative similarities among
 Fig.~\ref{fig:gaussbinary},
Fig.~\ref{fig:gaussuni}, and
Fig.~\ref{fig:gaussavg} suggest that
self-averaging is at work in the localization characteristics of
the eigenstates.  The averaging over many realizations of disorder
has the effect of removing much of the graininess due to
statistical fluctuations which would not appear in the bulk
limit and seem to be finite size artifacts, while preserving more
smoothly varying characteristics which appear to be associated
with the bulk limit. 

Similar results can be seen in the randomly disordered
tight-binding model with exponentially decaying hopping terms.
Figures \ref{fig:expbin} and \ref{fig:expuni} give the IPR as a
function of potential strength, $V$, and eigenstate number, $i$,
of the eigenstates of the exponential hopping tight-binding model
[i.e. $t_m = \exp(-p |m|)$] with a random, binary and random,
uniform on-site potential, respectively, and 5000 lattice sites.
Similar to the Gaussian decay model, we see that for small $p$,
the weakly localized states appear at higher energies compared to
the Anderson case, and for larger $p$, the IPR approaches
that in the Anderson case.  Moreover, just as we saw in
the $t_1-t_2$ and Gaussian models,
 evidence of self-averaging may be seen
by examining the average of the IPR over 100
realizations of a uniform random potential with 500 lattice sites
(Fig. \ref{fig:expavg}).

We now turn to the randomly disordered tight-binding
extended model with hopping terms that decay by
an inverse-power-law.  Since this slow form of decay essentially
allows for long-range hopping, the localization characteristics of this
model differ more from those of the Anderson model, where the
hopping scheme is short-ranged, than those of the other models we have
investigated.
In Figures  \ref{fig:polybin} and
\ref{fig:polyuni}, we show the IPR for a random, binary and
random, uniform on-site potential respectively as a function of
potential strength, $V$, and eigenstate number, $i$, of the
eigenstates of the tight-binding model with hopping terms that
fall off as $t_m = 1/|m|^r$.  In these figures, we see that the
states near the top band edge are weakly localized, and the top
band edge is possibly de-localized for $r=1/3$ and $r=1$.  This
supports earlier theoretical and numerical work which
suggests that there is, indeed, a mobility edge at the top band
edge for long-range hopping \cite{Moura, Malshev, Xiong}.
We revisit this question by examining the statistical
distribution of the IPR for successive system size doublings.  
As in the case of the previous models with short-range hopping, we
again find evidence of self-averaging
 in the surface IPR plots in
this model with hopping terms decaying slowly, as power laws,
by examining the average of the IPR over 100
realizations of a uniform random potential with 500 lattice sites
(Fig. \ref{fig:polyavg}). Again, averaging over
multiple realizations of disorder yields a smoother IPR plot by
removing minor noisy features which are essentially statistical
fluctuations about the bulk limit.

The surface plots show the structure of the inverse participation
ratio with respect to measures of the random
potential strength such as $V$ and the eigenstate number index.
Alternatively, one may concentrate on the frequency of particular
values of the IPR as a way to obtain a
statistical  description of the characteristics of the eigenstates
with respect to localization.  Although
we lose specific information for individual eigenstates,
we gain in return the ability to observe trends in the statistical
characteristics of the IPR distributions with respect to increasing
system size; in this way, we determine in a rigorous manner what portion of the states are localized and what portion, if any, have extended character. 

Localized states are associated with a finite IPR value, whereas
the IPR will tend to zero for extended
states. Hence, if in the bulk limit (i.e., in the limit of very
large $N$) all of the states for a particular random potential
strength $V$ are localized, the histogram will cease to change and
take the form of a constant profile independent of $N$, which is
determined only by the strength of the random disorder and the
extended coupling scheme.

On the other hand, if a finite portion of the states have genuine
extended character, the IPR will continue
to decrease for a fraction of the eigenstates, and a portion of
the histogram total weight will move steadily toward lower IPR
values. In the case of short ranged models such as the nn Anderson
model, we find the IPR histogram to eventually shift to a profile
which is converged with respect to increasing system size $N$, and
no further variation is seen in the shape or position of the IPR
histogram curve. With this in mind, we focus our efforts on the
long-range hopping model (i.e., power-law decay in hopping terms)
and examine the scaling of the IPR histograms with system size to
determine the presence of mobility edges, which are suggested by
the IPR surface plots.

In preparing the IPR histograms, it is important to average away
statistical fluctuations by sampling a sufficient number of
eigenvalues; for each system size we considered, we sample at
least $n_{\mathrm{eig}} = 4 \times 10^{5}$ eigenvalues.
 We obtain the required statistics by diagonalizing
$N_{\mathrm{samp}} = n_{\mathrm{eig}}/N$ Hamiltonian matrices; the
number of separate matrices to be considered decreases with
increasing system size $N$  as self-averaging within an individual
random potential configuration supplies more statistics for larger
systems. Hence, whereas $40,000$ matrices are analyzed for $N =
100$, for the largest system, $N = 6400$, we examine only $80$
random potential realizations.

\begin{figure}
\includegraphics[type=pdf,ext=.pdf,read=.pdf,width=.49\textwidth]{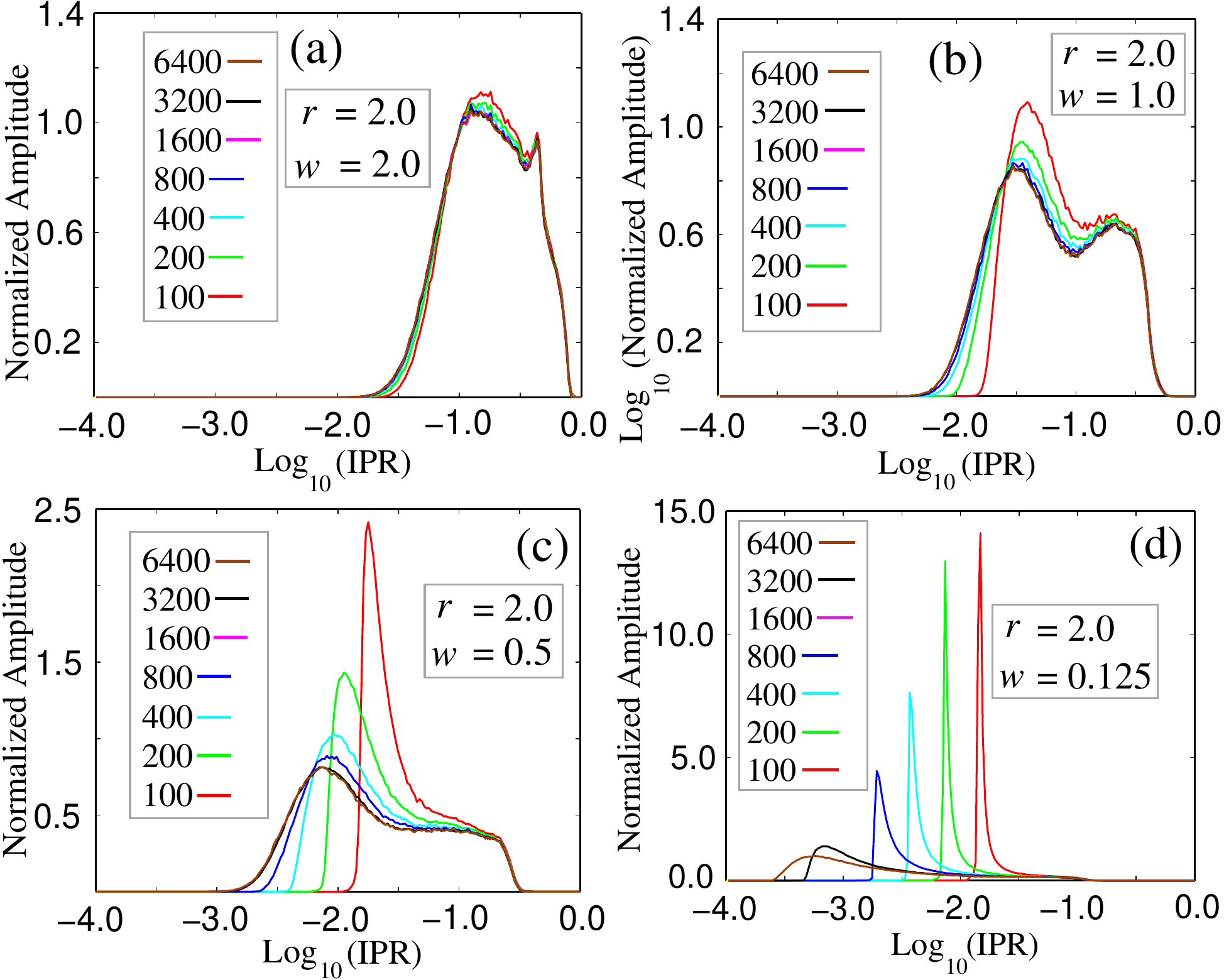}
\caption{\label{fig:hist2p0}  (Color online)
Histograms are shown versus $\log_{10} (\textrm{IPR})$ for the
decay power $r = 2.0$ for systems ranging in size (by successive
doubling) from $N = 100$ to $N = 6400$. Panels (a), (b), (c), and
(d) correspond to $V = 2.0~,1.0,~0.5~\textrm{and}~0.125$,
respectively.}
\end{figure}

In Fig.~\ref{fig:hist2p0}, the tunneling matrix element decreases
relatively rapidly with $r = 2.0$ for the decay power, and the
histograms shown in panels (a), (b), (c), and (d) are consistent
with the scenario in which all eigenstates are localized in the
bulk limit, even for small values of $V$ or weak random
potentials. In each of the panels (a), (b), (c), and (d) of
Fig.~\ref{fig:hist2p0}, histogram curves corresponding to various
systems sizes $N$ are shown, where the sequence of system sizes is
chosen to facilitate the study of the effect of successive
doubling of $N$ on the histogram profile. In panels (a) and (b),
convergence to an invariant histogram curve corresponding to
localized states is relatively swift, while the approach to the
limiting profile is more gradual in the curves shown in panel (c),
where $V = 0.5$, and panel (d) with $V = 0.125$. Nevertheless, the
graphs in Fig.~\ref{fig:hist2p0} indicate a stabilization of the
results with respect to doubling $N$ for each random potential
strength shown, and we conclude for the hopping term
decay exponent $r = 2.0$ that essentially any random on-site
potential strength (irrespective of the strength $V$)
is sufficient to localize states in the bulk limit).

\begin{figure}
\includegraphics[type=pdf,ext=.pdf,read=.pdf,width=.49\textwidth]{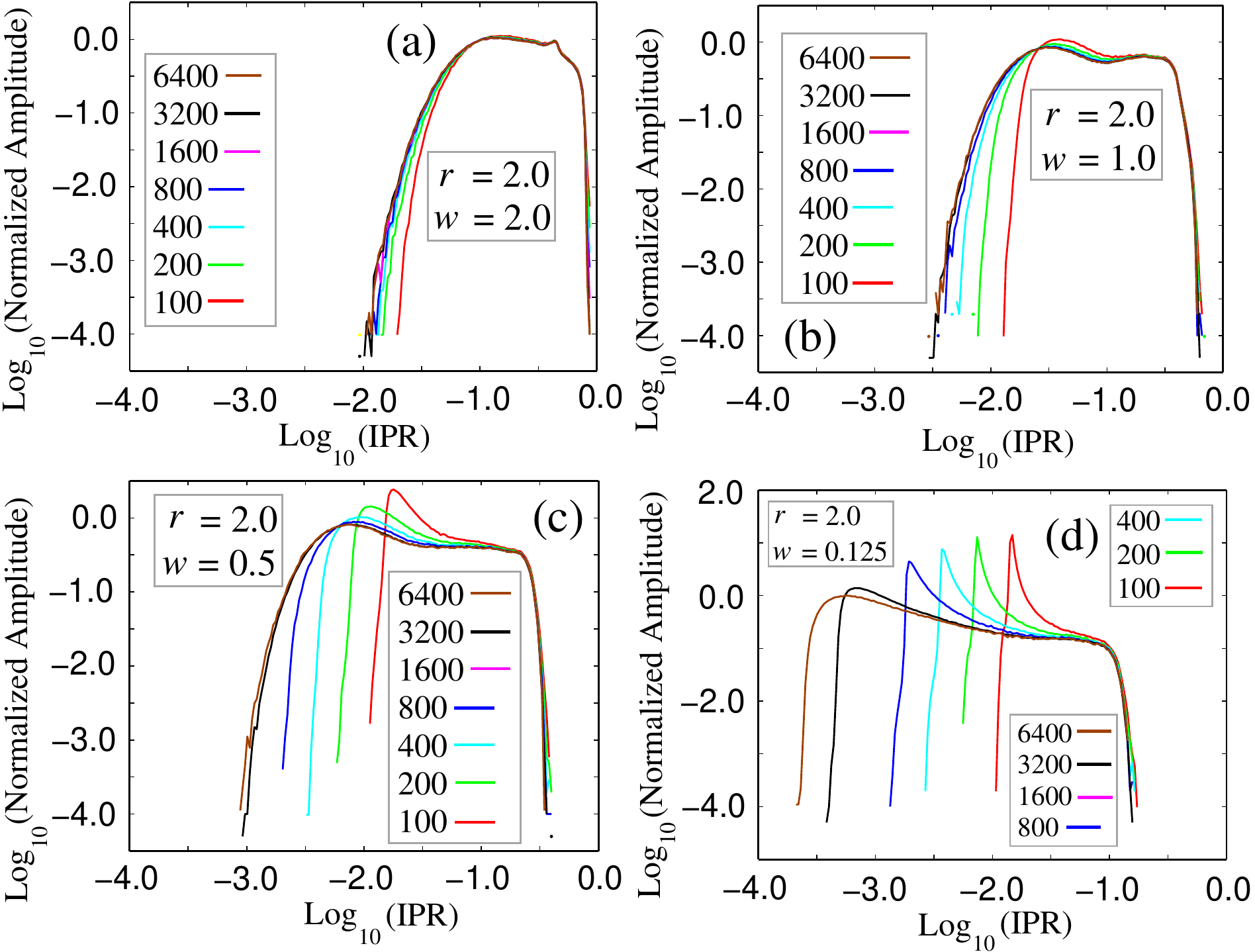}
\caption{\label{fig:hist2p0log} (Color online)
Histograms with the amplitude given as a base ten logarithm are
shown versus $\log_{10} (\textrm{IPR})$ for the decay power $r =
2.0$ for systems ranging in size (by successive doubling) from $N
= 100$ to $N = 6400$. Panels (a), (b), (c), and (d) correspond to
$V = 2.0,~1.0,~0.5,~\textrm{and}~0.125$, respectively.}
\end{figure}

For the graphs shown in Fig.~\ref{fig:hist2p0log}, the vertical
axis represents the base 10 logarithm of the histogram amplitude,
and the random potential strengths $V$ are identical to those of
the corresponding graphs in Fig.~\ref{fig:hist2p0}. A salient
feature of the curves is the convergence to a profile
which terminates for a particular IPR
 with no histogram weight above this upper limit
IPR value. For sufficiently large values of the decay exponent
$r$ (i.e. for at least $r \geq 2.0$), there is a
minimum value $\textrm{IPR}_{\mathrm{min}}$ where
the histogram amplitude abruptly falls to zero, and there is no
probability of finding states with a lower IPR, a condition indicating the absolute localization of all states in the bulk limit.

\begin{figure}
\includegraphics[type=pdf,ext=.pdf,read=.pdf,width=.49\textwidth]{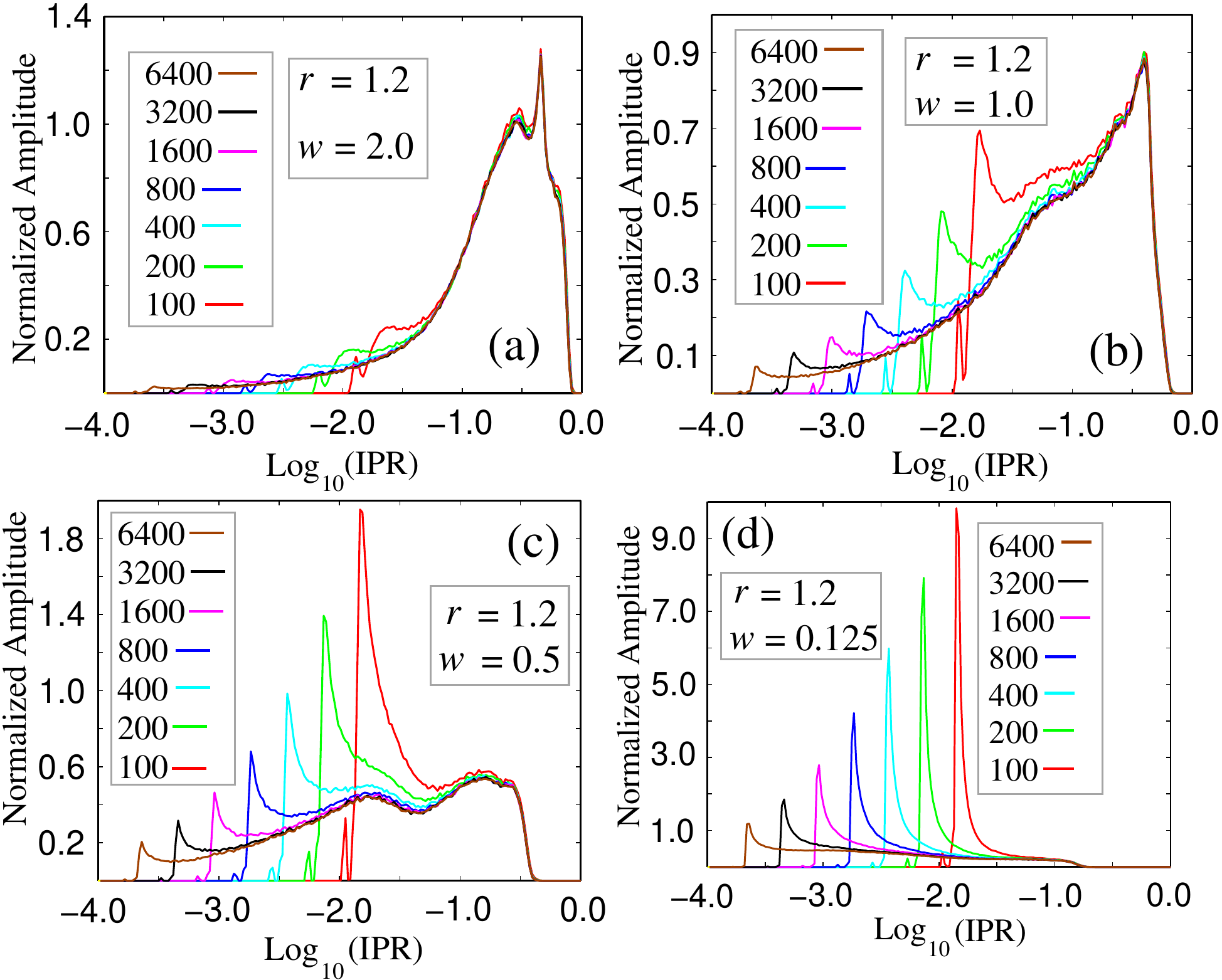}
\caption{\label{fig:hist1p2}  (Color online)
Histograms are shown versus $\log_{10} (\textrm{IPR})$ for the
decay power $r = 1.2$ for systems ranging in size (by successive
doubling) from $N = 100$ to $N = 6400$. Panels (a), (b), (c), and
(d) correspond to $V = 2.0,~ 1.0,~ 0.5,~\textrm{and}~0.125$,
respectively.}
\end{figure}

\begin{figure}
\includegraphics[type=pdf,ext=.pdf,read=.pdf,width=.49\textwidth]{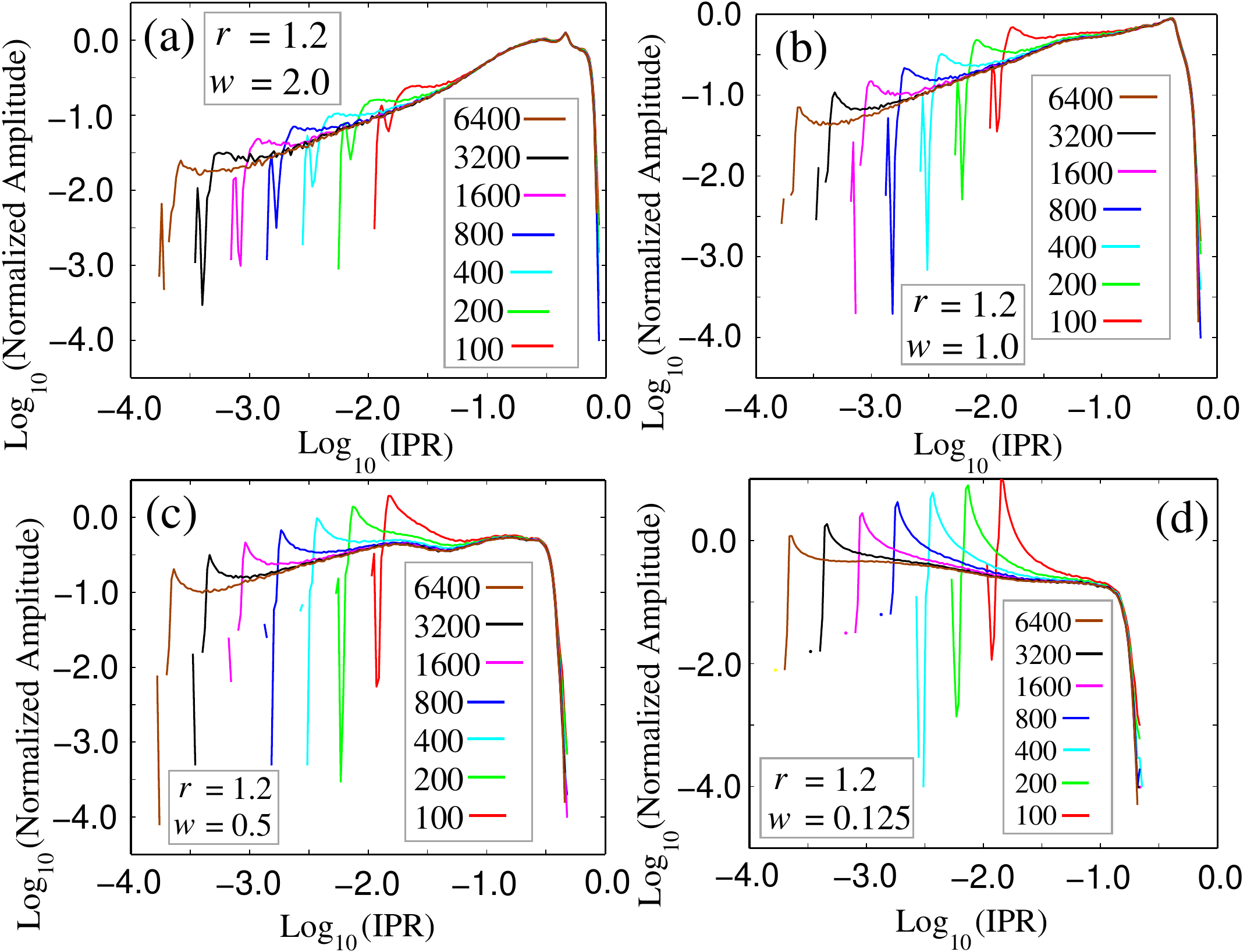}
\caption{\label{fig:hist1p2log} (Color online)
Histograms with the amplitude given as a base ten logarithm are
shown versus $\log_{10} (\textrm{IPR})$ for the decay power $r =
1.5$ for systems varying in
size (by successive doubling) from $N = 100$ to $N = 6400$. Panels
(a), (b), (c), and (d) correspond to $V = 2.0,~
1.0,~0.5,~\textrm{and}~0.125$, respectively.}
\end{figure}

\begin{figure}
\includegraphics[type=pdf,ext=.pdf,read=.pdf,width=.49\textwidth]{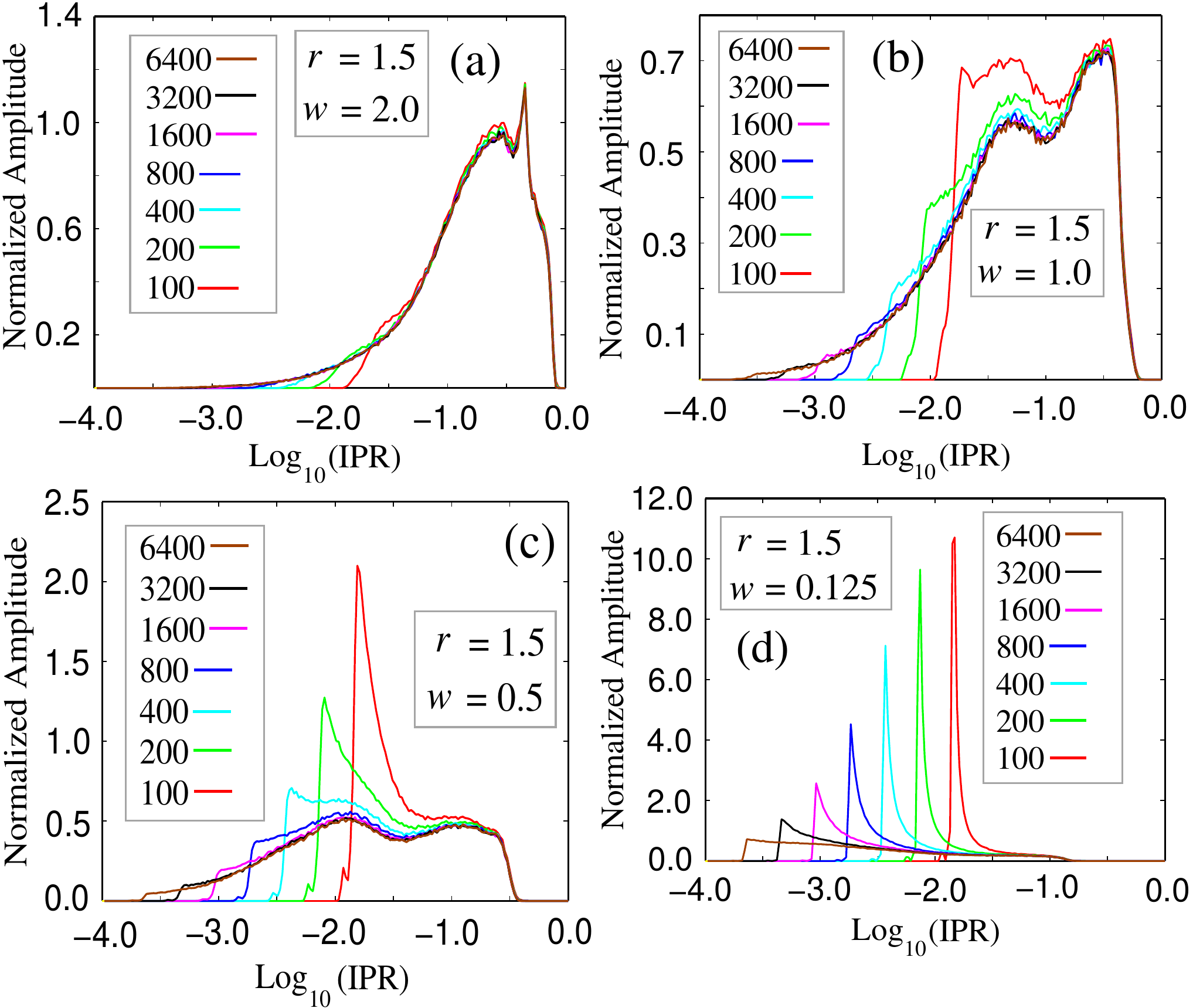}
\caption{\label{fig:hist1p5}  (Color online)
Histograms are plotted versus $\log_{10} (\textrm{IPR})$ for the
decay power $r = 1.5$ for systems ranging in size (by successive
doubling) from $N = 100$ to $N = 6400$. Panels (a), (b), (c), and
(d) correspond to $V = 2.0,~1.0,~0.5,~\textrm{and}~0.125$,
respectively.}
\end{figure}

\begin{figure}
\includegraphics[type=pdf,ext=.pdf,read=.pdf,width=.49\textwidth]{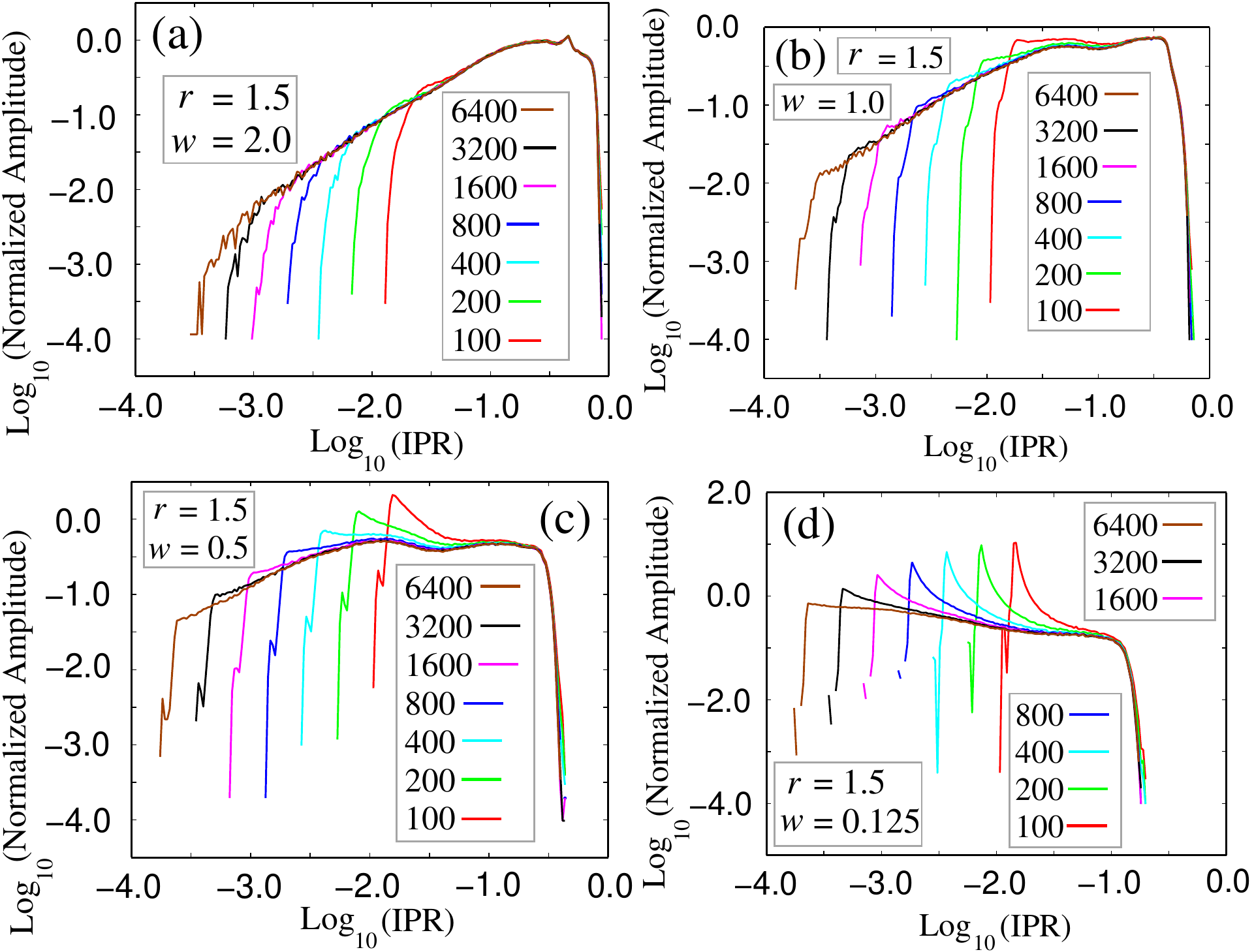}
\caption{\label{fig:hist1p5log} (Color online)
Histograms with the amplitude given as a base ten logarithm are
shown versus $\log_{10} (\textrm{IPR})$ for the decay power $r =
1.5$ for systems ranging in size (by successive doubling) from $N
= 100$ to $N = 6400$. Panels (a), (b), (c), and (d) correspond to
$V = 2.0,~1.0,~0.5,~\textrm{and}~0.125$, respectively.}
\end{figure}

As the decay exponent $r$ is decreased, there seems to be a
threshold value $r_{c}$ where the properties of the IPR histogram
change in a qualitative manner with respect to increasing $N$.
Fig.~\ref{fig:hist1p2} and Fig.~\ref{fig:hist1p2log} show
histogram curves for $r = 1.2$, whereas Fig.~\ref{fig:hist1p5} and
Fig.~\ref{fig:hist1p5log} display IPR histograms for the case $r =
1.5$ where the decay of the tunneling coefficients is faster.
A salient common characteristic in the graphs obtained for $r =
1.2$ and $r = 1.5$ suggests both decay exponents are below the
threshold value $r_{c}$. In contrast to the $r = 2.0$ case, where
the histogram profiles converge to a curve which does not change
with successive doubling of the size $N$  (a behavior compatible
with the localization of all states), for both the cases $r = 1.5$
and $r = 1.2$, there is a steady advance of the leftmost edge of
the histogram curve toward smaller IPR values in
the low-IPR regime.
 The size of the increment appears to be essentially the same each time
the size of the system is doubled.
 On the other hand, for larger IPR
values, the rightmost parts of the histogram converge and cease to
change with increasing $N$.

The shift of the leftward edge toward lower $\log_{10}
(\textrm{IPR})$ values occurs at a constant rate with
doubling of the size $N$ of the system,a phenomenon
seen for all of the histograms obtained for $r = 1.2$ and $r =
1.5$. While a bimodal structure may be seen in both the graphs
obtained for $r = 1.2$ and $r = 1.5$, dual peaked character is
most prominently manifest for the slower decay exponent $r = 1.2$
and for lower values of $V$ corresponding to weaker
random potentials. The two peaks behave very differently with
increasing $N$. Whereas the rightmost peak, corresponding to
relatively higher IPR values and hence more localized character
does not shift significantly in location, the peak on the left
migrates steadily toward lower IPR values
with successive doubling of $N$. In addition, the peak height
appears to decrease at a geometric rate each time the system size
is doubled.

The curves shown in Fig.~\ref{fig:hist1p2log} and
Fig.~\ref{fig:hist1p5log}, where the amplitude of the IPR
histogram is presented as a base 10 logarithm, highlight an
important feature for the histograms in the cases $r = 1.2$ and $r
= 1.5$ absent in the case of the more rapidly decaying scheme
where, for example, $r = 2.0$. For large (but finite) $N$ the log-log curves for $r < r_{c}$ may be
divided, crudely speaking, into three regimes. Moving  leftward
along the horizontal $\log_{10}(\textrm{IPR})$ axis, one first
sees a rapid rise to a maximum, and the curve then begins to
decrease with decreasing $\log_{10}(\textrm{IPR})$. For
intermediate values of the logarithm of the Inverse Participation
Ratio, the logarithm of the histogram amplitude
decreases linearly with decreasing $\log_{10}(\textrm{IPR})$.
Finally, the curve rises again to reach a second maximum
before beginning a rapid decline.

The intermediate regime where the logarithm of the histogram
amplitude decreases linearly is a salient common feature which
becomes broader as $N$ is increased (extending further and further
leftward). For $r = 1.2$ and $V = 0.125$, the region where the
dependence is approximately linear is more difficult to discern,
but would be more readily seen for systems sizes beyond the
largest ($N_{\mathrm{max}} = 6400$) we consider in the context of
this study.  However, even though the linear dependence may not be
readily visible, the leftward peaks in the log-log graph decrease
in height at a linear rate even for weaker potentials (e.g., $V =
0.125$) where the intermediate linear region is more
difficult to discern. (The linear decline of the peak height in
the log-log plot with successive system size doublings is
compatible with the geometric decline evident in the
semilogarithmic graphs.) Simple extrapolation would suggest that
as the bulk limit is approached, the leftward edge of the
histogram curve will continue to advance at a constant rate to the
left; ultimately, in the thermodynamic limit, the
asymptotically linear decrease in the logarithm of the histogram
amplitude would continue for arbitrarily small $\log_{10}
(\textrm{IPR})$.
 Hence, in terms of the histogram density $\phi_{\mathrm{IPR}}$, the support for
states decreases with decreasing IPR, ultimately vanishing as the IPR heads to zero (i.e., for \textit{bona fide} extended
states).  Although the statistical weighting decreases
with decreasing IPR for all values of $r$ considered, the decline is much less abrupt
if $r < r_{c}$.

In particular, we infer the explicit dependence for the IPR density $\phi_{\mathrm{IPR}}$ for $r < r_{c}$
will be
$\log_{10} ( \phi_{\mathrm{IPR}} ) = \alpha \log_{10} (\textrm{IPR}) +
\beta$, a relation which would hold in the bulk limit for
reasonably large values of $\log_{10} (IPR)$.  Inverting for
$\phi_{\mathrm{IPR}}$ yields $\phi_{\mathrm{IPR}} = f(r,v)
(\textrm{IPR})^{\alpha(r,v)}$, where $f(r,v)$ and $\alpha(r,v)$ depend
on the hopping decay exponent $r$ and the random potential
strength $V$ ($f$ would also depend on the specific type of random
disorder, be it generated in a binary fashion or sampled from a
uniform distribution).  
For fixed potential
strength $V$, we anticipate that $\alpha(r,v)$ will rise sharply
in the vicinity of $r = r_{c}$, where the decay of the histogram
weight becomes much more rapid than for $r < r_{c}$.

The sudden shift in the behavior of the IPR
histograms as the system size is successively doubled suggests an
abrupt transition from the condition where the asymptotic
dependence of the histogram density is a relatively slow decay
with decreasing IPR, $ \phi_{\mathrm{IPR}} =  f(r,V)
(\textrm{IPR})^{\alpha(r,v)}$ to a much more rapid decrease. The
transition is likely signaled by a divergence in the exponent
$\alpha$ at a critical value $r_{c}$ of the decay exponent in the
extended hopping scheme.

\section{Conclusion}

We have shown with numerical
calculations and analytical results that
(1D) tight-binding models with on-site disorder and higher-order
hopping terms exhibit interesting and non-trivial localization
phenomena that can vary considerably from the well-known results
of the nn tight-binding models. In the specific case
where the on-site potential is an incommensurate potential
(particularly the bi-chromatic problem), we have shown that for a
general expression for the decay in the hopping
terms with range, the energy-dependent mobility edges can be predicted
approximately based on the ratio of the nn and nnn hopping terms,
$t_1/t_2$, for sufficiently fast decay.

We have also considered the case of bounded,
uncorrelated disorder, where we have shown that in the examined models,
the higher order hopping terms can remove the symmetry
in the localization length about the energy band center compared
to the nn Anderson model. Broadly speaking, it appears that eigenstates with lower
energies tend to localize with shorter localization lengths (i.e.
higher IPR values) compared to higher-energy eigenstates due to
the presence of the higher-order hopping terms.

There is also the curious case of inverse-power-law hopping terms where a
mobility edge may manifest itself at the top band edge if the decay exponent is smaller than a certain critical value.

We have prepared histograms of the IPR to determine the
statistical characteristics of the IPR
distribution for the model with power law decay hopping. For
relatively short-range hopping schemes, the histogram weight falls
to zero for finite IPR values, suggesting that all states are
localized in the bulk limit.  However, if the hopping terms decay
sufficiently slowly with distance, the histogram density is still
zero in the low IPR limit, but the decay has a much slower
asymptotic dependence with the form
$f(r,v)(\textrm{IPR})^{\alpha(r,v)}$.

Our results are especially relevant to
current experimental efforts as we
consider alternatives to solid-state systems to study quantum
transport, such as cold atoms in shallow optical lattices. Such
systems may not be as strongly binding as solid-state systems, and
therefore, do not fit well with the nn tight-binding assumption.
Given this and the considerable degree of control given to
experimentalists in optical lattices, these results can be
observed experimentally within cold atomic systems.  In particular, the current experiments in cold atomic gases \cite{Billy08, Roati08} have already verified the basic (and long-established) features of 1D quantum localization properties in the Anderson and the Aubry model.  Experiments in shallow lattices allowing longer-range hopping should enable a deeper understanding of the localization properties discussed in the current work.

We conclude with some discussion of some of the open questions in this subject, which may be of importance for future studies.  One important issue completely beyond the scope of the current work is the effect of interaction on our predicted mobility edges in the generalized one-particle AA model.  In general the solid-state experimental systems are many-particle systems, and interaction is invariably present.   In atomic systems where the interaction is often short-ranged, it is possible to approach the noninteracting limit by having a very dilute system, and this is one reason behind the recent experimental success in studying localization properties in refs. \cite{Billy08,Roati08}.  It is also obvious that, while our one-electron localization theory applies both to fermions and bosons since quantum statistics are irrelevant in the single-particle limit, the many-particle interacting situation is different for fermions and bosons and must be studied independently.   In principle this is a formidable task, although in one dimension progress is feasible by combining numerical and analytical methods.

The interacting bosonic problem is easier to study theoretically, and for the strict nn-hopping AA model \cite{Aubry}, the conclusion --- based on extensive density-matrix-renormalization-group (DMRG) studies of the problem --- appears to be that the main features of the minimal AA model is preserved although a complex phase diagram now emerges in the presence of both interaction and incommensurate potential manifesting a complex interplay between superfluid and Mott insulating phases\cite{Roux08, *Cai10}.  Such numerical studies using DMRG or perhaps the time-evolving-block-decimation (TEBD) technique should, in principle, be possible for our extended AA model, and it will be an interesting future direction to pursue in this problem.  An alternative technique for studying the bosonic problem in the presence of both the incommensurate AA potential and interaction is to utilize the nonlinear Schrodinger equation approach using the so-called Gross-Pitaevski equation (GPE).  Such a study has recently been carried out\cite{Fishman2009, *Krivolapov2010, *Deissler2010, *Larcher2009} for the Anderson model in the bosonic case where the interplay of disorder and interaction was studied both numerically and analytically with the conclusion that the basic qualitative feature of the non-interacting model is not modified by the presence of interaction (i.e. all states remain localized even in the presence of interaction).  Again, such a GPE-based study should, in principle, be possible for our generalized AA model, although it is likely to be numerically challenging.  Based on the existing results, \cite{Roux08, *Cai10, Fishman2009, *Krivolapov2010, *Deissler2010, *Larcher2009} our best guess is that our conclusion in the current work about the existence of a mobility edge in the generalized AA model in the presence of non-nearest neighbor hopping will remain valid for bosonic systems even in the presence of interaction, but more work is necessary to establish this point.

Studying the interacting system becomes much more difficult and complex for fermions where the interplay of interaction and localization is notoriously difficult to study.  It is well known that in general (repulsive) interaction leads to effective delocalization since the interacting particles want to stay away from each other.  On the other hand, strong interaction also causes Mott transition and Wigner crystallization (where the system becomes pinned in the presence of disorder), and thus enhances localization effects in some situations.  Direct numerical diagonalization and other works for the fermionic Hubbard model in the presence of disorder \cite{Kotlyar2001, *Giamarchi1988} indicate that interaction tends to increase the localization length without modifying the basic localization properties of the 1D Anderson model.  Although no detailed investigation of the AA model for fermions in the presence of interaction has yet been undertaken in the literature, it is reasonable to assume, based on the results of the corresponding interacting Anderson model,  that the basic picture of the noninteracting AA  model would remain valid qualitatively even in the presence of interaction.  We therefore believe that the existence of mobility edges in the generalized AA model discussed in the current work will remain valid even in the presence of interaction for both bosons and fermions, but much more work will be needed to settle this issue decisively.  This remains an interesting and important problem for future studies.  
We also mention in this context a recent work \cite{Albert2010} which draws an interesting distinction between the AA model and the Anderson model with respect to the nature of the underlying localization properties, and it will be interesting to investigate whether such an analysis sheds insight into our discovery of a mobility edge in the generalized AA model in the presence of non-nearest-neighbor hopping.
\begin{acknowledgments}
This work is supported by ARO-DARPA-OLE  and JQI-NSF-PFC.
\end{acknowledgments}


\begin{thebibliography}{33}%
\makeatletter
\providecommand \@ifxundefined [1]{%
 \@ifx{#1\undefined}
}%
\providecommand \@ifnum [1]{%
 \ifnum #1\expandafter \@firstoftwo
 \else \expandafter \@secondoftwo
 \fi
}%
\providecommand \@ifx [1]{%
 \ifx #1\expandafter \@firstoftwo
 \else \expandafter \@secondoftwo
 \fi
}%
\providecommand \natexlab [1]{#1}%
\providecommand \enquote  [1]{``#1''}%
\providecommand \bibnamefont  [1]{#1}%
\providecommand \bibfnamefont [1]{#1}%
\providecommand \citenamefont [1]{#1}%
\providecommand \href@noop [0]{\@secondoftwo}%
\providecommand \href [0]{\begingroup \@sanitize@url \@href}%
\providecommand \@href[1]{\@@startlink{#1}\@@href}%
\providecommand \@@href[1]{\endgroup#1\@@endlink}%
\providecommand \@sanitize@url [0]{\catcode `\\12\catcode `\$12\catcode
  `\&12\catcode `\#12\catcode `\^12\catcode `\_12\catcode `\%12\relax}%
\providecommand \@@startlink[1]{}%
\providecommand \@@endlink[0]{}%
\providecommand \url  [0]{\begingroup\@sanitize@url \@url }%
\providecommand \@url [1]{\endgroup\@href {#1}{\urlprefix }}%
\providecommand \urlprefix  [0]{URL }%
\providecommand \Eprint [0]{\href }%
\@ifxundefined \urlstyle {%
  \providecommand \doi  [0]{\begingroup \@sanitize@url \@doi}%
  \providecommand \@doi [1]{\endgroup \@@startlink {\doibase
  #1}doi:\discretionary {}{}{}#1\@@endlink }%
}{%
  \providecommand \doi  [0]{doi:\discretionary{}{}{}\begingroup
  \urlstyle{rm}\Url }%
}%
\providecommand \doibase [0]{http://dx.doi.org/}%
\providecommand \Doi [0]{\begingroup \@sanitize@url \@Doi }%
\providecommand \@Doi  [1]{\endgroup\@@startlink{\doibase#1}\@@Doi}%
\providecommand \@@Doi [1]{#1\@@endlink}%
\providecommand \selectlanguage [0]{\@gobble}%
\providecommand \bibinfo  [0]{\@secondoftwo}%
\providecommand \bibfield  [0]{\@secondoftwo}%
\providecommand \translation [1]{[#1]}%
\providecommand \BibitemOpen [0]{}%
\providecommand \bibitemStop [0]{}%
\providecommand \bibitemNoStop [0]{.\EOS\space}%
\providecommand \EOS [0]{\spacefactor3000\relax}%
\providecommand \BibitemShut  [1]{\csname bibitem#1\endcsname}%
\bibitem [{\citenamefont {Kramer}\ and\ \citenamefont
  {MacKinnon}(1993)}]{Kramer93}%
  \BibitemOpen
  \bibfield  {author} {\bibinfo {author} {\bibfnamefont {B.}~\bibnamefont
  {Kramer}}\ and\ \bibinfo {author} {\bibfnamefont {A.}~\bibnamefont
  {MacKinnon}},\ }\href@noop {} {\bibfield  {journal} {\bibinfo  {journal}
  {Reports on Progress in Physics},\ }\textbf {\bibinfo {volume} {56}},\
  \bibinfo {pages} {1469} (\bibinfo {year} {1993})}\BibitemShut {NoStop}%
\bibitem [{\citenamefont {Anderson}(1958)}]{Anderson}%
  \BibitemOpen
  \bibfield  {author} {\bibinfo {author} {\bibfnamefont {P.~W.}\ \bibnamefont
  {Anderson}},\ }\Doi {10.1103/PhysRev.109.1492} {\bibfield  {journal}
  {\bibinfo  {journal} {Phys. Rev.},\ }\textbf {\bibinfo {volume} {109}},\
  \bibinfo {pages} {1492} (\bibinfo {year} {1958})}\BibitemShut {NoStop}%
\bibitem [{\citenamefont {Aubry}\ and\ \citenamefont {Andr\'e}(1980)}]{Aubry}%
  \BibitemOpen
  \bibfield  {author} {\bibinfo {author} {\bibfnamefont {S.}~\bibnamefont
  {Aubry}}\ and\ \bibinfo {author} {\bibfnamefont {G.}~\bibnamefont
  {Andr\'e}},\ }\href@noop {} {\bibfield  {journal} {\bibinfo  {journal} {Ann.
  Israel Phys. Soc},\ }\textbf {\bibinfo {volume} {3}},\ \bibinfo {pages} {133}
  (\bibinfo {year} {1980})}\BibitemShut {NoStop}%
\bibitem [{\citenamefont {Billy}\ \emph {et~al.}(2008)\citenamefont {Billy}
  \emph {et~al.}}]{Billy08}%
  \BibitemOpen
  \bibfield  {author} {\bibinfo {author} {\bibfnamefont {J.}~\bibnamefont
  {Billy}} \emph {et~al.},\ }\Doi {10.1038/nature07000} {\bibfield  {journal}
  {\bibinfo  {journal} {Nature},\ }\textbf {\bibinfo {volume} {453}},\ \bibinfo
  {pages} {891} (\bibinfo {year} {2008})}\BibitemShut {NoStop}%
\bibitem [{\citenamefont {Roati}\ \emph {et~al.}(2008)\citenamefont {Roati}
  \emph {et~al.}}]{Roati08}%
  \BibitemOpen
  \bibfield  {author} {\bibinfo {author} {\bibfnamefont {G.}~\bibnamefont
  {Roati}} \emph {et~al.},\ }\Doi {10.1038/nature07071} {\bibfield  {journal}
  {\bibinfo  {journal} {Nature},\ }\textbf {\bibinfo {volume} {453}},\ \bibinfo
  {pages} {895} (\bibinfo {year} {2008})}\BibitemShut {NoStop}%
\bibitem [{\citenamefont {Modugno}(2009)}]{Modugno2009}%
  \BibitemOpen
  \bibfield  {author} {\bibinfo {author} {\bibfnamefont {M.}~\bibnamefont
  {Modugno}},\ }\href@noop {} {\bibfield  {journal} {\bibinfo  {journal} {New
  Journal of Physics},\ }\textbf {\bibinfo {volume} {11}},\ \bibinfo {pages}
  {033023} (\bibinfo {year} {2009})}\BibitemShut {NoStop}%
\bibitem [{\citenamefont {Biddle}\ \emph {et~al.}(2009)\citenamefont {Biddle},
  \citenamefont {Wang}, \citenamefont {D.~J.~Priour},\ and\ \citenamefont
  {Sarma}}]{biddle}%
  \BibitemOpen
  \bibfield  {author} {\bibinfo {author} {\bibfnamefont {J.}~\bibnamefont
  {Biddle}}, \bibinfo {author} {\bibfnamefont {B.}~\bibnamefont {Wang}},
  \bibinfo {author} {\bibfnamefont {J.}~\bibnamefont {D.~J.~Priour}}, \ and\
  \bibinfo {author} {\bibfnamefont {S.~D.}\ \bibnamefont {Sarma}},\ }\href@noop
  {} {\bibfield  {journal} {\bibinfo  {journal} {Phys. Rev. A},\ }\textbf
  {\bibinfo {volume} {80}},\ \bibinfo {eid} {021603(R)} (\bibinfo {year}
  {2009})}\BibitemShut {NoStop}%
\bibitem [{\citenamefont {Boers}\ \emph {et~al.}(2007)\citenamefont {Boers},
  \citenamefont {Goedeke}, \citenamefont {Hinrichs},\ and\ \citenamefont
  {Holthaus}}]{Boers07}%
  \BibitemOpen
  \bibfield  {author} {\bibinfo {author} {\bibfnamefont {D.~J.}\ \bibnamefont
  {Boers}}, \bibinfo {author} {\bibfnamefont {B.}~\bibnamefont {Goedeke}},
  \bibinfo {author} {\bibfnamefont {D.}~\bibnamefont {Hinrichs}}, \ and\
  \bibinfo {author} {\bibfnamefont {M.}~\bibnamefont {Holthaus}},\ }\Doi
  {10.1103/PhysRevA.75.063404} {\bibfield  {journal} {\bibinfo  {journal}
  {Phys. Rev. A},\ }\textbf {\bibinfo {volume} {75}},\ \bibinfo {eid} {063404}
  (\bibinfo {year} {2007})}\BibitemShut {NoStop}%
\bibitem [{\citenamefont {Riklund}\ \emph {et~al.}(1986)\citenamefont
  {Riklund}, \citenamefont {Liu}, \citenamefont {Wahlstrom},\ and\
  \citenamefont {Zhao-bo}}]{Riklund86}%
  \BibitemOpen
  \bibfield  {author} {\bibinfo {author} {\bibfnamefont {R.}~\bibnamefont
  {Riklund}}, \bibinfo {author} {\bibfnamefont {Y.}~\bibnamefont {Liu}},
  \bibinfo {author} {\bibfnamefont {G.}~\bibnamefont {Wahlstrom}}, \ and\
  \bibinfo {author} {\bibfnamefont {Z.}~\bibnamefont {Zhao-bo}},\ }\href@noop
  {} {\bibfield  {journal} {\bibinfo  {journal} {J. Phys. C: Solid State
  Phys.},\ }\textbf {\bibinfo {volume} {19}},\ \bibinfo {pages} {L705}
  (\bibinfo {year} {1986})}\BibitemShut {NoStop}%
\bibitem [{\citenamefont {de~Moura}\ \emph {et~al.}(2005)\citenamefont
  {de~Moura}, \citenamefont {Malyshev}, \citenamefont {Lyra}, \citenamefont
  {Malyshev},\ and\ \citenamefont {Dom\'inguez-Adame}}]{Moura}%
  \BibitemOpen
  \bibfield  {author} {\bibinfo {author} {\bibfnamefont {F.~A. B.~F.}\
  \bibnamefont {de~Moura}}, \bibinfo {author} {\bibfnamefont {A.~V.}\
  \bibnamefont {Malyshev}}, \bibinfo {author} {\bibfnamefont {M.~L.}\
  \bibnamefont {Lyra}}, \bibinfo {author} {\bibfnamefont {V.~A.}\ \bibnamefont
  {Malyshev}}, \ and\ \bibinfo {author} {\bibfnamefont {F.}~\bibnamefont
  {Dom\'inguez-Adame}},\ }\Doi {10.1103/PhysRevB.71.174203} {\bibfield
  {journal} {\bibinfo  {journal} {Phys. Rev. B},\ }\textbf {\bibinfo {volume}
  {71}},\ \bibinfo {pages} {174203} (\bibinfo {year} {2005})}\BibitemShut
  {NoStop}%
\bibitem [{\citenamefont {Malyshev}\ \emph {et~al.}(2004)\citenamefont
  {Malyshev}, \citenamefont {Malyshev},\ and\ \citenamefont
  {Dom\'inguez-Adame}}]{Malshev}%
  \BibitemOpen
  \bibfield  {author} {\bibinfo {author} {\bibfnamefont {A.~V.}\ \bibnamefont
  {Malyshev}}, \bibinfo {author} {\bibfnamefont {V.~A.}\ \bibnamefont
  {Malyshev}}, \ and\ \bibinfo {author} {\bibfnamefont {F.}~\bibnamefont
  {Dom\'inguez-Adame}},\ }\Doi {10.1103/PhysRevB.70.172202} {\bibfield
  {journal} {\bibinfo  {journal} {Phys. Rev. B},\ }\textbf {\bibinfo {volume}
  {70}},\ \bibinfo {pages} {172202} (\bibinfo {year} {2004})}\BibitemShut
  {NoStop}%
\bibitem [{\citenamefont {Xiong}\ and\ \citenamefont {Zhang}(2003)}]{Xiong}%
  \BibitemOpen
  \bibfield  {author} {\bibinfo {author} {\bibfnamefont {S.-J.}\ \bibnamefont
  {Xiong}}\ and\ \bibinfo {author} {\bibfnamefont {G.-P.}\ \bibnamefont
  {Zhang}},\ }\Doi {10.1103/PhysRevB.68.174201} {\bibfield  {journal} {\bibinfo
   {journal} {Phys. Rev. B},\ }\textbf {\bibinfo {volume} {68}},\ \bibinfo
  {pages} {174201} (\bibinfo {year} {2003})}\BibitemShut {NoStop}%
\bibitem [{\citenamefont {Rodr\'iguez}\ \emph {et~al.}(2003)\citenamefont
  {Rodr\'iguez} \emph {et~al.}}]{Rodriguez03}%
  \BibitemOpen
  \bibfield  {author} {\bibinfo {author} {\bibfnamefont {A.}~\bibnamefont
  {Rodr\'iguez}} \emph {et~al.},\ }\Doi {10.1103/PhysRevLett.90.027404}
  {\bibfield  {journal} {\bibinfo  {journal} {Phys. Rev. Lett.},\ }\textbf
  {\bibinfo {volume} {90}},\ \bibinfo {pages} {027404} (\bibinfo {year}
  {2003})}\BibitemShut {NoStop}%
\bibitem [{\citenamefont {Das~Sarma}\ \emph {et~al.}(1986)\citenamefont
  {Das~Sarma}, \citenamefont {Kobayashi},\ and\ \citenamefont
  {Prange}}]{DasSarma86}%
  \BibitemOpen
  \bibfield  {author} {\bibinfo {author} {\bibfnamefont {S.}~\bibnamefont
  {Das~Sarma}}, \bibinfo {author} {\bibfnamefont {A.}~\bibnamefont
  {Kobayashi}}, \ and\ \bibinfo {author} {\bibfnamefont {R.~E.}\ \bibnamefont
  {Prange}},\ }\Doi {10.1103/PhysRevLett.56.1280} {\bibfield  {journal}
  {\bibinfo  {journal} {Phys. Rev. Lett.},\ }\textbf {\bibinfo {volume} {56}},\
  \bibinfo {pages} {1280} (\bibinfo {year} {1986})}\BibitemShut {NoStop}%
\bibitem [{\citenamefont {Biddle}\ and\ \citenamefont
  {Das~Sarma}(2010)}]{biddle-2009}%
  \BibitemOpen
  \bibfield  {author} {\bibinfo {author} {\bibfnamefont {J.}~\bibnamefont
  {Biddle}}\ and\ \bibinfo {author} {\bibfnamefont {S.}~\bibnamefont
  {Das~Sarma}},\ }\Doi {10.1103/PhysRevLett.104.070601} {\bibfield  {journal}
  {\bibinfo  {journal} {Phys. Rev. Lett.},\ }\textbf {\bibinfo {volume}
  {104}},\ \bibinfo {pages} {070601} (\bibinfo {year} {2010})}\BibitemShut
  {NoStop}%
\bibitem [{\citenamefont {Soukoulis}\ and\ \citenamefont
  {Economou}(1982)}]{Soukoulis82}%
  \BibitemOpen
  \bibfield  {author} {\bibinfo {author} {\bibfnamefont {C.~M.}\ \bibnamefont
  {Soukoulis}}\ and\ \bibinfo {author} {\bibfnamefont {E.~N.}\ \bibnamefont
  {Economou}},\ }\href@noop {} {\bibfield  {journal} {\bibinfo  {journal}
  {Phys. Rev. Lett.},\ }\textbf {\bibinfo {volume} {48}},\ \bibinfo {pages}
  {1043} (\bibinfo {year} {1982})}\BibitemShut {NoStop}%
\bibitem [{\citenamefont {Das~Sarma}\ \emph {et~al.}(1988)\citenamefont
  {Das~Sarma}, \citenamefont {He},\ and\ \citenamefont {Xie}}]{DasSarma88}%
  \BibitemOpen
  \bibfield  {author} {\bibinfo {author} {\bibfnamefont {S.}~\bibnamefont
  {Das~Sarma}}, \bibinfo {author} {\bibfnamefont {S.}~\bibnamefont {He}}, \
  and\ \bibinfo {author} {\bibfnamefont {X.~C.}\ \bibnamefont {Xie}},\ }\Doi
  {10.1103/PhysRevLett.61.2144} {\bibfield  {journal} {\bibinfo  {journal}
  {Phys. Rev. Lett.},\ }\textbf {\bibinfo {volume} {61}},\ \bibinfo {pages}
  {2144} (\bibinfo {year} {1988})}\BibitemShut {NoStop}%
\bibitem [{\citenamefont {Xie}\ and\ \citenamefont
  {Das~Sarma}(1988)}]{XieDasSarma88}%
  \BibitemOpen
  \bibfield  {author} {\bibinfo {author} {\bibfnamefont {X.~C.}\ \bibnamefont
  {Xie}}\ and\ \bibinfo {author} {\bibfnamefont {S.}~\bibnamefont
  {Das~Sarma}},\ }\Doi {10.1103/PhysRevLett.60.1585} {\bibfield  {journal}
  {\bibinfo  {journal} {Phys. Rev. Lett.},\ }\textbf {\bibinfo {volume} {60}},\
  \bibinfo {pages} {1585} (\bibinfo {year} {1988})}\BibitemShut {NoStop}%
\bibitem [{\citenamefont {Scarola}\ and\ \citenamefont
  {Das~Sarma}(2006)}]{Scarola06}%
  \BibitemOpen
  \bibfield  {author} {\bibinfo {author} {\bibfnamefont {V.~W.}\ \bibnamefont
  {Scarola}}\ and\ \bibinfo {author} {\bibfnamefont {S.}~\bibnamefont
  {Das~Sarma}},\ }\Doi {10.1103/PhysRevA.73.041609} {\bibfield  {journal}
  {\bibinfo  {journal} {Phys. Rev. A},\ }\textbf {\bibinfo {volume} {73}},\
  \bibinfo {pages} {041609} (\bibinfo {year} {2006})}\BibitemShut {NoStop}%
\bibitem [{\citenamefont {Thouless}(1972)}]{Thouless72}%
  \BibitemOpen
  \bibfield  {author} {\bibinfo {author} {\bibfnamefont {D.~J.}\ \bibnamefont
  {Thouless}},\ }\href@noop {} {\bibfield  {journal} {\bibinfo  {journal}
  {Journal of Physics C: Solid State Physics},\ }\textbf {\bibinfo {volume}
  {5}},\ \bibinfo {pages} {77} (\bibinfo {year} {1972})}\BibitemShut {NoStop}%
\bibitem [{\citenamefont {Bellissard}\ and\ \citenamefont
  {Simon}(1982)}]{Bellissard}%
  \BibitemOpen
  \bibfield  {author} {\bibinfo {author} {\bibfnamefont {J.}~\bibnamefont
  {Bellissard}}\ and\ \bibinfo {author} {\bibfnamefont {B.}~\bibnamefont
  {Simon}},\ }\Doi {DOI: 10.1016/0022-1236(82)90094-5} {\bibfield  {journal}
  {\bibinfo  {journal} {Journal of Functional Analysis},\ }\textbf {\bibinfo
  {volume} {48}},\ \bibinfo {pages} {408 } (\bibinfo {year}
  {1982})}\BibitemShut {NoStop}%
\bibitem [{\citenamefont {Sokoloff}(1985)}]{Sokoloff85}%
  \BibitemOpen
  \bibfield  {author} {\bibinfo {author} {\bibfnamefont {J.~B.}\ \bibnamefont
  {Sokoloff}},\ }\Doi {DOI: 10.1016/0370-1573(85)90088-2} {\bibfield  {journal}
  {\bibinfo  {journal} {Physics Reports},\ }\textbf {\bibinfo {volume} {126}},\
  \bibinfo {pages} {189 } (\bibinfo {year} {1985})}\BibitemShut {NoStop}%
\bibitem [{\citenamefont {Harper}(1955)}]{Harper55}%
  \BibitemOpen
  \bibfield  {author} {\bibinfo {author} {\bibfnamefont {P.~G.}\ \bibnamefont
  {Harper}},\ }\href@noop {} {\bibfield  {journal} {\bibinfo  {journal}
  {Proceedings of the Physical Society. Section A},\ }\textbf {\bibinfo
  {volume} {68}},\ \bibinfo {pages} {874} (\bibinfo {year} {1955})}\BibitemShut
  {NoStop}%
\bibitem [{\citenamefont {Hofstadter}(1976)}]{Hofstadter76}%
  \BibitemOpen
  \bibfield  {author} {\bibinfo {author} {\bibfnamefont {D.~R.}\ \bibnamefont
  {Hofstadter}},\ }\Doi {10.1103/PhysRevB.14.2239} {\bibfield  {journal}
  {\bibinfo  {journal} {Phys. Rev. B},\ }\textbf {\bibinfo {volume} {14}},\
  \bibinfo {pages} {2239} (\bibinfo {year} {1976})}\BibitemShut {NoStop}%
\bibitem [{\citenamefont {Roux}\ \emph {et~al.}(2008)\citenamefont {Roux} \emph
  {et~al.}}]{Roux08}%
  \BibitemOpen
  \bibfield  {author} {\bibinfo {author} {\bibfnamefont {G.}~\bibnamefont
  {Roux}} \emph {et~al.},\ }\href@noop {} {\bibfield  {journal} {\bibinfo
  {journal} {Phys. Rev. A},\ }\textbf {\bibinfo {volume} {78}},\ \bibinfo
  {pages} {023628} (\bibinfo {year} {2008})}\BibitemShut {NoStop}%
\bibitem [{\citenamefont {Cai}\ \emph {et~al.}(2010)\citenamefont {Cai},
  \citenamefont {Chen},\ and\ \citenamefont {Wang}}]{Cai10}%
  \BibitemOpen
  \bibfield  {author} {\bibinfo {author} {\bibfnamefont {X.}~\bibnamefont
  {Cai}}, \bibinfo {author} {\bibfnamefont {S.}~\bibnamefont {Chen}}, \ and\
  \bibinfo {author} {\bibfnamefont {Y.}~\bibnamefont {Wang}},\ }\href@noop {}
  {\bibfield  {journal} {\bibinfo  {journal} {Phys. Rev. A},\ }\textbf
  {\bibinfo {volume} {81}},\ \bibinfo {pages} {023626} (\bibinfo {year}
  {2010})}\BibitemShut {NoStop}%
\bibitem [{\citenamefont {Fishman}\ \emph {et~al.}()\citenamefont {Fishman},
  \citenamefont {Krivolapov},\ and\ \citenamefont {Soffer}}]{Fishman2009}%
  \BibitemOpen
  \bibfield  {author} {\bibinfo {author} {\bibfnamefont {S.}~\bibnamefont
  {Fishman}}, \bibinfo {author} {\bibfnamefont {Y.}~\bibnamefont {Krivolapov}},
  \ and\ \bibinfo {author} {\bibfnamefont {A.}~\bibnamefont {Soffer}},\
  }\href@noop {} {}\Eprint {http://arxiv.org/abs/0901.4951} {arXiv:0901.4951}
  \BibitemShut {NoStop}%
\bibitem [{\citenamefont {{Krivolapov}}\ \emph {et~al.}()\citenamefont
  {{Krivolapov}}, \citenamefont {{Fishman}},\ and\ \citenamefont
  {{Soffer}}}]{Krivolapov2010}%
  \BibitemOpen
  \bibfield  {author} {\bibinfo {author} {\bibfnamefont {Y.}~\bibnamefont
  {{Krivolapov}}}, \bibinfo {author} {\bibfnamefont {S.}~\bibnamefont
  {{Fishman}}}, \ and\ \bibinfo {author} {\bibfnamefont {A.}~\bibnamefont
  {{Soffer}}},\ }\href@noop {} {}\Eprint {http://arxiv.org/abs/0912.3906}
  {arXiv:0912.3906} \BibitemShut {NoStop}%
\bibitem [{\citenamefont {Deissler}\ \emph {et~al.}(2010)\citenamefont
  {Deissler} \emph {et~al.}}]{Deissler2010}%
  \BibitemOpen
  \bibfield  {author} {\bibinfo {author} {\bibfnamefont {B.}~\bibnamefont
  {Deissler}} \emph {et~al.},\ }\href
  {http://www.scopus.com/inward/record.url?eid=2-s2.0-77951994758&partnerID=40%
&md5=1db2f04fb50775cd5ef38c5b1df573b5} {\bibfield  {journal} {\bibinfo
  {journal} {Nature Physics},\ }\textbf {\bibinfo {volume} {6}},\ \bibinfo
  {pages} {354} (\bibinfo {year} {2010})}\BibitemShut {NoStop}%
\bibitem [{\citenamefont {Larcher}\ \emph {et~al.}(2009)\citenamefont
  {Larcher}, \citenamefont {Dalfovo},\ and\ \citenamefont
  {Modugno}}]{Larcher2009}%
  \BibitemOpen
  \bibfield  {author} {\bibinfo {author} {\bibfnamefont {M.}~\bibnamefont
  {Larcher}}, \bibinfo {author} {\bibfnamefont {F.}~\bibnamefont {Dalfovo}}, \
  and\ \bibinfo {author} {\bibfnamefont {M.}~\bibnamefont {Modugno}},\ }\Doi
  {10.1103/PhysRevA.80.053606} {\bibfield  {journal} {\bibinfo  {journal}
  {Phys. Rev. A},\ }\textbf {\bibinfo {volume} {80}},\ \bibinfo {pages}
  {053606} (\bibinfo {year} {2009})}\BibitemShut {NoStop}%
\bibitem [{\citenamefont {Kotlyar}\ and\ \citenamefont
  {Das~Sarma}(2001)}]{Kotlyar2001}%
  \BibitemOpen
  \bibfield  {author} {\bibinfo {author} {\bibfnamefont {R.}~\bibnamefont
  {Kotlyar}}\ and\ \bibinfo {author} {\bibfnamefont {S.}~\bibnamefont
  {Das~Sarma}},\ }\Doi {10.1103/PhysRevLett.86.2388} {\bibfield  {journal}
  {\bibinfo  {journal} {Phys. Rev. Lett.},\ }\textbf {\bibinfo {volume} {86}},\
  \bibinfo {pages} {2388} (\bibinfo {year} {2001})}\BibitemShut {NoStop}%
\bibitem [{\citenamefont {Giamarchi}\ and\ \citenamefont
  {Schulz}(1988)}]{Giamarchi1988}%
  \BibitemOpen
  \bibfield  {author} {\bibinfo {author} {\bibfnamefont {T.}~\bibnamefont
  {Giamarchi}}\ and\ \bibinfo {author} {\bibfnamefont {H.~J.}\ \bibnamefont
  {Schulz}},\ }\Doi {10.1103/PhysRevB.37.325} {\bibfield  {journal} {\bibinfo
  {journal} {Phys. Rev. B},\ }\textbf {\bibinfo {volume} {37}},\ \bibinfo
  {pages} {325} (\bibinfo {year} {1988})}\BibitemShut {NoStop}%
\bibitem [{\citenamefont {Albert}\ and\ \citenamefont
  {Leboeuf}(2010)}]{Albert2010}%
  \BibitemOpen
  \bibfield  {author} {\bibinfo {author} {\bibfnamefont {M.}~\bibnamefont
  {Albert}}\ and\ \bibinfo {author} {\bibfnamefont {P.}~\bibnamefont
  {Leboeuf}},\ }\Doi {10.1103/PhysRevA.81.013614} {\bibfield  {journal}
  {\bibinfo  {journal} {Phys. Rev. A},\ }\textbf {\bibinfo {volume} {81}},\
  \bibinfo {pages} {013614} (\bibinfo {year} {2010})}\BibitemShut {NoStop}%
\end{thebibliography}
%

\end{document}